\documentclass[reqno,11pt]{amsart}

\usepackage{fancyhdr,bbm}
\usepackage{datetime}
\settimeformat{ampmtime}

\usepackage[table,x11names,svgnames]{xcolor}
\usepackage{setspace,palatino}
\usepackage{pgf}
\usepackage{tikz}
\usepackage[scaled]{helvet} 
\usepackage{courier} 
\normalfont
\usepackage[T1]{fontenc}
\usepackage{booktabs}
\usepackage{arydshln}

\usetikzlibrary{patterns}
\usepackage{soul}
\usepackage{mathrsfs}
\usepackage{stmaryrd}
\usepackage[hidelinks=true]{hyperref}
\usepackage{natbib}
\usepackage{xfrac}
\usepackage[color=Orange]{todonotes}
\usepackage{layout}
\usepackage{microtype}
\usepackage{mathtools}
\usepackage{multirow}
\usepackage{floatrow}
\usepackage{subfig}
\usepackage[multiple]{footmisc}

\usepackage{color}

\definecolor{gainsboro}{RGB}{220,220,220}

\newtheorem{lemma}{Lemma}

\newtheorem{theorem}{Theorem}

\usepackage{cleveref}

\crefname{figure}{figure}{figures}
\creflabelformat{figure}{#2#1#3}
\crefname{equation}{equation}{equations}
\creflabelformat{equation}{#2(#1)#3}
\crefname{lemma}{lemma}{lemmas}
\creflabelformat{lemma}{#2#1#3}
\crefname{theorem}{theorem}{theorems}
\creflabelformat{lemma}{#2#1#3}
\crefname{condition}{condition}{conditions}
\creflabelformat{condition}{#2#1#3}
\crefname{assumption}{assumption}{assumptions}
\creflabelformat{assumption}{#2#1#3}
\crefname{appendix}{appendix}{appendices}
\creflabelformat{appendix}{#2#1#3}
\crefname{enumi}{}{}
\creflabelformat{enumi}{(#2#1#3)}

\newtheorem{assumption}{Assumption}

\usepackage{amssymb}

\renewcommand{\Pr}{\mathbb{P}}

\newcommand\abstraction{This paper considers a treatment effects model in which individual treatment effects may be heterogeneous, even among observationally identical individuals. Specifically, by extending the classical instrumental-variables (IV) model with an endogenous binary treatment, the heteroskedasticity of the error disturbance is allowed to depend upon the treatment variable so that treatment generates both mean and variance effects on the outcome. In this endogenous heteroskedasticity IV (EHIV) model, the standard IV estimator can be inconsistent for the average treatment effects (ATE) and lead to incorrect inference. After nonparametric identification is established, closed-form estimators are provided for the linear EHIV of the mean and variance treatment effects, the average treatment effect on the treated (ATT), and the full distribution of the individual treatment effects (ITE). Asymptotic properties of the estimators are derived. A Monte Carlo simulation investigates the performance of the proposed approach. An empirical application regarding the effects of fertility on female labor supply is considered, and the findings demonstrate the importance of accounting for endogenous heteroskedasticity.

\vspace{16pt}

\noindent
\textbf{Keywords:} Endogenous heteroskedasticity,  individual treatment effects, average treatment effects, local average treatment effects, instrumental variable\\

}

\begin{document}

\thispagestyle{empty}

\title{Estimation of treatment effects \\ under endogenous heteroskedasticity*}
\thanks{$^*$We thank Daniel Ackerberg, Sandra Black, Ivan Canay, Salvador Navarro, Max Stinchcombe, Quang Vuong, and Ed Vytlacil for useful comments. We also thank seminar participants at University of Iowa, University of Hong Kong, McMaster University, Western University, University of Texas at Austin, Xiamen University, Monash University, University of Melbourne, USC, the 2017 Shanghai workshop of econometrics at SUFE,  the 2018 Texas Econometrics Camp, and the 2018 CEME conference at Duke University.
}

\author{
\href{mailto:abrevaya@austin.utexas.edu}{Jason Abrevaya$^\dag$}}
\thanks{$^\dag$Department of Economics, University of Texas at Austin, Austin, TX, 78712,
\href{mailto:abrevaya@austin.utexas.edu}{abrevaya@austin.utexas.edu}}
\author{
\href{mailto:h.xu@austin.utexas.edu}{Haiqing Xu$^{\ddag}$}}
\thanks{$^{\ddag}$Department of Economics, University of Texas at Austin, Austin, TX, 78712, \href{mailto:h.xu@austin.utexas.edu}{h.xu@austin.utexas.edu}}

\date{\today}

\maketitle

\begin{abstract}
 \abstraction
\end{abstract}

\vspace{5ex}

\clearpage


\section{Introduction}

The empirical literature on program evaluation limits its scope almost exclusively to models where treatment effects are homogenous for observationally identical individuals. When treatment effects are heterogeneous among observationally identical individuals, the causal inference required for policy evaluation is considerably more difficult \citep[see e.g.][]{heckman2005structural}. In practice, researchers adopt the linear IV approach by switching their object of interest from  the population-level treatment effect to \cite{imbens1994identification}'s local average treatment effect (LATE), a concept that relies upon the monotonicity condition of the selection into treatment and also the choice of instrumental variable. If the population-level treatment effect (i.e., the average treatment effect (ATE)) is essential to understand the driving mechanism behind a particular program, the standard instrumental-variables (IV) approach can lead to inconsistency and incorrect inference.



In this paper, we propose a model  that allows for heterogeneous treatment effects by extending the classical  IV model to include both mean and variance effects rather than just mean effects:
\begin{equation}
\label{eq1}
Y=\mu(D,X)+\sigma(D,X)\times  \epsilon,
\end{equation}where $Y\in\mathbb R$ is the outcome variable of interest, $X\in\mathbb R$ is a vector of observed covariates, $D\in\{0,1\}$ denotes the binary treatment status, and $\epsilon\in\mathbb R$ is the model disturbance. Under an additional normalization assumption that $\epsilon$ has zero mean and unit variance (given $X$), the structural functions $\mu(\cdot,X)$ and $\sigma(\cdot,X)$ are the mean and standard deviation of the (potential) outcome, respectively, under different treatment statuses.  Hence, $\mu(1,X)-\mu(0,X)$ and $\sigma(1,X)-\sigma(0,X)$ measure the \textit{mean effects} and \textit{``variance'' effects} of the treatment, respectively. 

Our model  parsimoniously introduces heterogeneous   treatment effects across the population. The fact that the heteroskedasticity term $\sigma(\cdot,\cdot)$ depends on the endogenous treatment $D$ implies that treatment effects can differ across individuals even after  $X$ has been controlled for. As such, we say that model \eqref{eq1} exhibits {\it endogenous heteroskedasticity}, and we will call our instrumental-variables method the {\it endogenous heteroskedasticity IV} (or EHIV) approach.  As emphasized in \cite{heckman2005structural}, the absence of heterogeneous responses to treatment implies that different treatment effects collapse to the same parameter. If $\sigma(D,X)$ depends upon $D$ in \eqref{eq1}, however, heterogeneous treatment effects arise in general, and we show that the standard IV approach is generally inconsistent for estimating the mean effects  in the presence of endogenous heteroskedasticity.

On the other hand, if the heteroskedasticity is exogenous, the treatment effects are homogeneous across individuals (after covariates have been controlled for), which can be consistently estimated by the standard IV approach. Therefore, to apply the IV method for the mean effects of the  treatment, the exogeneity of heteroskedasticity serves as a key assumption, which should be justified from economic theory and/or statistical tests. By using a regression of squared IV estimated residuals on covariates as well as the treatment status, instrumented by the same instrumental variable, one can easily test the null hypothesis of exogenous heteroskedasticity (or equivalently, the homogeneous treatment effects hypothesis). If the heteroskedasticity is not exogenous, the standard IV estimator becomes a mixture of the mean and variance effects, interpreted as LATE under \cite{imbens1994identification}'s monotonicity condition. As a matter of fact, our model nests the classical IV model with exogenous heteroskedasticity as a special case. 

This paper builds upon several strands in the existing literature.  The literature on heterogeneous treatment effects \citep[e.g.][among many others]{imbens1994identification,heckman1997making,heckman2005structural} is an important antecedent. Within the LATE context, Abadie (2002, 2003) has considered the estimation of the variance and the distribution of treatment effects, but the causal interpretation is limited to compliers. The main difference of our approach from that literature is that we consider additional assumptions on the structural outcome model rather than additional assumptions on a selection equation and/or variation of the instrumental variable. Our approach does not restrict causal interpretation to compliers. As far as we know, the only other paper that explicitly considers a structural treatment-effect model with endogenous heteroskedasticity is \cite{chen2014semi}. Under the monotone selection assumption, \cite{chen2014semi} focus on identification and estimation of the ratio of the heteroskedasticity term under different treatment statuses, i.e., $\sigma(1,x)/\sigma(0,x)$.

Another important related literature concerns the identification and estimation of nonseparable models with binary endogeneity \citep[e.g.][among many others]{chesher2005nonparametric,chernozhukov2005iv,jun2011tighter}. In particular, \cite{chernozhukov2005iv} establish nonparametric (local and global) identification of quantile treatment effects under a rank condition. Extending \cite{chernozhukov2005iv}'s results, \cite{vuong2014counterfactual} develop a constructive identification strategy for the nonseparable structural model by assuming  monotonicity of the selection. This paper also derives closed-form identification for the mean and variance effects of the treatment, but the additional assumptions on the structural outcome equation lead to an estimation strategy that should be considerably simpler for practitioners to use.

While identification does not require additional parametric specification of $\mu(D,X)$, we take a semiparametric approach to estimation that imposes linearity of $\mu(D,X)$, in line with nearly all empirical work, and leaves $\sigma(D,X)$ unspecified. This specification allows for heterogeneous individual treatment effects, but it is quite tractable in the sense that the heterogenous individual treatment effects can be decomposed into mean and variance effects. 
On the other hand,  nonparametric estimation of fully nonseparable models is challenging. See e.g. \cite{cherno2004401kplan,chernozhukov2005iv} and \cite{vuong2015nonparametric}, who develop nonparametric estimation of quantiles and density functions of individual treatment effects, respectively, in fully nonseparable frameworks.

The structure of the paper is organized as follows. Section 2 formally introduces the notation and assumptions underlying the endogenous heteroskedasticity model in \eqref{eq1}, focusing on the case of a binary instrumental variable. Section 3 provides a constructive approach to nonparametric identification of the mean and variance functions in \eqref{eq1}. Section 4 considers a semiparametric version of \eqref{eq1} in which the mean function is a linear index of $X$ and $D$. An estimator (the EHIV estimator) of the coefficient parameters is proposed, and its asymptotic properties ($\sqrt{n}$-consistency and asymptotic normality) are established. Combining this estimator with a nonparametric estimator of the heteroskedasticity function $\sigma(\cdot,\cdot)$ allows us to consistently estimate the (conditional) distribution of the heterogeneous treatment effects. Section 5 provides Monte Carlo evidence to illustrate the performance of the proposed estimator. Section 6 applies the approach to an empirical application, where the effects of having a third child on female labor supply are estimated \citep[as previously considered by][]{angrist1998children}. Section 7 concludes. Proofs are collected in the Appendix.

\section{Assumptions and Motivation}

To deal with the endogeneity of treatment status, we consider the canonical case in which a binary instrumental variable $Z\in\{0,1\}$ exists. The case of binary-valued instruments has been emphasized in the treatment effect literature, in particularly in the applications using natural and social experiments. For each $(x,z)\in\mathscr S_{XZ}$, let $p(x,z)=\mathbb P(D=1|X=x,Z=z)$ denote the propensity score.  The following assumptions are maintained throughout the paper. 
\begin{assumption}
\label{ass1}
(Normalization) Let  $\mathbb E (\epsilon|X)=0$ and $\mathbb E(\epsilon^2|X)=1$.
\end{assumption}

\begin{assumption}
\label{ass2}
(i) (Instrument relevance) For every $x\in\mathscr S_X$, $\mathscr S_{Z|X=x}=\{0,1\}$ and  $p(x,0)\neq p(x,1)$; (ii) (Instrument exogeneity) $\mathbb E (\epsilon|X,Z)=\mathbb E (\epsilon|X)$ and $\text{Var} (\epsilon|X,Z)=\text{Var} (\epsilon|X)$.
\end{assumption}
\noindent
\noindent 
\Cref{ass1} is a normalization on the first two moments of the error term $\epsilon$. Clearly, the scale normalization on $\mathbb E(\epsilon^2|X)$ is indispensable  for identification of $\sigma(\cdot,\cdot)$. \Cref{ass2} contains the instrument relevance and instrument exogeneity conditions. In particular, (ii) is implied by the conditional independence of $Z$ and $\epsilon$ given $X$, i.e., $Z\bot \epsilon|X$, which is usually motivated by the choice of the instrumental variable \citep[see e.g.][]{angrist1991does}.  Combining \Cref{ass1,ass2}(ii), we have $\mathbb E (\epsilon|X,Z)=0$ and $\text{Var} (\epsilon|X,Z)=1$.  For expositional simplicity, we will assume throughout the paper that $p(x,0)<p(x,1)$ for all $x\in\mathscr S_X$.

Motivated by the fully nonseparable model approach \citep[see e.g.][]{chesher2005nonparametric,chernozhukov2005iv}, our model \eqref{eq1} parsimoniously introduces heterogeneous treatment effects across individuals.  In particular, model parameters $\mu(\cdot,\cdot)$ and $\sigma(\cdot,\cdot)$, respectively, capture the mean and variance effects of the treatment. Therefore, individual treatment effects can be written as
\[
\mu(1,X)-\mu(0,X)+[\sigma(1,X)-\sigma(0,X)]\times \epsilon,
\] which varies across individuals even with the same value of covariates $X$.  Such a semi-nonseparable specification makes our model tractable for estimation and inference. 

With non-degenerate variance effects,  the standard IV estimator is generally inconsistent for estimating the model parameter $\mu$. In particular, a closed-form expression for the bias of the IV estimator can be derived under our model specification. For expositional simplicity, the covariates $X$ are suppressed in the following discussion. Under \Cref{ass2}(i), define the quantities $r_0$ and $r_1$ as follows
\begin{align*}
&r_0= \mu(0)+[\sigma(1)-\sigma(0)]\times\frac{\mathbb E (D\epsilon|Z=0)p(1)-\mathbb E (D\epsilon|Z=1)p(0)}{p(1)-p(0)},\\
&r_1= \mu(1)-\mu(0)+[\sigma(1)-\sigma(0)]\times \frac{\mathbb E (D\epsilon|Z=1)-\mathbb E \left(D\epsilon|Z=0\right)}{p(1)-p(0)}.
\end{align*}
Then, model \eqref{eq1} can be represented by the following linear IV projection:
\[
Y=r_0+r_1D+ \tilde \epsilon,
\]where  $\tilde \epsilon\equiv \mu(D)+\sigma(D)\epsilon-r_0-r_1D$. By definition, $\tilde \epsilon$ measures the discrepancy between the structural model  and its linear IV projection, which satisfies $\mathbb E (\tilde \epsilon|Z)=0$ under \Cref{ass1,ass2}.  Therefore, the standard IV regression would estimate the coefficient $r_1$, which is a linear mixture of the mean effect, $\mu(1)-\mu(0)$, and the variance effect, $\sigma(1)-\sigma(0)$.

%
%
%

The seminal paper by \cite{imbens1994identification} show that the coefficient $r_1$ from the above linear IV projection has a LATE interpretation. Specifically, suppose that the selection to treatment satisfies the monotonicity condition, e.g., 
\begin{equation}
\label{selection}
D= \mathbbm 1 [\eta\leq m(Z)],
\end{equation} where  $\eta\in\mathbb R$ is a scalar-valued latent variable and $m(\cdot)$ is a real-valued function with $m(0)<m(1)$.\footnote{See \cite{vytlacil2002independence} for a proof of the observational equivalence between \eqref{selection} and the monotone selection condition.}  Under this selection assumption, the LATE can be written as
\[
r_1=\mu(1)-\mu(0)+[\sigma(1)-\sigma(0)]\times\mathbb E [\epsilon|m(0)<\eta\leq m(1)].
\]  The bias term of the LATE, i.e. $[\sigma(1)-\sigma(0)]\times\mathbb E [\epsilon|m(0)<\eta\leq m(1)]$, depends on the degree to which heteroskedasticity depends upon treatment, as well as  the average error disturbance for the compliers.

When treatment effects are homogeneous after a vector of covariates $X$ has been controlled for, i.e. the heteroskedasticity is exogenous,  the ATE can be estimated by the LATE. Therefore, it can be worthwhile to test the homogeneous treatment effects hypothesis via testing for exogenous heteroskedasticity. Since the IV estimator consistently estimates homogeneous treatment effects under the null hypothesis, a direct test can be conducted by determining whether the squared IV estimated residuals depend upon the instrumental variable $Z$ or not. One could simply apply e.g. \cite{fan1996consistent} for testing such a hypothesis. Although the IV estimator may be inconsistent under the alternative hypothesis, we show in Section~7 that such a test is surprisingly consistent.




%

\section{Nonparametric Identification}
In this section, we provide a constructive identification  that involves two steps. First, we identify $\sigma(\cdot,X)$ up-to-scale. Second, we transform \eqref{eq1}  into a model with exogenous heteroskedasticity, from which both $\mu(\cdot,\cdot)$ and $\sigma(\cdot,\cdot)$ are identified.

Some additional notation is required. For $d=0,1$, let
\begin{align}
\label{eq_delta}
&\delta_d(X)= \frac{\mathbb E [Y   \times  \mathbbm 1(D=d)|X,Z=1]-\mathbb E [Y  \times  \mathbbm 1(D=d)|X, Z=0]}{\mathbbm P(D=d|X,Z=1)-\mathbbm P(D=d|X,Z=0)};\\
\label{eq_v}
&V_{d}(X)   = \frac{\mathbb E [Y^2 \times \mathbbm 1(D=d)|X,Z=1]-\mathbb E [Y^2\times \mathbbm 1(D=d)|X, Z=0]}{\mathbbm P(D=d|X,Z=1)-\mathbbm P(D=d|X,Z=0)}-\delta^2_d(X).
\end{align}
Under \Cref{ass2}(i),  both $\delta_d(X)$ and $V_d(X)$ are well defined. Similarly to \cite{imbens1994identification}, $\delta_d(X)$ and $V_d(X)$ can be written in terms of covariances of the observables:
\begin{align*}
&\delta_d(X)=\frac{\text{Cov}\big(Y \times \mathbbm 1 (D=d),Z|X\big)}{\text{Cov}(\mathbbm 1 (D=d),Z|X)};\\
&V_d(X)=\frac{\text{Cov}\big(Y^2 \times \mathbbm 1 (D=d),Z|X\big)}{\text{Cov}(\mathbbm 1 (D=d),Z|X)}-\delta^2_d(X).
\end{align*}
Note that both $\delta(\cdot)$ and $V_d(\cdot)$ are identified from the data.

Moreover, for $\ell=1,2$, denote 
\[
\xi_\ell(x)= \frac{\mathbb E (\epsilon^\ell \times D|X=x,Z=1) -\mathbb E (\epsilon^\ell \times D|X=x, Z=0)}{p(x,1)-p(x,0)}.
\] By definition, $\xi_\ell(x)$ depends on the (unknown) distribution of $F_{\epsilon D|XZ}$. Then, model \eqref{eq1} and \Cref{ass1} imply
\begin{align*}
&\delta_d(X)= \mu(d,X)+\sigma(d,X)\times \xi_1(X),\\
&V_d(X)=\sigma^2(d,X) \times \left[\xi_2(X)-\xi^2_1(X)\right].
\end{align*}
Let $C(X)=\xi_2(X)-\xi^2_1(X)$.  Thus, the vector $(V_0(X),V_1(X))'$ identifies the heterogeneity component $\sigma(\cdot,X)$ up to the scale  $C(X)$.  The above discussion is summarized by the following lemma. 
\begin{lemma}
\label{lemma0}
Suppose \Cref{ass1,ass2} hold.  Then 
\[
V_d(X)=\sigma^2(d,X) \times C(X), \ \ \text{for}\  d=0,1.
\]
\end{lemma}
\noindent 
\noindent 
\Cref{lemma0} implies that $\text{sign}(V_0(X))=\text{sign}(V_1(X))$, which is a testable model restriction. As a matter of fact,   \Cref{lemma0} provides a basis for the identification of our model. Before proceeding, however, an assumption ruling out zero-valued variances is needed:
 \begin{assumption}
\label{rank}
$C(X)\neq 0$ almost surely.
\end{assumption}
\noindent 
\Cref{rank} is verifiable since $C(X)\neq 0$ if and only if $V_d(X)\neq 0$.  Moreover, note that if \eqref{selection} holds,  $C(X)$ is interpreted as the (conditional) variance of $\epsilon$ given $X$ and the ``complier group''. In this case, $C(X)>0$ if and only if the (conditional) distribution of $\epsilon$ is non-degenerate. 

Model \eqref{eq1} can now be transformed to deal with the issue of endogenous heteroskedasticity. Defining
\[
S=|V_0(X)|^{\frac{1}{2}} \times (1-D)+|V_1(X)|^{\frac{1}{2}} \times D,
\] one can show  that  $S= \sigma(D,X)\times |C(X)|^{\frac{1}{2}}$ by \Cref{lemma0}. Dividing the original model \eqref{eq1} by $S$ yields the transformed model
\begin{equation}
\label{trans_model}
\frac{Y}{S}= \frac{\mu(D,X)}{S} +\frac{\epsilon}{|C(X)|^{\frac{1}{2}}},
\end{equation}
for which $Z$ satisfies the instrument exogeneity condition with the (transformed) error disturbance ${\epsilon}/{|C(X)|^{\frac{1}{2}}}$.

Closed-form expressions for $\mu(\cdot,x)$ and $\sigma(\cdot,x)$ are now provided. Fixing $x\in\mathscr S_{X}$, note that
\[
\mathbb E \Big(\frac{Y}{S}\Big|X=x,Z=z\Big)= \frac{\mu(1,x)}{|V_1(x)|^{\frac{1}{2}}}\times p(x,z)+\frac{\mu(0,x)}{|V_0(x)|^{\frac{1}{2}}}\times [1-p(x,z)], \text{ for }  \ z=0,1,
\]which is a linear equation system in  $\mu(0,x)$ and $\mu(1,x)$.
\Cref{ass2} implies
\begin{align}
\label{mu1}
&\mu(1,x)=\frac{\mathbb E\big(\frac{Y}{S}\big|X=x,Z=1\big)[1-p(x,0)]-\mathbb E\big(\frac{Y}{S}\big|X=x,Z=0\big)[1-p(x,1)]}{p(x,1)-p(x,0)}\times|V_1(x)|^{\frac{1}{2}};\\
\label{mu0}
&\mu(0,x)=\frac{\mathbb E\big(\frac{Y}{S}\big|X=x,Z=1\big)p(x,0)-\mathbb E\big(\frac{Y}{S}\big|X=x,Z=0\big)p(x,1)}{p(x,0)-p(x,1)}\times |V_0(x)|^{\frac{1}{2}}.
\end{align}
Moreover, it is straightforward to show that
\[
\sigma^2(d,x)=|V_d(x)|\times \mathbb E\left\{\left[\frac{Y-\mu(D,X)}{S}\right]^2\big|X=x\right\}.
\]which can be equivalently rewritten as
\begin{align*}
&\sigma^2(d,x) 
= \left|\frac{V_d(x)}{V_{1}(x)}\right|\times \mathbb E\big[ D(Y-\mu(D,X))^2\big|X=x\big]+  \left|\frac{V_d(x)}{V_{0}(x)}\right|\times \mathbb E\big[ (1-D)(Y-\mu(D,X))^2\big|X=x\big].
\end{align*}

It should also be noted that one could further obtain identification of the \textit{average treatment effect on the treated} \citep[ATT, see e.g.][]{heckman2005structural}.  Specifically, 
\begin{eqnarray*}
\text{ATT}&=& \mathbb E\left[\mu(1,X)-\mu(0,X)|D=1\right]+\mathbb E\big\{[\sigma(1,X)-\sigma(0,X)]\times\epsilon |D=1\big\}\\
&=&\mathbb E\left[\mu(1,X)-\mu(0,X)|D=1\right]+\mathbb E \left\{\Big[1-\frac{|V_0(X)|^{\frac{1}{2}}}{|V_1(X)|^{\frac{1}{2}}}\Big]\times \mathbb E [Y-\mu(1,X)|X,D=1]\right\}.
\end{eqnarray*}

Interestingly, once $\mu(\cdot,\cdot)$ and $\sigma(\cdot,\cdot)$ have been identified, \cite{vuong2014counterfactual}'s {\it counterfactual mapping} approach can be used to identify counterfactual outcomes for each individual. Let $Y_d\equiv \mu(d,X)+\sigma(d,X)\times\epsilon$ be the ``potential outcome'' under the treatment status $d$. By definition, $Y_d$ is observed in the data if and only if $D=d$. The endogeneity issue arises due to the missing observations of $Y_{1-d}$ when $D=d$. Given model \eqref{eq1}, the unobserved potential outcomes (counterfactuals) can be explicitly constructed by the distribution of the observables: Suppose w.l.o.g. $D=1$. Then, $Y_1=Y$, and by \Cref{lemma0}, 
\[
Y_{0}=\mu(0,X)+[Y-\mu(1,X)]\times   \frac{\sigma(0,X)}{\sigma(1,X)}=\delta_{0}(X)+[Y-\delta_1(X)]\times \frac{|V_0(X)|^{\frac{1}{2}}}{|V_1(X)|^{\frac{1}{2}}},
\]which is constructively identified from the data. This also suggests an alternative expression for ATT:
\[
\text{ATT}=\mathbb E \left\{Y-\delta_{0}(X)-[Y-\delta_1(X)]\times \frac{|V_0(X)|^{\frac{1}{2}}}{|V_1(X)|^{\frac{1}{2}}}\Big|D=1\right\}.
\]

\subsection{Interpretations under monotone selection and misspecification}
If the linear  outcome equation is misspecified,  \cite{imbens1994identification} points out that the usual IV estimator should be interpreted as LATE (under an additional monotone selection assumption) rather than ATE. Though our model is less restrictive, it is still useful to interpret the EHIV estimators when the underlying structure for the data generating process is  fully nonseparable. 

Specifically, suppose the outcome equation is given as follows:
\[
Y=h(D,X,\epsilon)
\] where $h$ is nonseparable in the error term $\epsilon$, and in addition  equation \eqref{selection} holds with $m(X,0)<m(X,1)$. First, we argue that $V_d(X)$ can be interpreted as the (conditional) variance of  the corresponding potential outcome given the ``compliers group''. To fix ideas, define  
\[
\text{Complier}(X)\equiv \{\eta\in\mathbb R: m(X,0)<\eta\leq m(X,1)\}
\] as the group of compliers  who switch their treatment participation decision with the realization of $Z$. Specifically, a complier chooses $D=0$ if and only if $Z=0$. Moreover, define
\begin{align*}
&\text{Always-Taker}(X)\equiv \{\eta\in\mathbb R: \eta\leq m(X,0)\};\\
&\text{Never-Taker}(X)\equiv \{\eta\in\mathbb R: \eta> m(X,1)\},
\end{align*} as the group of individuals who always participate in the treatment and the group of individuals who never participate, respectively, regardless the realization of $Z$; see \cite{imbens1994identification} for a detailed discussion on these three groups. By a similar argument to \cite{imbens1994identification}, one can show that $\delta_d(X)$ can be interpreted as the (conditional) mean of the potential outcome $Y_d$ given $X$ and the group of compliers: 
\[
\delta_d(X)=\mathbb E(Y_d|X, \text{Complier}(X)).
\] In addition,  $V_d(X)$ is the (conditional) variance of potential outcome $Y_d$ given $X$ and the group of compliers: 
\[
V_d(X)=\text{Var}(Y_d|X, \text{Complier}(X)).
\] 
It is worth pointing out that such a ``local variance'' interpretation does not depend on the functional form specification in model \eqref{eq1}.

Furthermore,  denote $R(X)=\sqrt{V_0(X)/V_1(X)}$.  Let further $Q_1(X)=1-p(X,0)+R(X)p(X,0)$ and $Q_0(X)=p(X,1)+R^{-1}(X)[1-p(X,1)]$. By definition, $Q_1(X)=R(X)Q_0(X)+[1-R(X)]\times[p(X,1)-p(X,0)]$, and both $Q_0(X)$ and $Q_1(X)$ are positive. Using eqs. \eqref{mu1} and \eqref{mu0}, we have
\begin{eqnarray*}
&&\mu(1,X)-\mu(0,X)\\
&=&\mathbb E [h(1,X,\epsilon)|X, \text{Complier}(X)]\times Q_1(X)+\mathbb E [h(1,X,\epsilon)|X, \text{Always-Taker}(X)]\times [1-Q_1(X)]\\
&-& \mathbb E [h(0,X,\epsilon)|X, \text{Complier}(X)]\times Q_0(X)- \mathbb E [h(0,X,\epsilon)|X, \text{Never-Taker}(X)]\times [1-Q_0(X)],
\end{eqnarray*}
which we call the ``adjusted'' LATE if model \eqref{eq1} is indeed misspecified.  Note that the LATE uses  information contained only in the complier group. The ``adjusted'' LATE, however, depends upon information contained in all three groups. Moreover, if $V_0(X)=V_1(X)$, i.e. the case of exogenous heteroskedasticity, we have $Q_0(X)=Q_1(X)=1$, then  $\mu(1,X)-\mu(0,X)$ becomes the (conditional) LATE. Alternatively, suppose $p(X,0)=0$ and $p(X,1)=1$. Then we also have $Q_0(X)=Q_1(X)=1$. Our ``adjusted'' LATE  extrapolates information from the three groups to the whole population, depending on the relative variance of potential outcomes in the complier groups as well as the probability masses of the three groups. It should also be noted that under misspecification, our model can provide a ``better'' approximation to the underlying data generating structure than the standard IV model with exogenous heteroskedasticity since the latter is nested in our model. 
\section{Semiparametric Estimation}
For ease of implementation and in line with empirical practice, a linear specification for the $\mu(\cdot,\cdot)$ is considered here. Specifically, the following model with $\mu(D,X)=X'\beta_1 + \beta_2D$ is considered:
\begin{equation}
\label{parametric model}
Y=X'\beta_1+ \beta_2 D + \sigma(D,X) \times  \epsilon
\end{equation}where  $\beta_1\in\mathbb R^{d_X}$ and $\beta_2\in\mathbb R$. Such a specification is parsimonious, with the average treatment effects measured by the scalar parameter $\beta_2$. 
This semiparametric model is a natural extension of the standard linear IV model with (exogenous) heteroskedasticity. While it is possible to estimate $\mu(\cdot,\cdot)$ in model \eqref{eq1} nonparametrically, such an approach would suffer from the curse of dimensionality.

 For notational simplicity, let $W=(X',Z)'\in\mathbb R^{d_X}\times\{0,1\}$ and $\beta=(\beta_1',\beta_2)'\in\mathbb R^{d_X+1}$.  Let  $\{(Y_i,D_i,W'_i)': i\leq n\}$ be an i.i.d. random sample  of $(Y,D,W')'$ generated from \eqref{parametric model}, where $n\in\mathbb N$ is the sample size.  To simplify the theoretical development, all the components of $X$ are assumed to be continuously distributed, with $f_X(\cdot)$ denoting the density function. In practice, if $X$ contains discrete variables which are ordered with rich support, then the discrete components can be simply treated as continuous random variables or a smoothing method \citep[see e.g.][]{racine2004nonparametric} can be applied. Denote $\Delta_\sigma(X)\equiv\sigma(1,X)-\sigma(0,X)$ and $\Delta_p(X)\equiv p(X,1)-p(X,0)$. 

First, we nonparametrically estimate $\delta_d(X_i)$ and $V_d(X_i)$ for each $i\leq n$. Let $K:\mathbb R^{d_X}\rightarrow \mathbb R$ and $h$ be a Nadaraya-Watson kernel and bandwidth, respectively. Conditions on $K$ and $h$  will be formally introduced in the asymptotic analysis below. For a generic random variable $A\in\mathbb R$, denote $\phi_A(X_i)\equiv f_X(X_i)\times \mathbb E (A_i|X_i)$. Following the standard kernel estimation literature, $\phi_A(X_i)$ is estimated by 
\[
\hat \phi_A(X_i)=\frac{1}{(n-1)h^{d_X}}\sum_{j\neq i} A_jK\big(\frac{X_j-X_i}{h}\big).
\]  In particular, when $A$ is a constant, e.g. $A=1$, we have
\[
\hat \phi_1(X_i)=\frac{1}{(n-1)h^{d_X}}\sum_{j\neq i} K\big(\frac{X_j-X_i}{h}\big),
\]which is a kernel density estimator of $f_X(X_i)$. Note that the estimation of $\phi_A(X_i)$ leaves the $i$-th observation out to improve its finite sample performance.  Moreover, for $d=0,1$, let   
\begin{align*}
&\hat \delta_{d}(X_i)= (-1)^{1+d}\times \frac{\hat \phi_1(X_i)\hat \phi_{Y\mathbbm 1 (D=d)Z}(X_i)-\hat \phi_{Y\mathbbm 1 (D=d)}(X_i)\hat \phi_{Z}(X_i)}{\hat \phi_1(X_i)\hat \phi_{DZ}(X_i)-\hat \phi_{D}(X_i)\hat \phi_{Z}(X_i)},\\
&\hat V_d(X_i)=(-1)^{1+d}\times \frac{\hat \phi_1(X_i)\hat \phi_{Y^2\mathbbm 1(D=d)Z}(X_i)-\hat \phi_{Y^2\mathbbm 1 (D=d)}(X_i)\hat \phi_{Z}(X_i)}{\hat \phi_1(X_i)\hat \phi_{DZ}(X_i)-\hat \phi_{D}(X_i)\hat \phi_{Z}(X_i)}-\hat \delta^2_d(X_i).
\end{align*}In the above expressions, the term $(-1)^{1+d}$ is introduced due to the fact that 
\[
\text{Cov}(\mathbbm 1(D=d),Z|X)=(-1)^{1+d}\times \text{Cov}(D,Z|X), \ \text{for } d=0,1.
\]
Thereafter, we estimate $S_i$  by the plug-in method:
\[
\hat S_i\equiv |\hat V_0(X_i)|^{\frac{1}{2}} \times (1-D_i)+ |\hat V_1(X_i)|^{\frac{1}{2}} \times D_i.
\]

Let  $\varphi_{ni}=\hat \phi_1(X_i)\hat \phi_{DZ}(X_i)-\hat \phi_{D}(X_i)\hat \phi_{Z}(X_i)$ be the denominator from the estimators above. Clearly, small values of $\varphi_{ni}$ could lead to a denominator issue. Moreover, it is well known that the above kernel estimators will be biased at the boundaries of the support. Therefore, attention is restricted to nonparametric estimation on an inner support $\mathscr X_n\equiv \{x\in\mathscr S_X: \mathcal B_x(h)\subseteq \mathscr S_X \}$, where $\mathcal B_x(h)\equiv \left\{\tilde x\in\mathbb R^{d_X}: \|\tilde x-x\|\leq h\right\}$. 

In the second step of estimation, $\beta$ is estimated.  Note that the conventional IV regression model with exogenous heteroskedasticity is a special case of \eqref{parametric model}. When $\sigma(1,\cdot)\neq \sigma(0,\cdot)$, however, the standard IV estimator of $\beta$ is inconsistent:
\begin{multline*}
\hat \beta_{IV}=\Big[\sum_{i=1}^nW_i(X'_i,D_i)\Big]^{-1}\sum_{i=1}^nW_iY_i = \beta+\Big[\sum_{i=1}^nW_i(X'_i,D_i)\Big]^{-1}\sum_{i=1}^nW_i\sigma(D_i,X_i)\epsilon_i\\
\overset{p}{\rightarrow}\beta+\mathbb E^{-1} [W(X',D)]\times \mathbb E \left[W\Delta_\sigma(X) D\epsilon\right]
\end{multline*} under standard conditions for applying the WLLN in the last step.  Clearly,  the bias term is equal to zero if and only if $\mathbb E \left[W\Delta_\sigma(X) D\epsilon\right]=0$. (The Monte Carlo experiments of Section~5 provide empirical evidence of the inconsistency of $\hat\beta_{IV}$). The proposed {\it endogenous heteroskedasticity IV (EHIV)} estimator is defined as follows:
\[
\hat \beta= \left[\frac{1}{n}\sum_{i=1}^n\frac{T_{ni} W_i(X'_i, D_i)}{\hat S_i}\right]^{-1}\times \frac{1}{n}\sum_{i=1}^n \frac{T_{ni}W_iY_i}{\hat S_i},
\]where $ \{T_{ni}: i\leq n\}$ is a trimming sequence  for dealing with the denominator issue and the boundary issue in the nonparametric estimation. Specifically, 
\[
T_{ni}=\mathbbm 1\big(|\varphi_{ni}|\geq \tau_n; \ |\hat {V}_0(X_i)|\geq \kappa_{0n}; \ |\hat {V}_1(X_i)|\geq \kappa_{1n};\ X_i\in\mathscr X_n\big)
\]
for  positive deterministic sequences  $\tau_n\downarrow 0,\kappa_{0n}\downarrow 0$, and $\kappa_{1n}\downarrow 0$ as $n\rightarrow \infty$. Conditions on $\tau_n$, $\kappa_{0n}$, and $\kappa_{1n}$ will be introduced later for the asymptotics properties of $\hat\beta$. Note that it is possible to apply more sophisticated trimming mechanisms used in the nonparametric regression literature \citep[see, e.g.,][]{klein1993efficient}.

Next, the heteroskedasticity function $\sigma(\cdot,\cdot)$ is estimated, which immediately leads to estimates of the variance effects of the treatment. Fix $x\in\mathscr X_n$. For  $d=0,1$, let $d'=1-d$,  and then define
\[
\hat \sigma^2(d,x)= \frac{|\hat V_d(x)|}{|\hat V_1(x)|}\times  \frac{ \sum_{i=1}^n D_i\hat u_i \times K\big(\frac{X_i-x}{h}\big)}{ \sum_{i=1}^n K\big(\frac{X_i-x}{h}\big)}
+ \frac{|\hat V_d(x)|}{|\hat V_0(x)|} \times \frac{ \sum_{i=1}^n (1-D_i)\hat u^2_i \times K\big(\frac{X_i-x}{h}\big)}{ \sum_{i=1}^nK\big(\frac{X_i-x}{h}\big)}
\] where $\hat u_i = Y_i-X'_i\hat\beta_1-\hat\beta_2D_i$. Under additional conditions, it is shown below that $\hat\beta$ converges to $\beta$ at the parametric rate, and therefore $\hat u_i$ converges to $u_i\equiv \sigma(D_i,X_i)\times \epsilon_i$ at the same rate.  Therefore, the estimation errors associated with $\hat u_i$ are asymptotically negligible in the estimation of $\sigma^2(d,x)$ under some regularity conditions. The variance effects of the treatment are estimated by $\hat \sigma(1,x)-\hat \sigma(0,x)$ for all $x\in\mathscr X_n$, and also the median of the variance effects, denoted as $\text{MVE}$, is estimated  by
$\text{Median}\left\{T_{ni}\left[\hat \sigma(1,X_i)-\hat \sigma(0,X_i)\right]\right\}$.
Note that the MVE differs from the variance of the treatment effects. 

In conducting program evaluation, decision-makers might also be interested in the distributional effects of the treatment \citep[see e.g.][]{heckman2007econometric}. From the model in \eqref{parametric model}, the {\it individual treatment effect} (ITE) is given by
\[
\text{ITE}=\beta_2+\Delta_\sigma(X) \times \epsilon,
\]which takes a non-degenerate probability distribution as long as $\Delta_\sigma(X)\neq 0$ with strict positive probability.  By  \Cref{lemma0} and $S= \sigma(D,X)\times |C(X)|^{\frac{1}{2}}$, the ITE can be re-written as
\[
ITE = \beta_2+\Delta_\sigma(X) \times \epsilon=
\beta_2+\frac{|V_1(X)|^{\frac{1}{2}}-|V_0(X)|^{\frac{1}{2}}}{S}\times [Y-(X',D)\beta].
\]
Based upon this expression, we estimate the ITE for observation $i$ (if $T_{ni}\neq 0$)  by
\[
\widehat{\text{ITE}}_i=\hat \beta_2+\frac{|\hat V_1(X_i)|^{\frac{1}{2}}-|\hat V_0(X_i)|^{\frac{1}{2}}}{\hat S_i}\times \hat u_i.
\]
Then, to estimate the distribution of ITE (conditional on covariates), we follow \cite{guerre2000optimal} by using the pseudo-sample of $\widehat{\text{ITE}}_i$'s estimated above:
\[
\hat f_{\text{ITE}|X}(e|x)=\frac{h^{-(d_X+1)}_f\sum_{i=1}^nT_{ni}K_f\Big(\frac{X_i-x}{h_f},\frac{\widehat{\text{ITE}}_i-e}{h_f}\Big)}{h^{-d_X}_X\sum_{i=1}^nT_{ni}K_X\Big(\frac{X_i-x}{h_X}\Big)}, \ \ \forall \  e\in\mathbb R,
\]where $K_f:\mathbb R^{d_X+1}\rightarrow \mathbb R$ and  $K_X:\mathbb R^2\rightarrow \mathbb R$  are Nadaraya-Watson kernels; $h_f\in \mathbb R^+$ and $h_X\in \mathbb R^+$ are bandwidths. By a similar argument to  \cite{guerre2000optimal}, conditions for the choice of $h_f$ (see below) imply oversmoothing due to the fact that the ITE is estimated rather than directly observed.

\subsection{Discussion}


It is worth noting that our model \eqref{parametric model} fits \cite{ai2003efficient}'s general framework of \textit{sieve minimum distance (SMD)} estimation. Therefore, given the identification of structural functions established in Section 3, \cite{ai2003efficient}'s SMD approach could apply here to construct a $\sqrt n$-consistent estimator for $\beta$. The SMD approach would estimate the finite-dimensional parameter $\beta$ and nonparametric functions $\sigma(\cdot,\cdot)$ simultaneously from the following conditional moments:
\begin{align*}
&\mathbb E \Big[\frac{Y-X'\beta_1-\beta_2D}{\sigma(D,X)}\Big|W\Big]=0,\\
&\mathbb E \Big[\frac{(Y-X'\beta_1-\beta_2D)^2}{\sigma^2(D,X)}\Big|W\Big]=1.
\end{align*}
In contrast to SMD, the EHIV approach described above leads to closed-form expressions for all of the estimators of interest.

In addition, suppose one assumes the following parametric variance model:
\[
\sigma(D,X)=\exp\left[(1,X') \times\pi_1+ \pi_2D\right],
\] 
where $\pi_1\in\mathbb R^{d_X+1}$ and $\pi_2\in\mathbb R$ are coefficients.  In particular, $\pi_2$ characterizes the endogenous heteroskedasticity. Thus, we can estimate $\beta_1$, $\beta_2$ and $\pi_2$  from the following moment equations:
\begin{align*}
&\mathbb E \Big[\frac{Y-X'\beta_1-\beta_2D}{\exp(\pi_2D)}\Big|W\Big]=0,\\
&\mathbb E \Big[\frac{(Y-X'\beta_1-\beta_2D)^2}{\exp(2\pi_2D)}\Big|W\Big]=\mathbb E \Big[\frac{(Y-X'\beta_1-\beta_2D)^2}{\exp(2\pi_2D)}\Big|X\Big].
\end{align*}A standard GMM approach applies. Note that the first moment equation provide a closed-form solution of $\beta_1$ and $\beta_2$ depending on the scalar parameter $\pi_2$.

\subsection{Asymptotic properties} In this subsection, we  establish  asymptotic properties for the EHIV estimator by following the semiparametric two-step estimation literature \citep[e.g.][among many others]{bierens1983uniform,powell1989semiparametric,andrews1994asymptotics,newey1994large}.  Before we proceed, it is worth pointing out that the EHIV estimator $\hat\beta$ is $\sqrt n$-consistent if the heteroskedasticity is exogenous, i.e., $\sigma(d,\cdot)=\tilde\sigma(\cdot)$ for some $\tilde \sigma$, without additional conditions on the first-stage estimation. In the presence of endogeneity, however,  the following consistency (resp. $\sqrt n$-consistency) argument of $\hat\beta$ requires that  the  first-stage   estimation error, i.e. $\hat V_d(X_i)-V_d(X_i)$, uniformly converges to zero (resp. uniformly converges to zero faster than $n^{-1/4}$).   

To begin with, we make the following assumptions. Most of them are weak and standard in the  literature. 
\begin{assumption}
\label{ass3}
(i)  Eq. \eqref{parametric model} holds; (ii) The data $\{(Y_i,D_i,W'_i)': i\leq n\}$ is an i.i.d. random sample; (iii) The support $\mathscr S_X$ is compact with nonempty interior; (iv) The density  of $X$ is bounded and bounded away from zero on $\mathscr S_X$; (v) The function $\mathbb P(Z=0|X=x)$ is bounded away from 0 and 1 on $\mathscr S_X$; (vi)  The parameter space $ \mathbb B\subseteq \mathbb R^{d_X+1}$ of $\beta$ is compact.
\end{assumption}
 \begin{assumption}
 \label{ass4}
For each $x\in\mathscr S_X$, $|\Delta_p(x)|\geq C_0$ for some $C_0\in\mathbb R_+$.
\end{assumption}
\begin{assumption}
\label{ass5}
For some integer $R\geq 2$, the functions $\sigma(d,\cdot)$, $p(\cdot,z)$, $f_{XZ}(\cdot,z)$, $\mathbb E (\epsilon |D=d,X=\cdot ,Z=z)$ and $\mathbb E (\epsilon^2|D=d,X=\cdot,Z=z)$ are $R$-times continuously differentiable on  $\mathscr S_X$.
\end{assumption}

\begin{assumption}
\label{ass6}
Let $K:\mathbb R^{d_X}\rightarrow \mathbb R$ be a kernel function satisfying: (i) $K(\cdot)$ has bounded support; (ii); $\int k(u) du =1$; (iii) $K(\cdot)$ is an $R$-th order kernel, i.e.,
\[
\begin{array}{cll}\int u_1^{r_1}\cdots u_{d_X}^{r_{d_X}} K(u)du &=0, & \text{ if }\ 1\leq \sum_{\ell=1}^{d_X} r_\ell\leq R-1; \\
 & <\infty,& \text{ if }\ \sum_{\ell=1}^{d_X} r_\ell= R,\end{array}
\] where $(r_1,\cdots, r_{d_X})\in \mathbb N^{d_X}$; (iv) $K(\cdot)$ is differentiable with bounded first derivatives on $\mathbb R^{d_X}$. 
\end{assumption}
\begin{assumption}
\label{ass7}
 As $n\rightarrow \infty$, (i) $h\rightarrow 0$; (ii) $nh^{d_X}/\ln n\rightarrow \infty$.
\end{assumption}


\noindent
\Cref{ass3}  can be relaxed to some extent:  \Cref{ass3}-(ii) could be extended to allow for weak time/spatial dependence across observations. Regarding \Cref{ass3}-(iii) , unbounded regressors can be accommodated by using high order moment restrictions on the tail distribution of $X$ at the expense of longer proofs.  \Cref{ass4} is introduced for expositional simplicity. \Cref{ass6,ass5,ass7} are standard in the kernel regression literature. See e.g. \cite{PaganUllah1999}. In particular, \Cref{ass5} is a  smoothness  condition that can be further relaxed by a Lipschitz condition. \Cref{ass4,ass5} imply that for $d,z=0,1$, the functions $\delta_d(\cdot)$, $ V_d(\cdot)$, $\mathbb E [Y\mathbbm 1(D=d)|X=\cdot,Z=z]$ and $\mathbb E [Y^2\mathbbm 1(D=d)|X=\cdot,Z=z]$ are $R$-times continuously differentiable on $\mathscr S_X$ with bounded $R$-th partial derivatives. In \Cref{ass7}, the $\ln n$ arises because we drive uniform consistency for the first-stage nonparametric estimation. 
\begin{lemma}
\label{lemma2}
Under \Cref{ass3,ass4,ass5,ass6,ass7}, we have
\[
\sup_{x\in\mathscr S_X}  \left|\hat {V}_d(x)- V_d(x)\right|=O_p\Big(h^R+ \sqrt{\frac{\ln n}{nh^{d_X}}}\Big).
\]
\end{lemma}
\noindent
The uniform convergence result in \Cref{lemma2} is  standard  in the kernel estimation literature \citep[see e.g.][]{andrews1995nonparametric}, and therefore proofs are omitted. In particular, the choice of $h$ should balance the bias and variance in the nonparametric estimation.  Suppose $h=\lambda_0(n/\ln n)^{-\gamma}$ for some $\lambda_0>0$ and $\gamma\in(0,1/d_X)$. Note that  such a choice of $h$ satisfies  \Cref{ass7}. Then, the convergence rate in \Cref{lemma2} becomes  $\big({n}/{\ln n}\big)^{- (R\gamma\wedge \frac{1-\gamma d_X}{2})}$. 


\begin{assumption}
\label{ass8}
The random matrix $\frac{W(X',D)}{S}$ has finite second moments and $\frac{W\epsilon}{\sqrt{|C(X)|}}$ has finite forth moments, i.e.,
\[
\mathbb E \Big\|\frac{W(X',D)}{S}\Big\|^2<+\infty; \text{ and }\ \mathbb E\Big\|\frac{W\epsilon}{\sqrt{|C(X)|}}\Big\|^4<+\infty.
\]
\end{assumption}

\begin{assumption}
\label{ass9}
The matrix  $\mathbb E\big [\frac{W(X',D)}{S}\big]$ is invertible. 
\end{assumption}

\begin{assumption}
\label{ass10}
For each $x\in\mathscr S_X$ and $d=0,1$, let  $|V_d(x)|\geq C_1$ for some $C_1\in\mathbb R_+$.
\end{assumption}

\begin{assumption}
\label{ass11}
As $n\rightarrow +\infty$, the trimming parameters satsify (i) $\tau_n\downarrow 0$, $\kappa_{0n}\downarrow 0$, and $\kappa_{1n}\downarrow 0$; (ii)~$\tau^{-1}_n \big(h^{2R}+ {\frac{\ln n}{nh^{d_X}}}\big)\downarrow 0$, $\kappa_{01}^{-1} \big(h^{2R}+ {\frac{\ln n}{nh^{d_X}}}\big)\downarrow 0$, and $\kappa_{1n}^{-1} \big(h^{2R}+ {\frac{\ln n}{nh^{d_X}}}\big)\downarrow 0$.
\end{assumption}

\noindent 
\Cref{ass8} is standard,  allowing us to apply the WLLN and CLT. \Cref{ass9} is a testable rank condition, given that $S_i$ can be consistently estimated. Similar to \Cref{ass4}, \Cref{ass10} is introduced for expositional simplicity, dealing with the denominator issue. Such a condition can be relaxed at the expense of a longer proof and exposition.  \Cref{ass11} imposes mild restrictions on the choice of the trimming parameters.   

\begin{theorem}
\label{theorem1}
Suppose all the assumptions in \Cref{lemma2} and \Cref{ass8,ass9,ass10,ass11} hold. Then, $\hat\beta\overset{p}{\rightarrow} \beta$.
\end{theorem}

\noindent
\Cref{theorem1} shows that if the first-stage nonparametric estimation is uniformly consistent, then the EHIV converges to the true parameter in probability.

With consistency,  we are now ready to establish the  limiting distribution of $\hat\beta$. Following \cite{powell1989semiparametric}, we impose conditions on the kernel function and the bandwidth such that the first-stage estimation bias vanishes faster than $\sqrt n$. It is worth pointing out that our model fits the general framework in the semiparametric two-step estimation literature \citep[e.g.][]{andrews1994asymptotics,andrews1995nonparametric}. Thus, the $\sqrt n$-consistency of $\hat\beta$ requires that  the first-stage estimator $\hat V_d(\cdot)$ converges to $V_d(\cdot)$  faster than $ n^{-1/4}$.  
\begin{assumption}
\label{ass12}
As $n\rightarrow +\infty$,  (i) $n^{\frac{1}{2}}h^R\rightarrow 0$; (ii) $n^{\frac{1}{4}}\sqrt{\frac{\ln n}{nh^{d_X}}}\rightarrow 0$.
\end{assumption}
\noindent
\Cref{ass12} strengthens \Cref{ass7} by requiring that both the first-stage estimation bias $\mathbb E [\hat V_d(\cdot)]-V_d(\cdot)$ and variance of $\hat V_d(\cdot)$ vanish faster than $n^{-1/2}$. Note that this assumption implies that $R\geq d_X$. For instance, one could choose e.g. $h=\lambda\times (n/\ln n)^{1/(2R-\iota)}$ for some positive constants $\lambda$ and $\iota$ to satisfy \Cref{ass12}, as long as $d_X-R+\frac{1}{2}\iota>0$ and $\iota<2R$.

To derive $\hat \beta$'s limiting distribution, we plug \eqref{parametric model} into the expression of $\hat \beta$, which gives us
\begin{multline*}
\hat \beta =\beta+\Big[\frac{1}{n}\sum_{i=1}^n\frac{T_{ni} W_i(X'_i, D_i)}{\hat S_i}\Big]^{-1}\times \frac{1}{n}\sum_{i=1}^n \frac{T_{ni}W_i\epsilon_i}{\sqrt{|C(X_i)|}}\\
+\Big[\frac{1}{n}\sum_{i=1}^n\frac{T_{ni} W_i(X'_i, D_i)}{\hat S_i}\Big]^{-1}\times \frac{1}{n}\sum_{i=1}^n\left[\frac{T_{ni}W_i\epsilon_i}{\sqrt{|C(X_i)|}} \Big(\frac{S_i}{\hat S_i}-1\Big)\right].
\end{multline*} Note that the last term on the right-hand side comes from the first-stage estimation error. Unlike the semiparametric weighted least squares estimator \citep[see e.g.][]{andrews1994asymptotics}, the last term  on the right hand side converges in distribution to a limiting normal distribution under additional assumptions, instead of being $o_p(n^{-1/2})$.  This is because the weighting function  used for transformation (i.e. $1/S_i$) depends on the endogenous variable $D_i$.



Define
\[
\psi(Y,D,X)= \frac{D[Y-\delta_1(X)]^2}{V_1(X)}+\frac{(1-D)[Y-\delta_0(X)]^2}{V_0(X)}
\] and let $\Psi=\psi(Y,D,X)$ be a random variable.  By  \Cref{lemma0}, we have $\Psi={[\epsilon-\xi_1(X)]^2}/{C(X)}$, which is uncorrelated with $Z$ conditional on $X$, i.e., $\text{Cov}(\Psi,Z|X)=0$. Thus, $\mathbb E (\Psi|W)=\mathbb E (\Psi|X)$. Let further
\[
\zeta= \frac{[\Psi-\mathbb E (\Psi|X)]\times [Z-\mathbb E (Z|X)]}{2\text{Cov}(D,Z|X)} \times \left[\frac{\mathbb E(D\epsilon|X)}{|C(X)|^{1/2}}X',\frac{\mathbb E(ZD\epsilon|X)}{|C(X)|^{1/2}}\right]'.
\]
 By definition, $\zeta$ is a  random vector of $d_X+1$-dimensions and  $\mathbb E (\zeta|W)=0$. 
\begin{theorem}
\label{theorem2}
Suppose  \Cref{ass1,ass2,rank,ass3,ass4,ass5,ass6,ass7,ass9,ass8,ass9,ass10,ass11,ass12} hold. Then we have $\sqrt n(\hat\beta-\beta)\overset{d}{\rightarrow} N\big(0,\Omega\big)$, where 
$\Omega\equiv \mathbb E^{-1} \big[\frac{(X',D)'W'}{S}\big] \times\text{Var} \big[\frac{W\epsilon}{\sqrt{|C(X)|}}-\zeta\big]\times \mathbb E^{-1} \big[\frac{W(X',D)}{S}\big]$.
\end{theorem}
\noindent
In the asymptotic variance matrix $\Omega$,  the term $\zeta$  accounts for the first-stage estimation error.


For inference based on Theorem~2, it's necessary to estimate the variance matrix $\Omega$. First, we estimate $\mathbb E \big[\frac{(X',D)'W'}{S}\big]$  by    
\[
 \mathbb E_n \Big[\frac{(X',D)'W'}{S} \Big]= \frac{1}{\sum_{i=1}^n T_{ni}}\times \sum_{i=1}^n T_{ni}\frac{(X'_i,D_i)'W'_i}{\hat S_i}.
\] Next, we  construct a pseudo sample of $\{\zeta_i: i\leq n; T_{ni}=1\}$. Let 
\begin{align*}
& \mathbb E_n \big(\frac{X_iD_i\epsilon_i}{\sqrt{|C(X_i)|}}\big|X_i\big)=\frac{X_i}{\sqrt {|\hat V_1(X_i)|}}\times \frac{\sum_{j\neq i}D_j\hat u_jK\big(\frac{X_j-X_i}{h}\big)}{ \sum_{j\neq i} K\big(\frac{X_j-X_i}{h}\big)},\\
& \mathbb E_n \big(\frac{Z_iD_i\epsilon_i}{\sqrt{|C(X_i)|}}\big|X_i\big)=\frac{1}{\sqrt {|\hat V_1(X_i)|}}\times \frac{\sum_{j\neq i}Z_jD_j\hat u_jK\big(\frac{X_j-X_i}{h}\big)}{\sum_{j\neq i} K\big(\frac{X_j-X_i}{h}\big)},
\end{align*} be estimators of $\mathbb E\big(\frac{X_iD_i\epsilon_i}{\sqrt{|C(X_i)|}}\big|X_i\big)$ and $\mathbb E \big(\frac{Z_iD_i\epsilon_i}{\sqrt{|C(X_i)|}}\big|X_i\big)$, respectively. For all $i,j\leq n$ satisfying $T_{ni}=1$, let further
\[
\hat\Psi_{ji}= \frac{D_j[Y_j-\hat \delta_1(X_i)]^2}{\hat V_1(X_i)}+\frac{(1-D_j)[Y_j-\hat \delta_0(X_i)]^2}{\hat V_0(X_i)}
\] and $\hat\Psi_i=\hat \Psi_{ii}$.
Thus, we  construct $\zeta_i$ by
\begin{multline*}
\hat\zeta_i= \frac{\frac{1}{n-1}\sum_{j\neq i}(\hat \Psi_i-\hat\Psi_{ji})K_h\big({X_j-X_i}\big)\times \frac{1}{n-1}\sum_{j\neq i}(Z_i-Z_j)K_h\big(X_j-X_i\big)}{2\big[\hat \phi_1(X_i)\hat \phi_{DZ}(X_i)-\hat \phi_{D}(X_i)\hat \phi_{Z}(X_i)\big]}\\
 \times \Big\{\mathbb E_n \Big[\frac{X'_iD_i\epsilon_i}{\sqrt{|C(X_i)|}}\big|X_i\Big],\mathbb E_n \Big[\frac{Z_iD_i\epsilon}{\sqrt{|C(X_i)|}}\big|X_i\Big]\Big\}',
\end{multline*}where $K_h(\cdot)=K(\cdot/h)/h^{d_X}$. Hence, we obtain a pseudo sample $\{\hat \zeta_i: i\leq n, T_{ni}=1\}$ of $\zeta$. Furthermore, because $\frac{W\epsilon}{\sqrt{|C(X)|}}=\frac{Wu}{S}$, we estimate  ${\text{V}}{\text{ar}}(\frac{W\epsilon}{\sqrt{|C(X)|}}-\zeta)$ by the sample variance of  $\left\{\frac{W_i\hat u_i}{\hat S_i}-\hat \zeta_i: i\leq n, T_{ni}=1\right\}$,  denoted as $\hat{\text{V}}{\text{ar}}(\frac{W\epsilon}{\sqrt{|C(X)|}}-\zeta)$.


We are now ready to define an estimator of $\Omega$ as follows:
\[
\hat \Omega\equiv \mathbb E_n^{-1} \left[\frac{(X',D)'W'}{S}\right]\times\hat{\text{V}}{\text{ar}}\left(\frac{W\epsilon}{\sqrt{|C(X)|}}-\zeta\right)\times \mathbb E_n^{-1} \left[\frac{W(X',D)}{S}\right].
\]The consistency is given by a similar argument to \Cref{theorem1}. In practice, one could also obtain the standard errors of $\hat \beta$  by the bootstrap \citep[see e.g.][]{abadie2002bootstrap} and/or by simulation methods \citep[see e.g.][]{barrett2003consistent}. 

Finally, we provide the asymptotic properties of $\hat\sigma(\cdot,\cdot)$. Note that $
\hat u_i=u_i-(X'_i,D_i)(\hat\beta-\beta)=u_i+O_p(n^{-1/2})$, where the $O_p(n^{-1/2})$  holds uniformly. Therefore, we have
\begin{multline*}
\hat \sigma^2(d,x)= \frac{|\hat V_d(x)|}{|\hat V_1(x)|}\times  \frac{ \sum_{i=1}^n D_i u^2_i \times K\big(\frac{X_i-x}{h}\big)}{ \sum_{i=1}^nK\big(\frac{X_i-x}{h}\big)}\\
+ \frac{|\hat V_d(x)|}{|\hat V_0(x)|} \times \frac{ \sum_{i=1}^n (1-D_i) u^2_i \times K\big(\frac{X_i-x}{h}\big)}{ \sum_{i=1}^nK\big(\frac{X_i-x}{h}\big)}+O_p(n^{-1/2}),
\end{multline*}
provided that the conditions in \Cref{theorem2} hold.  Following the standard nonparametric literature \citep[e.g.][]{PaganUllah1999}, we obtain the asymptotic properties of $\hat\sigma(\cdot,\cdot)$.
 
%

 \begin{theorem}
 \label{theorem3}
Suppose all the assumptions in \Cref{theorem2} hold. Then for
any compact subset $\mathbb C$ of $\mathbb R^{d_X}$,
 \[
\sup_{x\in\mathbb C}\left| \hat\sigma(d,x)- \sigma(d,x)\right|=O_p\Big(\sqrt{\frac{\ln n}{nh^{d_X}}}\Big), \ \text{for } d=0,1.
 \] 
 \end{theorem}
\noindent 
\Cref{theorem3} establishes the uniform convergence of $\hat\sigma(d,\cdot)$ on any compact subset $\mathbb C$.  Note that  \Cref{ass12} implies that the bias in the estimation of $\sigma(d,\cdot)$ vanishes faster than $\sqrt n$. Therefore, the convergence rate of $\hat \sigma(d,\cdot)$ is fully determined by the asymptotic variance of the nonparametric estimator $\hat V_d(x)$.

By a  similar argument to \cite{guerre2000optimal}, one can also establish the uniform convergence of $\hat f_{\text{ITE}|X}(\cdot|\cdot)$  to   $f_{\text{ITE}|X}(\cdot|\cdot)$ under their conditions.

\section{Monte Carlo Evidence}
To illustrate our two-step semiparametric procedure, we conduct a Monte Carlo study. In particular, we consider the following triangular model as the data generating process:
\begin{align*}
&Y=\beta_0+\beta_1 X+\beta_2D+(0.1+0.25|X|+\lambda_0  D)\epsilon,\\
&D= \mathbbm 1 \big[\Phi(\eta) \geq 0.2|X|+r_0Z \big]
\end{align*} where $X\sim N(0,1)$, $Z\sim Bernoulli(0.5)$,  $(\epsilon,\eta)$ has a bivariate normal distribution with unit variance and correlation coefficient $\rho_0\in(-1,1)$, and $\Phi(\cdot)$ denotes the CDF of the standard normal distribution. Moreover, $\lambda_0\in\mathbb R_+$ and $r_0\in\mathbb R_+$ are two positive constants to be specified, with the former measuring the level of endogenous heteroskedasticity and the latter capturing the size of the ``complier group''.  Let $(X,Z)\bot (\epsilon,\eta)$ to satisfy \Cref{ass1,ass2}. For simplicity, let further $X\bot Z$. \Cref{rank} holds trivially. Regarding conditions for asymptotics,     Assumptions \ref{ass3}-(iv) and \ref{ass4} are not satisfied in our setting, but note that these conditions are imposed for the simplicity of proofs and expositions. 

For each replication, we draw an i.i.d. random sample $\{(W_i,\epsilon_i,\eta_i): i\leq n\}$ and then generate a random sample $\{(Y_i,D_i,W_i): i\leq n\}$ of size  $n=1000, 2000, 4000$ from the data generating process. Next, we apply our estimation procedure for each replication. All reported results are based on 500 replications.

To assess the finite sample behavior of the estimators, we set $\beta=(0,1,1)'$ and $(\lambda_0,r_0,\rho_0)=(0.5,0.5,0.5)$ and then compare EHIV's performance with the standard IV estimator. For the first stage estimation of $V_d(\cdot)$,  we consider two kernel functions of order $R=4$, i.e., the  Gaussian kernel and the  Epanechnikov kernel: 
\begin{align*}
&K_G(u)=\frac{1}{2}(3-u^2)\times \frac{1}{\sqrt {2\pi}}\exp(-\frac{u^2}{2});\\
&K_{E}(u)=\frac{15}{8}(1-\frac{7}{3}u^2)\times \frac{3}{4}(1-u^2)\times \mathbbm 1(|u|\leq 1).
\end{align*} 
Note that the bounded support condition in \Cref{ass6}-(i) is satisfied by $K_E(\cdot)$, but not by $K_G(\cdot)$.   Moreover, we follow Silverman's rule of thumb to choose the bandwidth, i.e.,  $h=1.06\times n^{-1/5}$.  Clearly, \Cref{ass12} is satisfied.  For the trimming sequence $T_{ni}$, we choose $\tau_n=\kappa_{0n}=\kappa_{1n}=0.1$. We also considered other values for the trimming parameters (e.g., $\tau_n=\kappa_{0n}=\kappa_{1n}=0.05$ and  $0.01$), for which the results are qualitatively similar. 

\Cref{est: tab1} in the Appendix reports the finite performance of the EHIV estimator in terms of the Mean Bias (MB),  Median Bias (MEDB), Standard Deviation (SD), and Root Mean Square Error (RMSE). For comparison, we also provide summary statistics of the  IV estimates.   In particular, the MB and MEDB  of  the IV estimates of $\beta_2$  do not shrink with the sample size, which provides evidence for inconsistency of the IV estimation. In contrast, both the bias (MB, MEDB) and the variance (SD) of the EHIV estimator decrease at the expected $\sqrt n$-rate. Moreover, the summary statistics show that the EHIV behaves similarly for the difference choices of kernel functions.

\Cref{fig1} in the Appendix illustrates the performance of the nonparametric estimates of the endogenous heteroskedasticity  $\sigma(\cdot,\cdot)$.  The figures on the left side display the true functions $\sigma(d,\cdot)$ and the averages of $\hat \sigma(d,\cdot)$ over 500 replications for different sample sizes. As sample size increases, the bias of $\hat \sigma(d,\cdot)$ converges to zero quickly. Note that there is a positive finite-sample bias, in particular when the endogenous heteroskedasticity is small. The figures on the right side of \Cref{fig1}  provide 95\% confidence intervals for $\sigma(d,x)$ for a sample size of $4000$.

Next, we estimate $f_{\text{ITE}|X}(\cdot|x)$ at $x=-0.6745$, $0$, and $0.6745$, which are  the first, second, and third quartiles of the distribution of $X$, respectively.  Note that  our specification   implies that  the conditional  ITE follows a normal distribution with mean $\beta_0$ and variance $\lambda_0^2$, regardless of the value of $x$. \Cref{fig2} in the Appendix shows that $\hat f_{\text{ITE}|X}(\cdot|x)$ behaves well for all sample sizes.

As a robustness check,  we also consider  different sizes of the compliers group (varying $r_0$), degrees of endogeneity (varying $\rho_0$), and levels of heteroskedasticity (varying $\lambda_0$). For different values of $r_0$,  we use $\tau_n=0.2\times r_0$ for the trimming mechanism; otherwise, more observations would be trimmed out as $r_0$ decreases.  \Cref{est: tab2} in the Appendix  reports the summary statistics for $n=4000$. The results are qualitatively similar across different settings.  The EHIV performs worse as $r_0$ decreases to zero, in line with the asymptotic results in \Cref{theorem3}.

\section{Empirical application}  \label{sec:empirical}
In this section, we apply the EHIV estimation approach to an empirical application, specifically studying the causal effects of fertility on female labor supply. Motivated by \cite{angrist1998children}, we investigate the effects of having a third child on hours worked per week. Having a third child might be expected to affect a mother's labor supply heterogeneously,  given that fertility and labor supply are determined simultaneously and some latent variables may interact with the presence of a third child. Following \cite{angrist1998children}, we use the gender mix of the first two children to instrument for the decision of having a third child.\footnote{There is also a sizable literature that use twins at first birth as an IV to estimate the relationship between childbearing and female labor supply; see e.g. \cite{rosenzweig1980life,rosenzweig1980testing}, \cite{bronars1994economic}, and \cite{gangadharan1996effects}, and references therein. Relatedly, \cite{maurin2009social} consider the peer mechanism and suggest neighbors' children sex mix as an IV to identify peer effects in female labor market participation.} There is a strong argument for the validity of this instrument since child gender is randomly assigned and families with first two children of the same gender are significantly more likely to have a third child. Given households' (heterogenous) preferences over consumption, leisure and childrearing, female labor supply is mainly determined by  financial and time constraints.  Having a third child  might cause time constraints to become more stringent and therefore reduce the role of preference heterogeneity, which implies variance effects in the labor supply model.

For our application, the sample is drawn from the 2000 Census data (5-percent public-use microdata sample (PUMS)). The outcome of interest ($Y$) is hours worked per week of the mother worked in 1999, the binary endogenous explanatory variable ($D$) is the presence of a third child, and the instrument ($Z$) is whether the mother's first two children were of the same gender. The specifications considered below include mother's education, mother's age at first birth, and age of first child as exogenous covariates ($X$). To have the units of education in years, we recode some of the Census education classifications as detailed in \Cref{table recode}. \Cref{table 3} provides descriptive statistics  for the observable realizations of $(Y,D,Z,X)$ in our sample.
\begin{table}[!ht]
   \centering
      \caption{Re-coding of mother's education based upon Census classifications}
   \begin{tabular}{lcc}
\hline \hline
        Education level        &  Coded value & Recoded value  \\\hline
       No schooling completed       & \ 1                                                              &  \ \ \ 0   \\
       Nursery school to 4th grade & \ 2                                                              & \ \ \ 2   \\
       5th grade or  6th grade        & \ 3                                                              &   \ 5.5   \\
       7th grade or  8th grade        & \ 4                                                              &  \ 7.5   \\  
       9th grade                              & \ 5                                                              & \ \ \ 9   \\ 
       10th grade                            & \ 6                                                              & \ \ 10   \\ 
        11th grade                           & \ 7                                                              & \ \ 11   \\ 
         12th grade, No diploma      & \ 8                                                              &  11.5   \\ 
       High school graduate           & \ 9                                                               & \ \ 12   \\ 
       Some college credit, but less than 1 year              &   10                            &  12.5   \\ 
      1 or more years of college, no degree                    &  11                            & \ \ 14   \\ 
     Associate degree                    & 12                                                              & \ \ 14   \\ 
     Bachelor's degree                    & 13                                                            & \ \ 16   \\ 
     Master's degree                    & 14                                                                & \ \ 18   \\ 
    Professional degree                    & 15                            & \ \ 18   \\ 
    Doctorate degree      & 16                            & \ \ 21   \\\hline \hline
 \end{tabular}
   \label{table recode}
     \end{table}

\begin{table}[!ht]
   \centering
      \caption{Descriptive statistics}
   \begin{tabular}{llcccc}
\hline \hline
        Variable          &  Description& Mean & Median & SD \\\hline
        Hours       & Hours worked per week in 1999                            & 23.291& 25&18.755   \\
        Had third child         & 1 if had third child, 0 otherwise                                          & 0.257 &0&  0.437\\
        Same-sex                & 1 if first two children are same gender, 0 otherwise    & 0.502 &1& 0.500 \\
        Education                 & Mother's education level (in years)                                   &  13.951&14& 2.228\\
        Age at first birth       & Mother's age when first child was born                              & 26.364  &26& 5.034  \\
        1st child's age          & Age of first child in 2000                                                    & 7.550    &8& 3.032\\
        2nd child's age         & Age of second child in 2000                                              & 4.548    &4&  3.061\\
 \hline
 Sample Size & 293,771\\\hline \hline
   \end{tabular}
   \label{table 3}
  \end{table}

In our estimation, we assume $R=6$ for \Cref{ass5} and use the 6th order Gaussian kernel, i.e.,
\[
k_j(u)=\frac{1}{8}(15-10u^2+u^4)\times \frac{1}{\sqrt {2\pi}}\exp(-\frac{u^2}{2}), \ \ \forall u\in\mathbb R,
\] and $K(u)=k_1(u)k_2(u)k_3(u)k_4(u)$. The bandwidth is chosen by 
\[
h_z=1.06\times \hat \sigma_X\times (\hat c_z\times n)^{-1/9},
\]where $\hat \sigma_X$ is the sample standard deviation of the covariates and  $\hat c_z=n^{-1}\sum_{i=1}^n\mathbbm 1(Z_i=z)$. With these choices, one can verify that \Cref{ass7,ass12} are satisfied. Moreover, to specify our trimming sequence $T_{ni}$, we set $\tau_n=10^{-10}$ and $\kappa_{0n}=\kappa_{1n}=10^{-2}$. For this trimming sequence, 75,654 observations (roughly 26\% of the whole sample) are ``trimmed away.''

\begin{table}[!ht]
\small
\caption{Estimation Results}
\label{table4}
{\begin{center}
\begin{tabular}{l|cccccc}\hline \hline
Hours worked per week  & OLS & IV&EHIV \\\hline
Has a third child        &-7.597**                                      &-4.226**                   & -5.343**\\
                                 &(0.084)                                       &(1.123)                  & (1.401)\\ 
Education                 &\ 1.046**                                       &\ 1.005**                    &\ 0.685**\\
                                 & (0.017)                                      &(0.023)                 & (0.033)\\
Age at first birth        &  -0.341**                                    &-0.282**                 & -0.368**\\
                                 &(0.007)                                       &(0.023)                   & (0.010)\\
1st child's age         &\ 0.635**                                        &\ 0.740**                    &\ 0.731**\\
                                &(0.022)                                        &(0.044)                  & (0.043)\\
2nd child's age        &\ 0.022                                        &-0.225**                      & -0.045\\
                                 &(0.022)                                        &(0.093)                   & (0.047)\\
Constant                   & 14.761**                                       &13.219**                 & 19.163**\\
                                 &(0.271)                                          & (0.625)                 & (0.830)\\  \hline
\rowcolor{gainsboro}    ATT                         &                                               &                                   & -4.861\\
\rowcolor{gainsboro}                                 &                                               &                             & (2.980) \\ 
\hline
\end{tabular}
\end{center}}

\end{table}

\Cref{table4} reports the main results from EHIV estimation along with the results obtained from OLS and IV. Across the three methods, there is consistently a negative relationship between having a third child and labor supply. In looking at the OLS and IV results, a similar finding to that in \cite{angrist1998children} is obtained, with the LATE effect of a third child being considerably lower in magnitude (4.226 hour reduction) than the OLS estimate (7.597 hour reduction). As we've shown previously, the IV estimate of $-4.226$ may be an inconsistent estimate of the ATE in the presence of endogeneous heteroskedasticity. The EHIV, in contrast, is consistent for the ATE under our model of endogenous heteroskedasticity. In this application, the EHIV estimate is more negative ($-5.343$) than the IV estimate, although it is still within a standard deviation of the latter. It is interesting to note that, despite the non-parametric estimates that play a role in EHIV estimation, the EHIV standard error is less than 30\% larger than the IV estimator, and this difference is likely to be largely driven by the trimming described above. For the exogenous covariates, EHIV estimates are all of the same sign as the IV estimates, with the largest difference in magnitudes seen for the education and age-at-first-birth covariates. Moreover, the estimate of ATT is $-4.861$, though this estimate is not significant at a 5\% level. 

\begin{figure}[!ht] 
   \centering
   \includegraphics[width=5in]{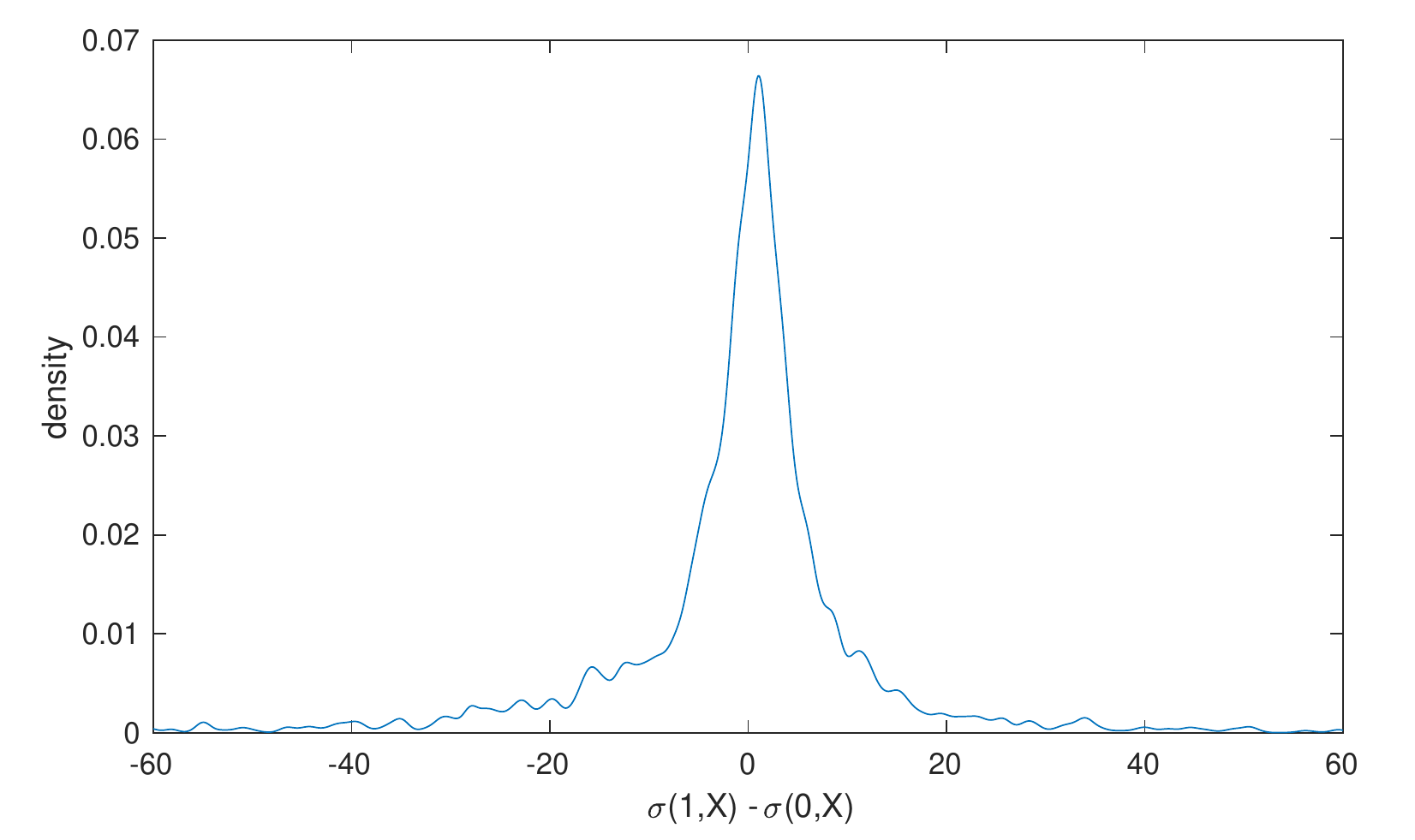} 
   \caption{Density of EHIV variance effects}
   \label{figdensity}
\end{figure}

Next,  we estimate $\sigma(1,X_i)$ and $\sigma(0,X_i)$ for each observation in the sample.  Using the kernel approach, we show the density function of variance effects (i.e., $\sigma(1,X)-\sigma(0,X)$) in \Cref{figdensity}. Overall, variance effects are distributed around zero. This means, having a third child could either increase or decrease the standard deviation of the mother's labor supply, depending on the value of covariates.  

\begin{figure}[!ht] 
   \centering
   \includegraphics[width=2.5in]{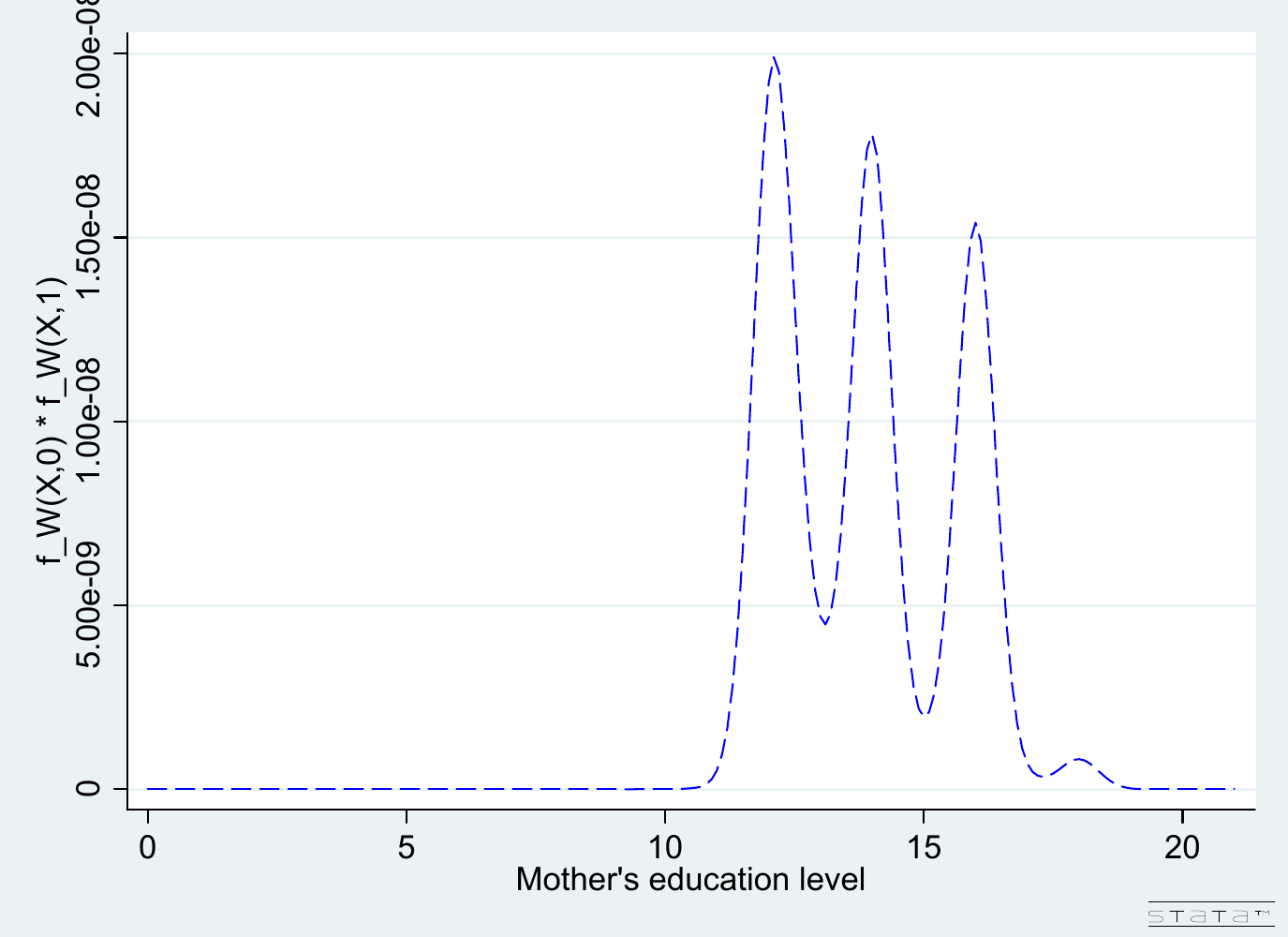} 
     \includegraphics[width=2.5in]{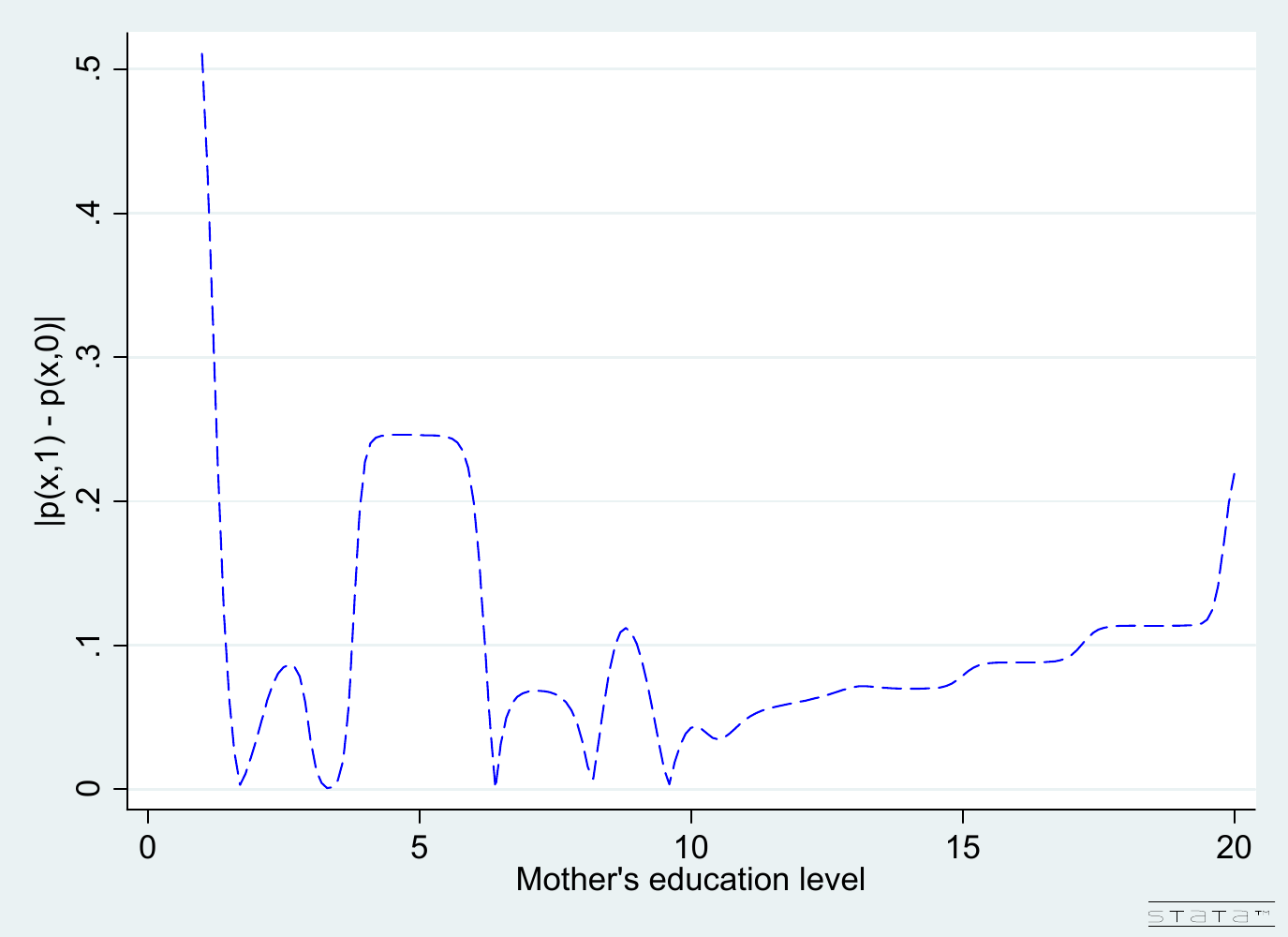} \\
        \includegraphics[width=2.5in]{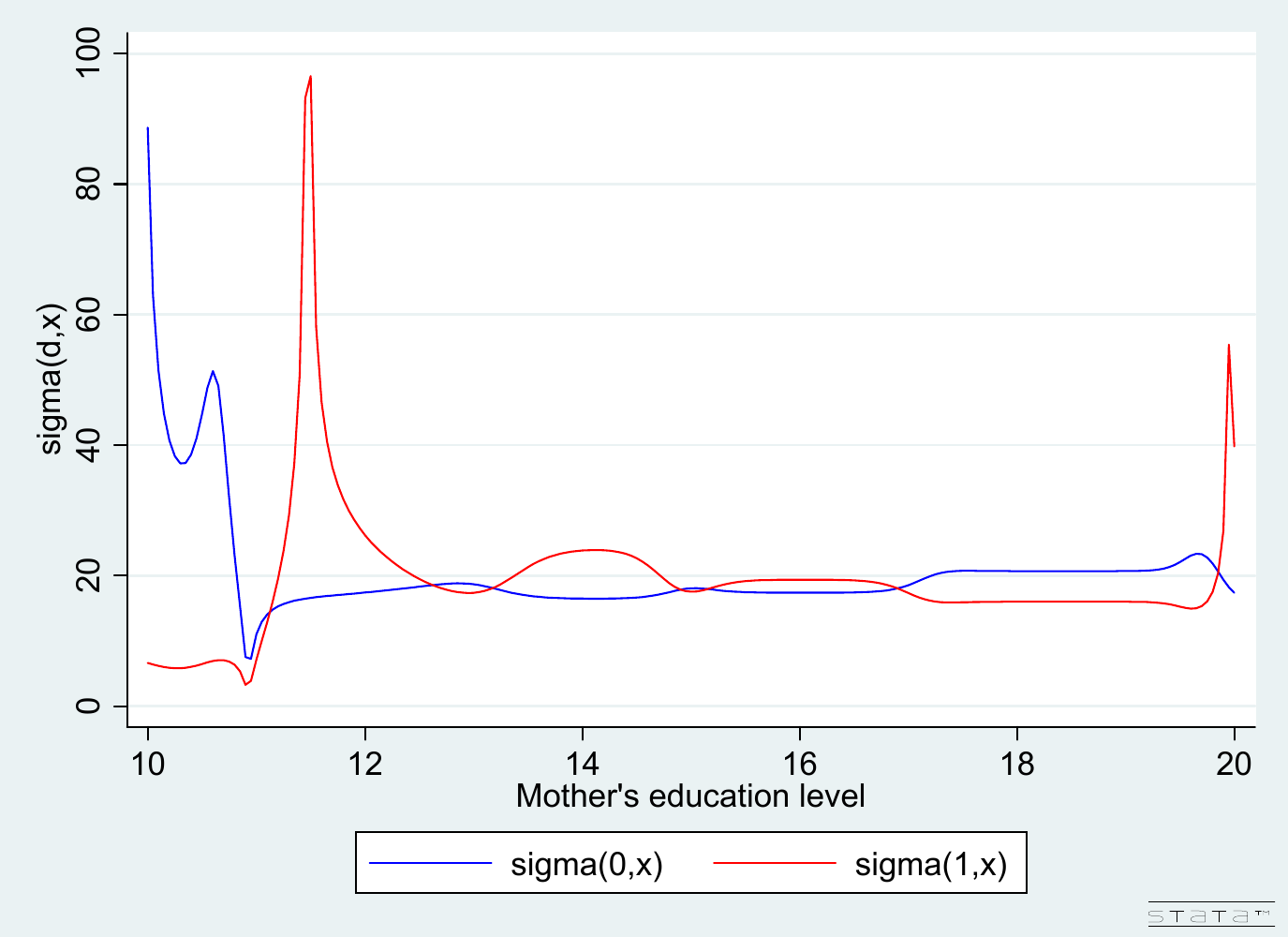} 
               \includegraphics[width=2.5in]{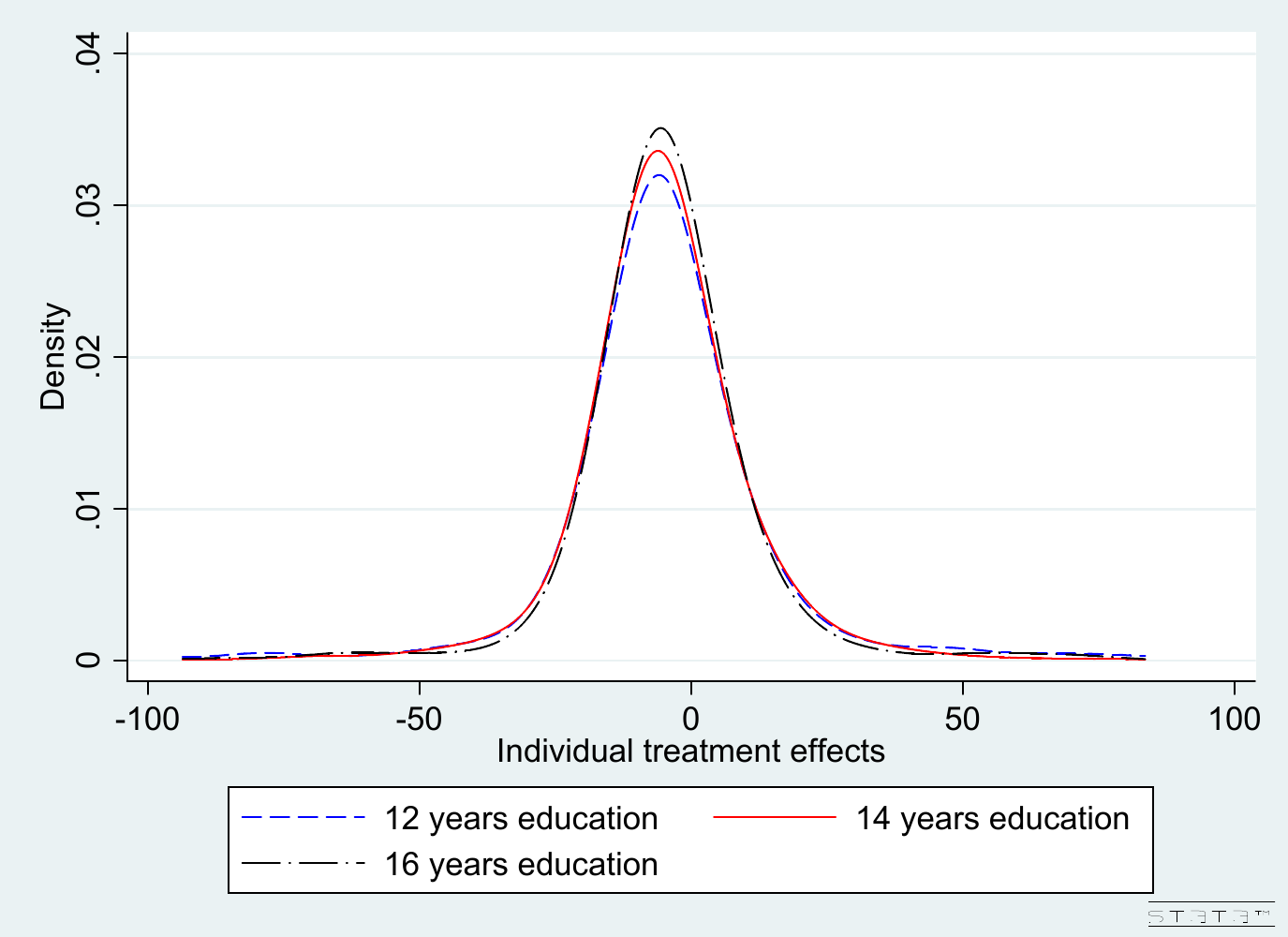} 
   \caption{EHIV variance effects and ITE distributions (education)}
   \label{fig5}
\end{figure}

We also plot $\sigma(d,x)$ at different values of $x$. Fixing age at first birth, 1st child's age, and 2nd child's age at their median values, we first estimate $\sigma(d,x)$ as a function of the treatment variable and the mother's education level. The top-left figure in \Cref{fig5} shows the density of the education variable, which leads us to focus our estimation of $\sigma(d,x)$ on the range between 10 and 20 years of education. The estimated $\sigma(d,0)$ and $\sigma(d,1)$ functions (i.e., as a function of education) are shown in the bottom-left figure of \Cref{fig5}. The top-right figure of \Cref{fig5} gives a sense of the size of the complier group, as it shows $|\hat{p}(x,1)-\hat{p}(x,0)|$ as a function of education (again fixing other covariates at their median). Finally, we provide the estimated ITE distributions for three different levels of education (12 years, 14 years, 16 years) in the bottom-right figure of \Cref{fig5}. The most notable feature of the ITE distributions is the large amount of heterogeneity in the ITE's. Although the center of these ITE distributions lines up with the EHIV coefficient estimate ($-5.343$) from \Cref{table4}, the region of non-negligible positive weight includes positive ITE's of up to 20 hours and negative ITE's as low as -30 hours. 

\begin{figure}[!ht] 
   \centering
   \includegraphics[width=2.5in]{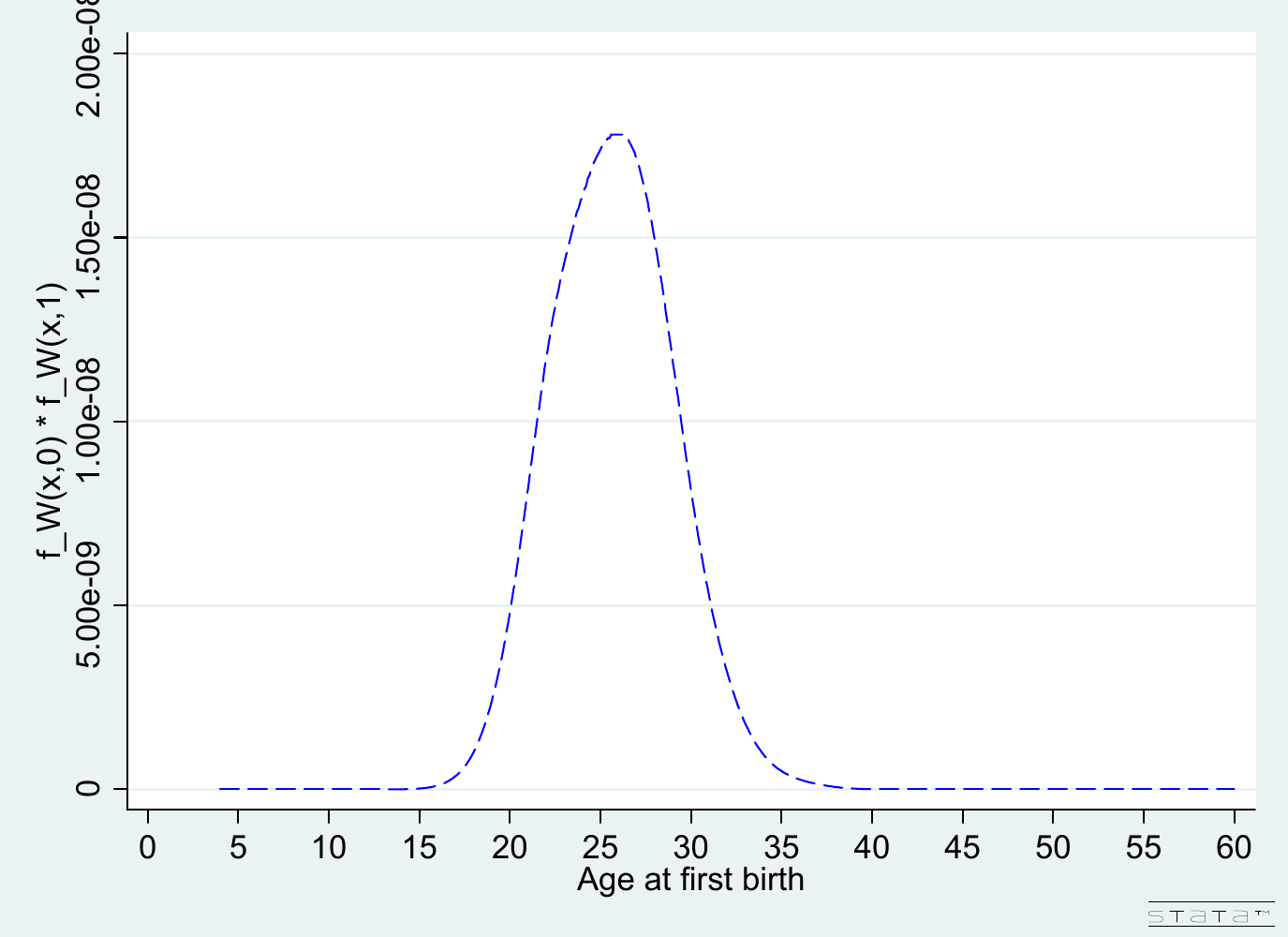} 
     \includegraphics[width=2.5in]{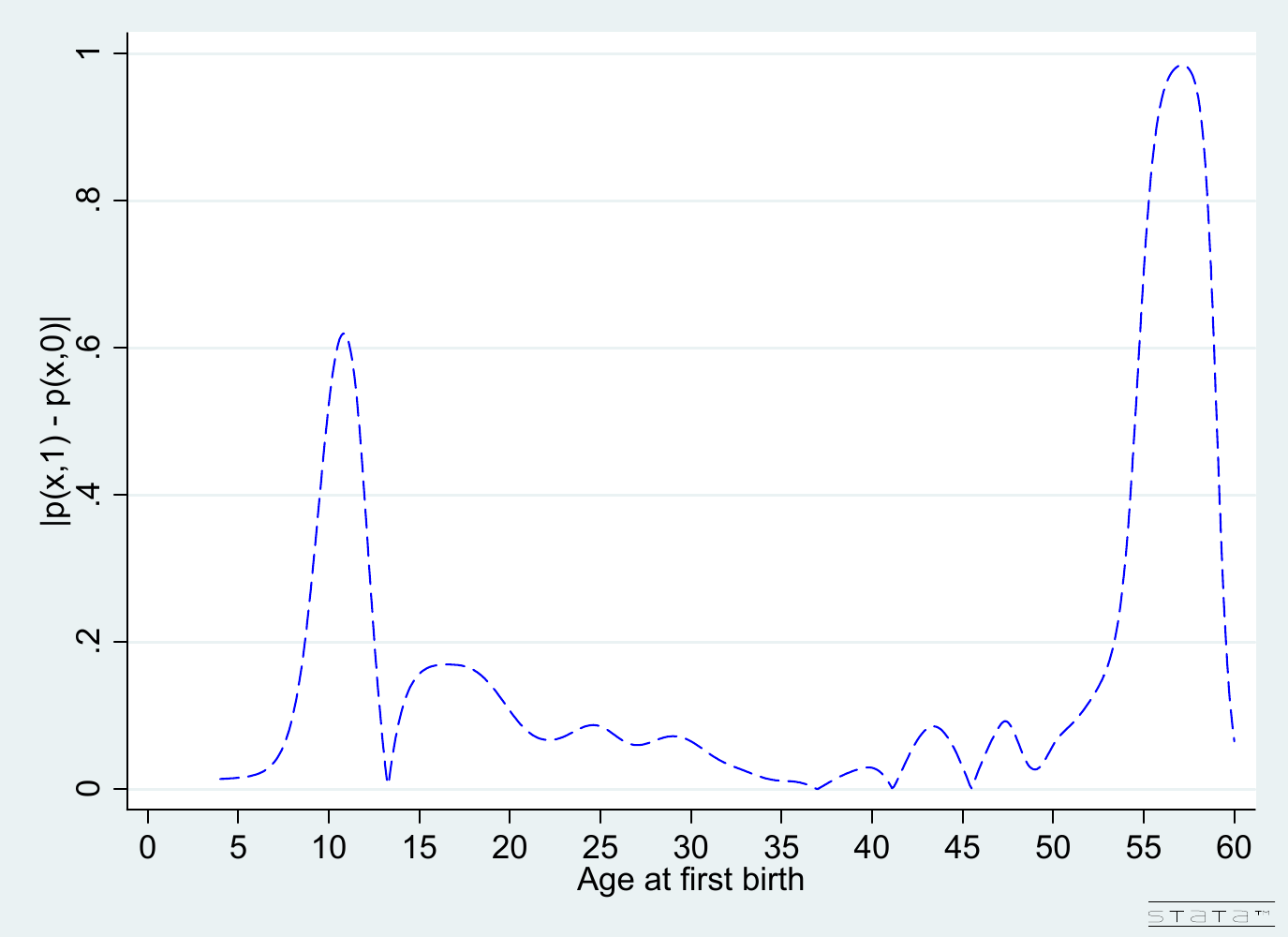} \\
      \includegraphics[width=2.5in]{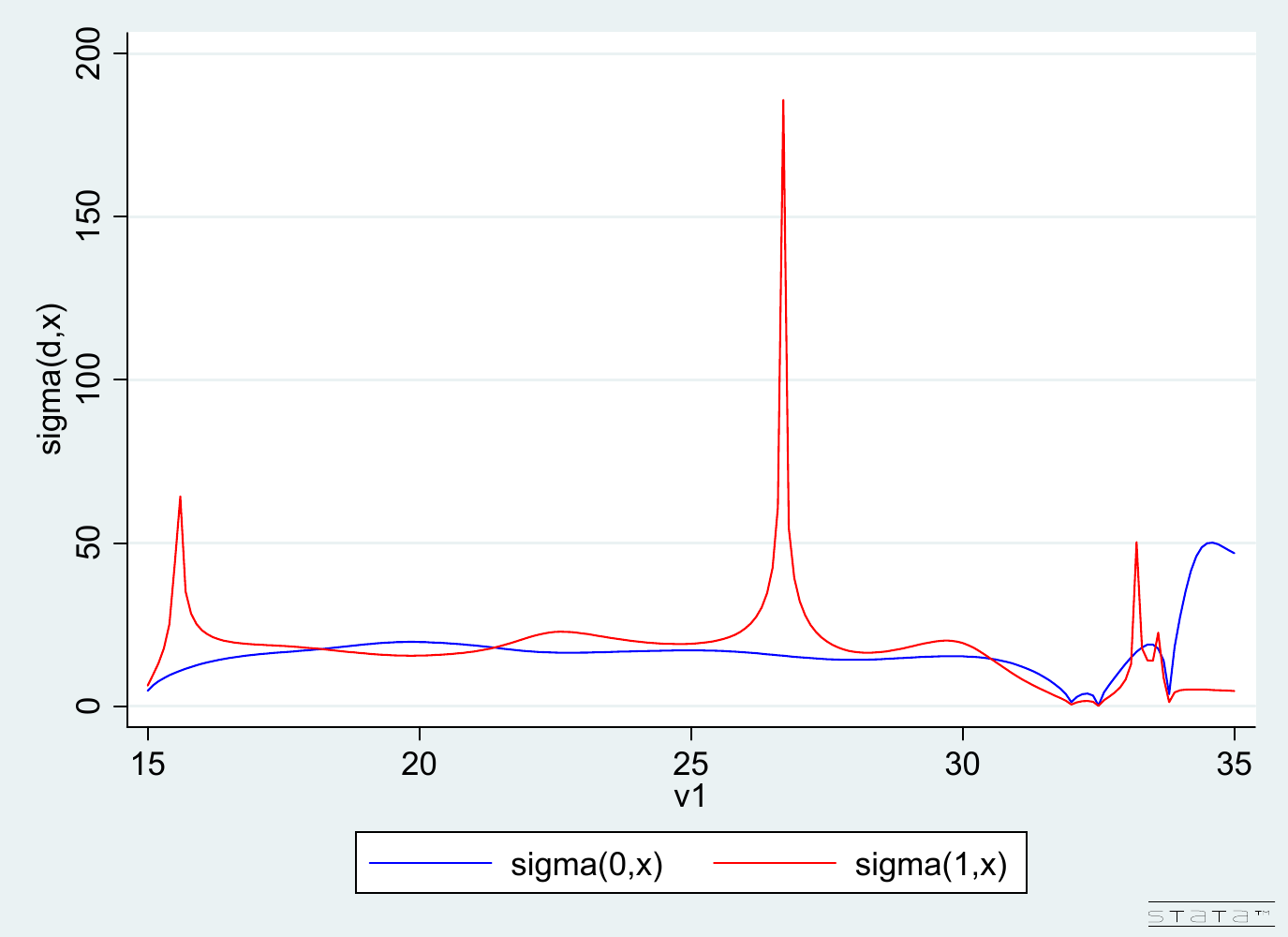} 
             \includegraphics[width=2.5in]{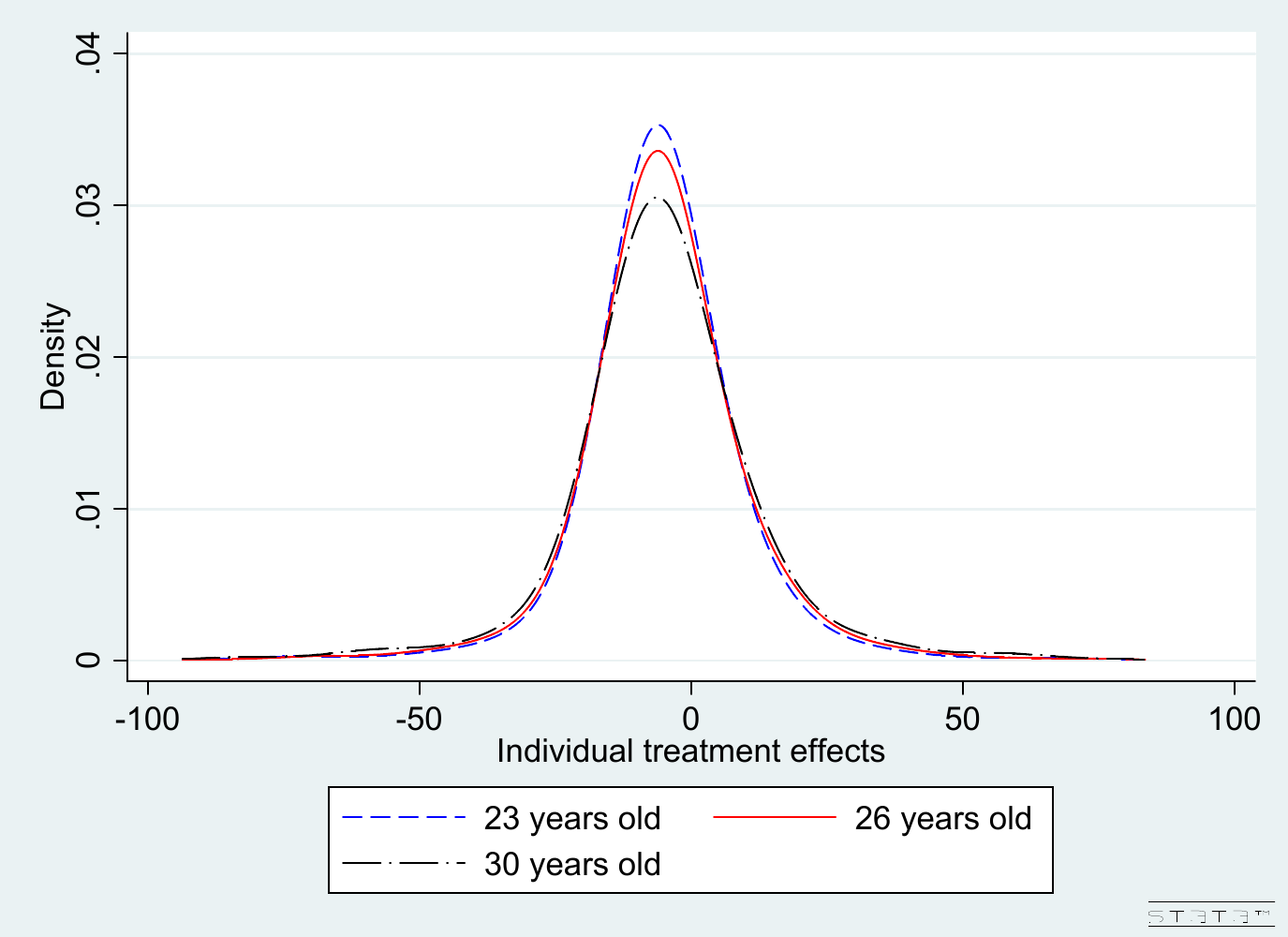} 
   \caption{EHIV variance effects and ITE distributions (age at first birth)}
   \label{fig6}
\end{figure}

\begin{figure}[!ht] 
   \centering
   \includegraphics[width=2.5in]{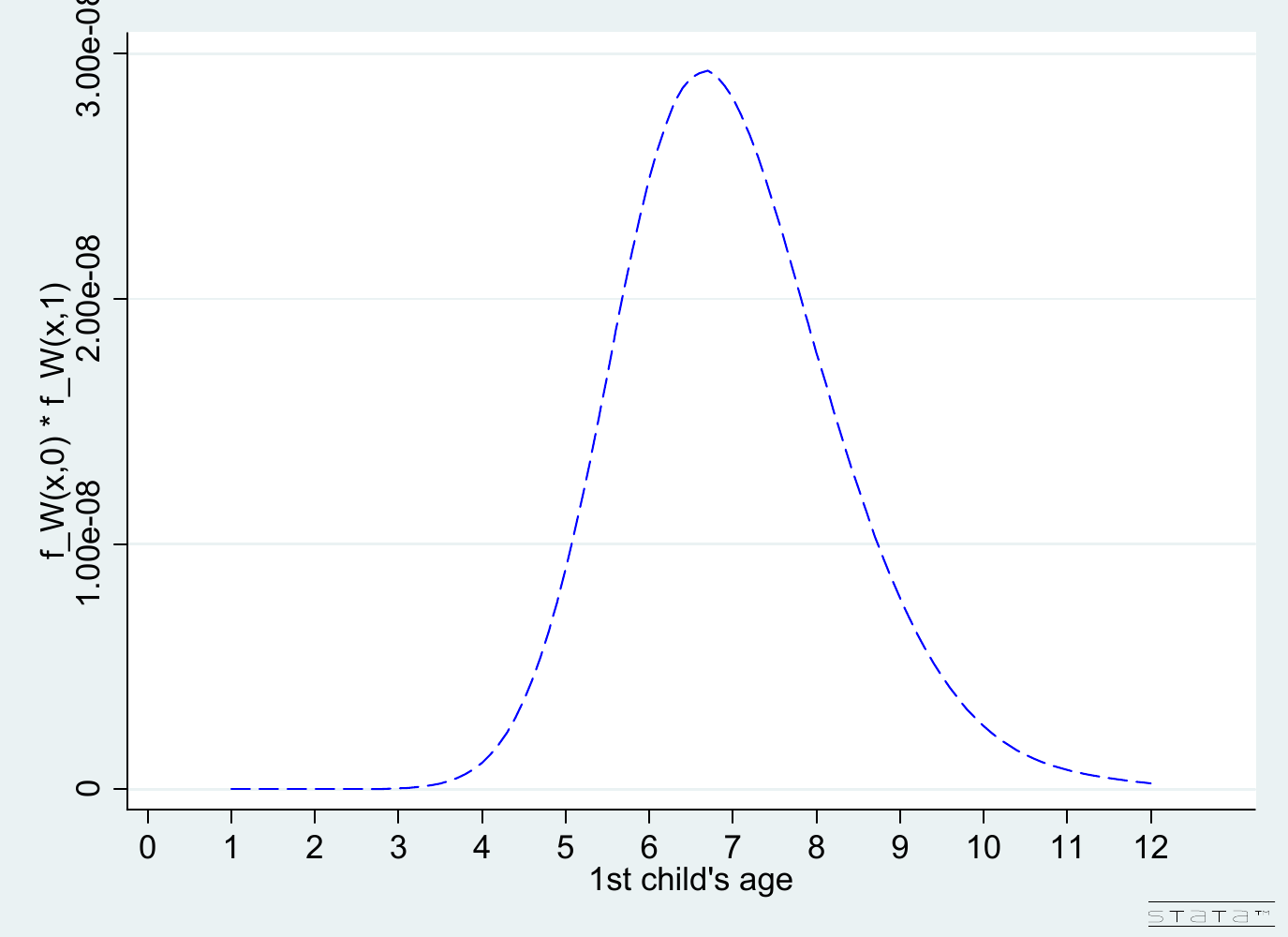} 
     \includegraphics[width=2.5in]{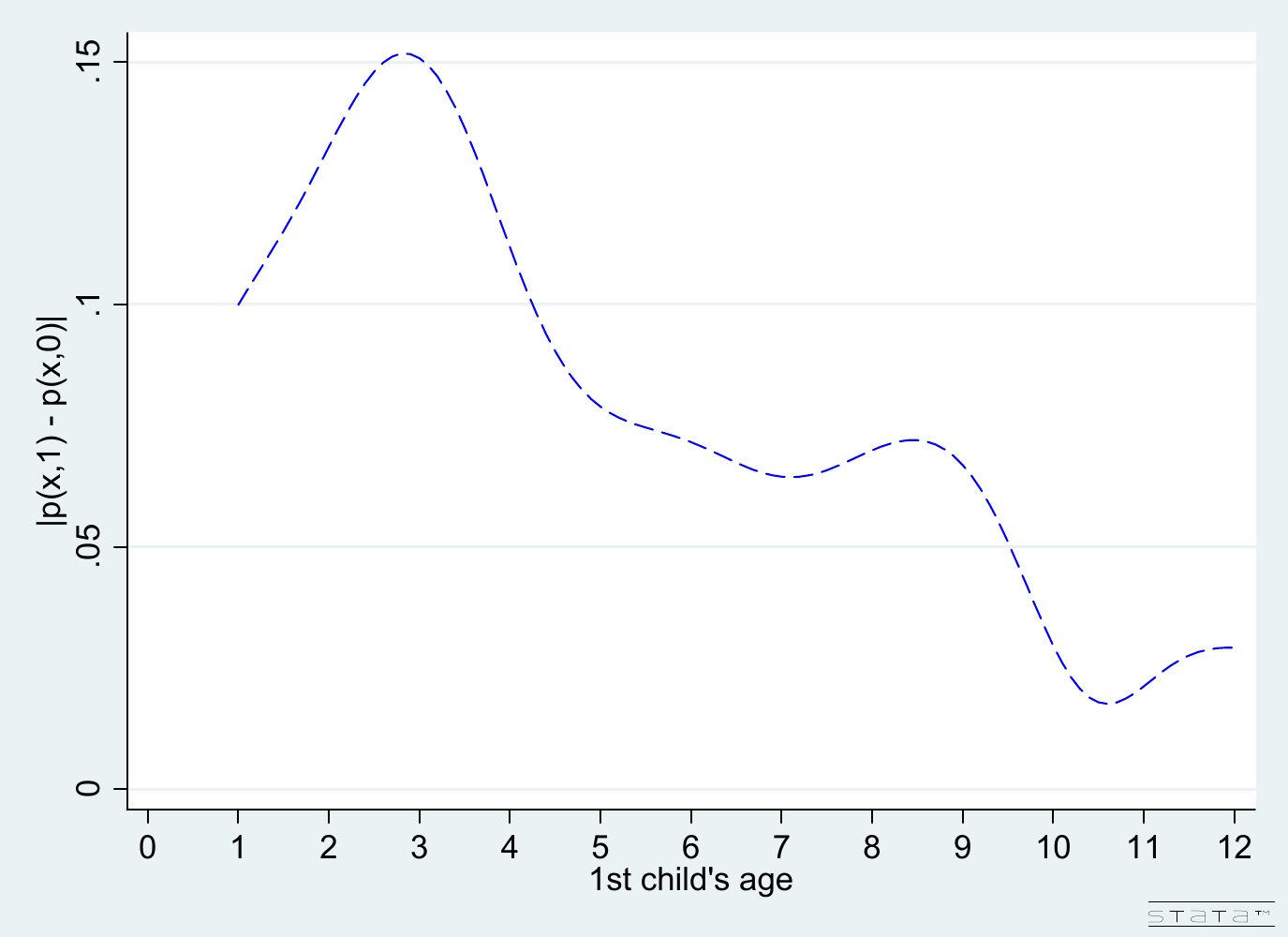} \\
   \includegraphics[width=2.5in]{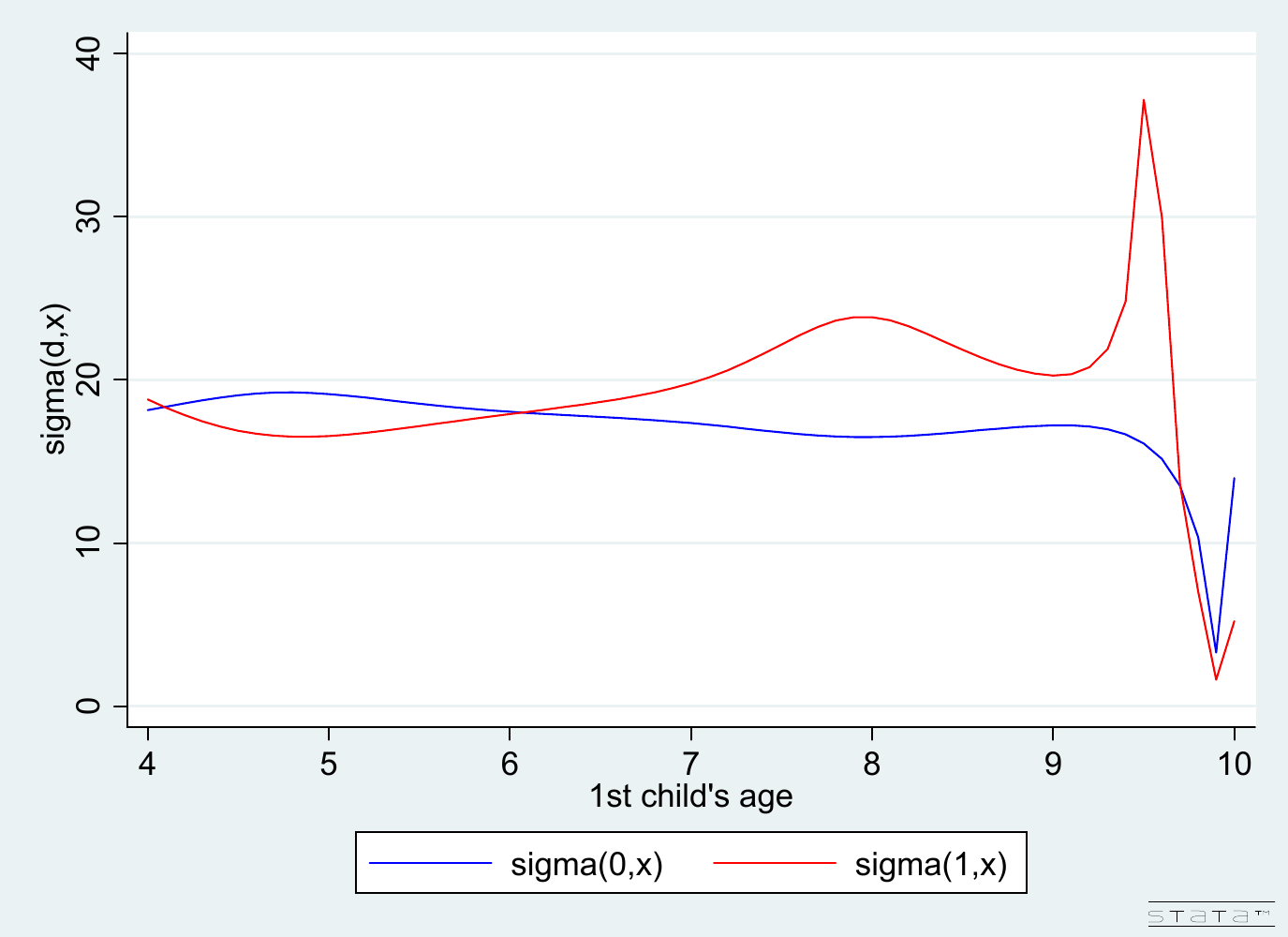} 
            \includegraphics[width=2.5in]{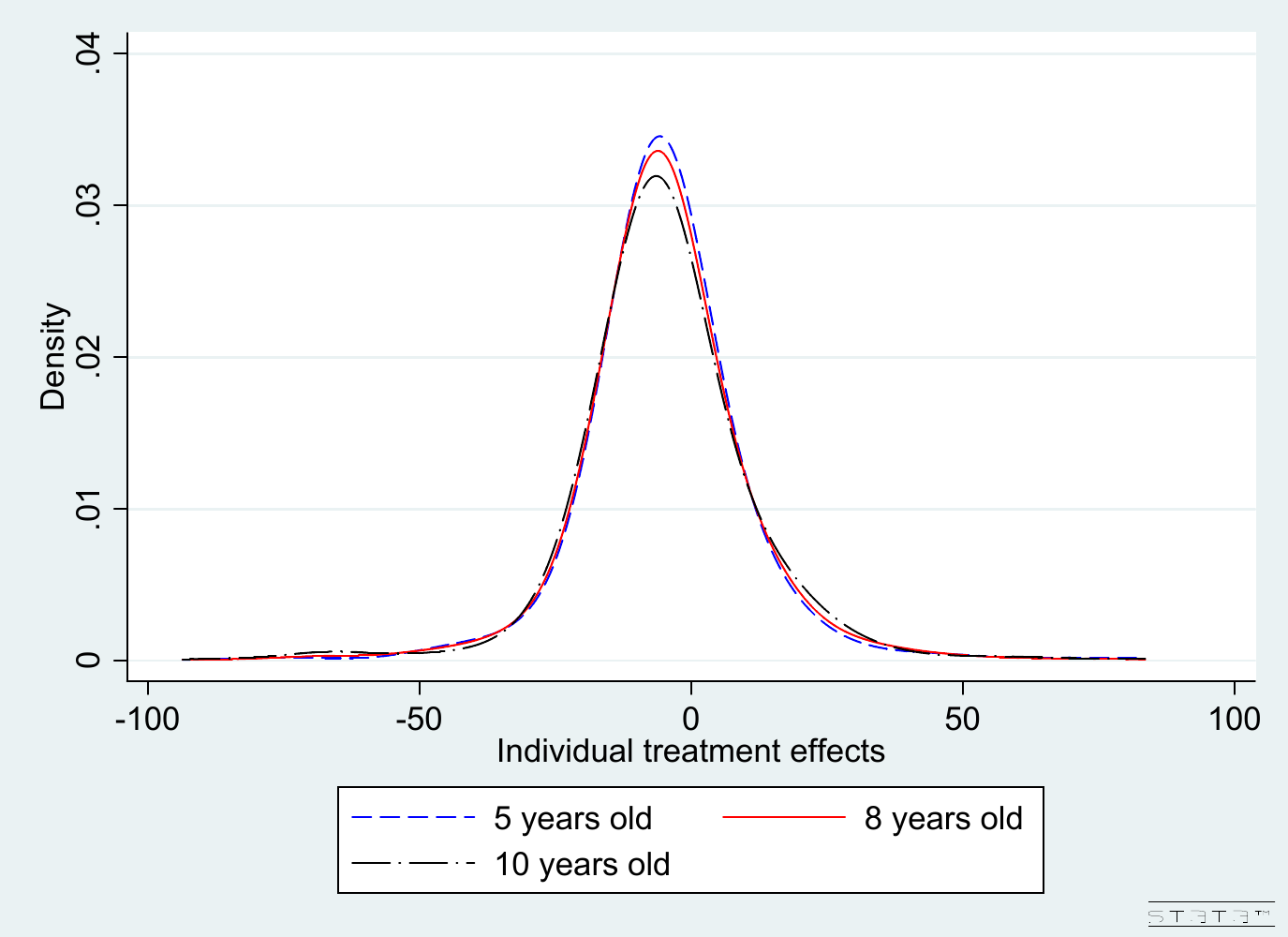} 
   \caption{EHIV variance effects and ITE distributions (1st child's age)}
   \label{fig7}
\end{figure}

\begin{figure}[!ht] 
   \centering
   \includegraphics[width=2.5in]{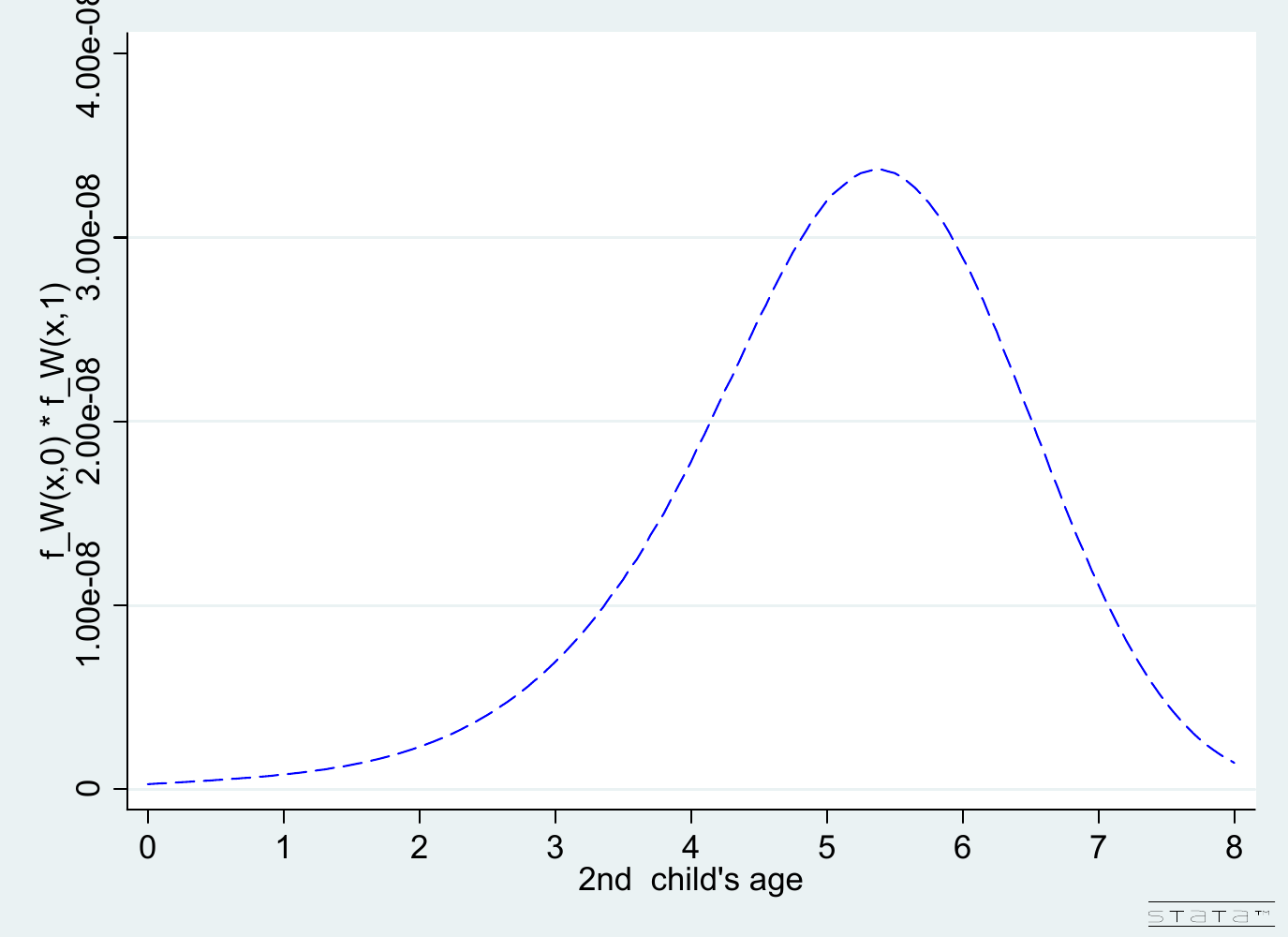} 
     \includegraphics[width=2.5in]{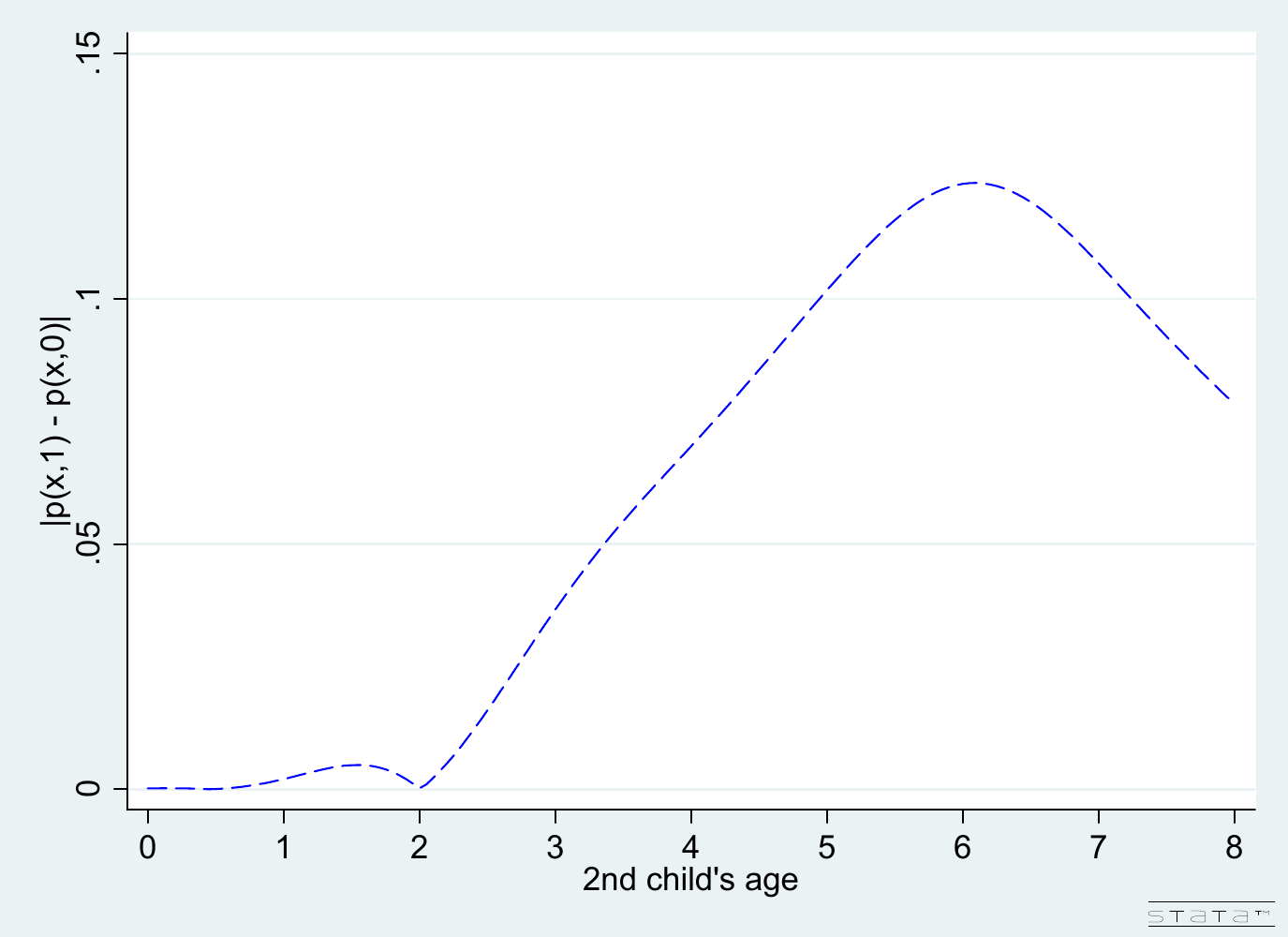} \\
        \includegraphics[width=2.5in]{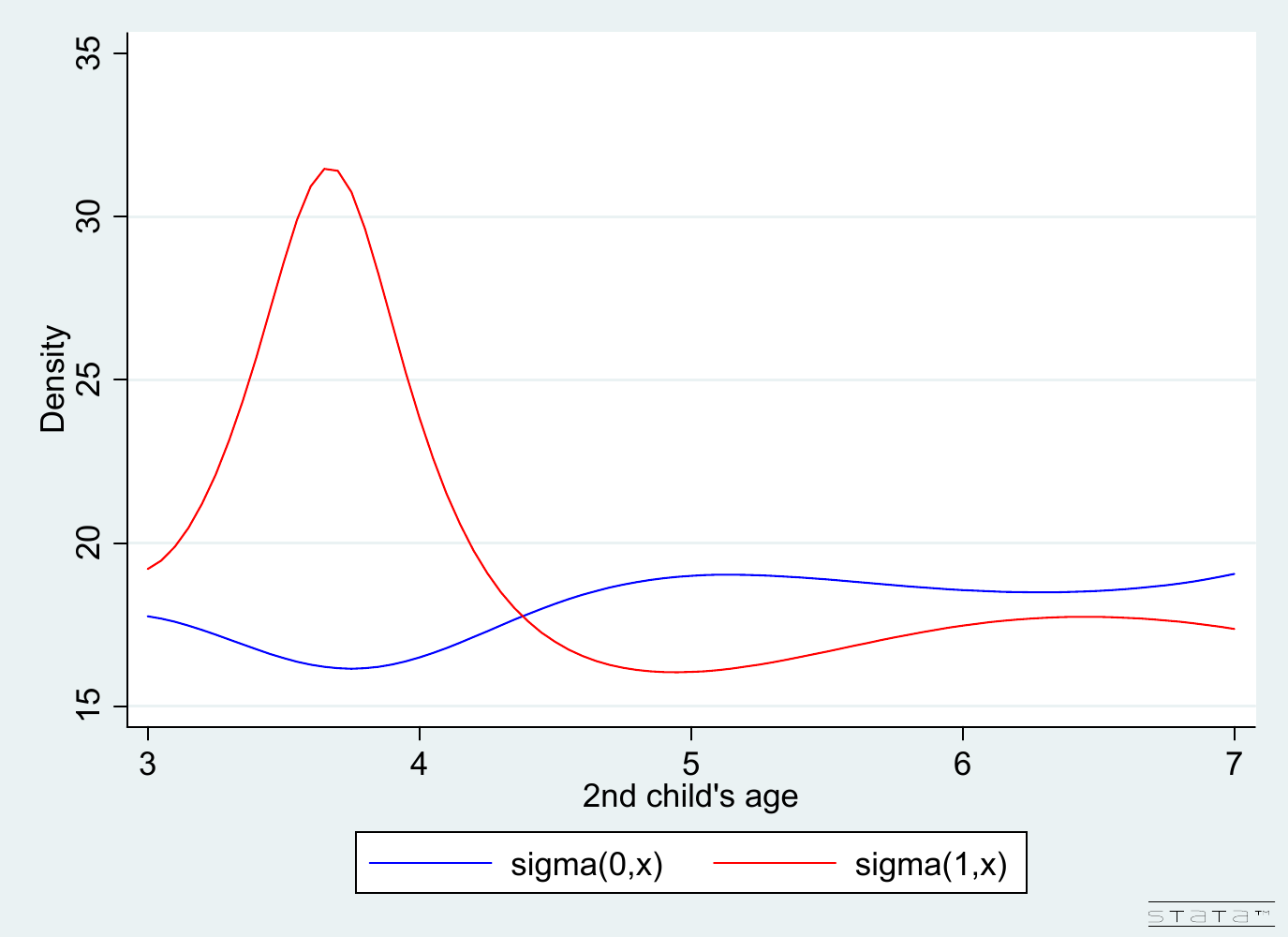} 
       \includegraphics[width=2.5in]{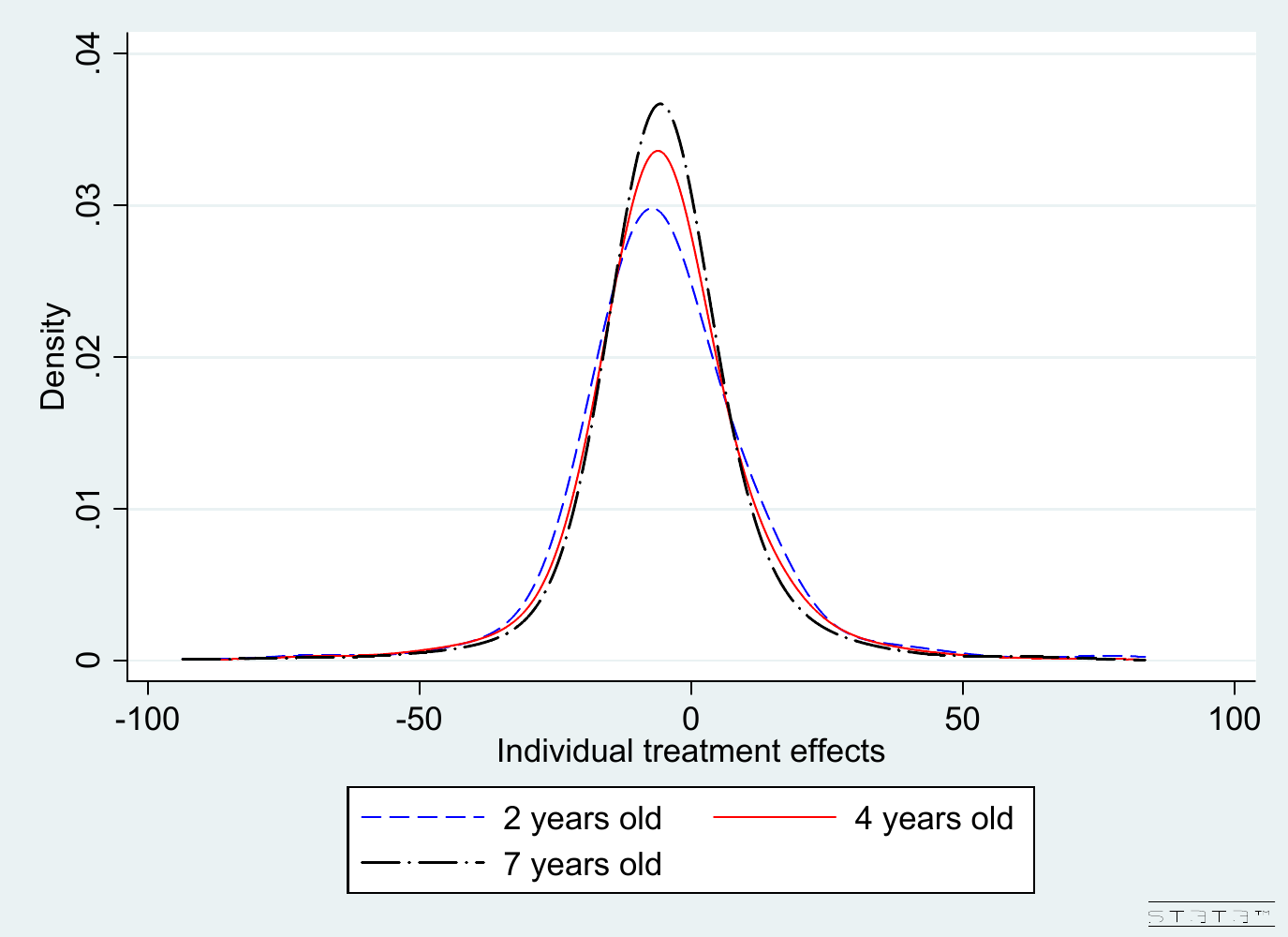} 
   \caption{EHIV variance effects and ITE distributions (2nd child's age)}
   \label{fig8}
\end{figure}

\Cref{fig6,fig7,fig8} are similar to \Cref{fig5}, except that they consider the other three exogenous variables (age at first birth, 1st child's age, and 2nd child's age, respectively). For example, \Cref{fig6} provides estimates of $\sigma(d,x)$ and the ITE distributions as functions of age at first birth, with the other exogenous covariates fixed at their median values. Not surprisingly, the large heterogeneity found in the ITE distributions (each in the lower-right of the corresponding figure) is similar to that seen in \Cref{fig5}. In terms of how these distributions vary for different covariate values, it appears that the largest differences are found for age at first birth (\Cref{fig6}) and 2nd child's age (\Cref{fig8}).
 
\clearpage

\section{Extensions and Conclusion}

This paper has considered identification and estimation of a linear model with endogenous heteroskedasticity. Our model assumes that the treatment variable has both mean and variance effects on the outcome variable, which implies heterogenous treatment effects even among observationally identical individuals. Because of the endogenous heteroskedasticity, the standard IV estimator is inconsistent. We then propose a consistent estimation procedure, modified from the IV approach,  which has a closed-form expression and is simple to implement.  Under appropriate conditions, we establish the $\sqrt n$-consistency and the limiting normal distribution for the proposed  estimator.  Monte Carlo simulations show that the EHIV estimator works well even in moderately sized samples. 

An issue briefly discussed within our empirical application is how to test for endogenous heteroskedasticity. If the heteroskedasticity is indeed exogenous, there are efficiency gains to using the usual IV methods (rather than EHIV), which can be attractive especially for smaller sample sizes. While we conducted a parametric test of exogeneity in \Cref{sec:empirical}, it would be interesting to develop a nonparametric test of $\mathbb H_0: \sigma(0,\cdot)=\sigma(1,\cdot)=\tilde \sigma(\cdot)$ for some $\tilde \sigma: \mathbb R^{d_X}\rightarrow \mathbb R_+$. Under \Cref{ass1,ass2,rank},   \Cref{lemma0} implies that $\mathbb H_0$ holds if and only if  $V_0(X)=V_1(X)$ holds a.s., which suggests that a  test could be developed based upon nonparametric model specification tests developed in the statistics and econometrics literature \citep[see e.g.][]{fan1996consistent,lavergne1996nonparametric,blundell2007non}. Given the widely used  IV method, however,  it's more convenient to develop an IV-residual-based test procedure for exogenous heteroskedasticity. Difficulties arise from the inconsistency of IV under the alternative hypothesis, which brings concern that the  IV-residual-based test might not have  power against some alternatives. In the next lemma,  we show that the  endogenous heteroskedasticity can be consistently detected by the IV residuals.   
\begin{lemma}
\label{lemma1}
Suppose \eqref{eq1} and \Cref{ass1,ass2,rank} hold. Then $\sigma(X,0)=\sigma(X,1)$ if and only if 
\begin{equation}
\label{nonpara_IVreg}
\mathbb E ( \tilde\epsilon^2|X,Z=0)=\mathbb E ( \tilde\epsilon^2|X,Z=1)
\end{equation}where $\tilde \epsilon=Y-\tilde r_0(X)-\tilde r_1(X)D$, in which $\tilde r_1(X)= \text{Cov}(Y,Z|X)/\text{Cov}(D,Z|X)$ and $\tilde r_0 (X)=[\text{Cov}(YZ,D|X)-\text{Cov}(Y,DZ|X)]/\text{Cov}(D,Z|X)$. In addition, suppose the semiparametric model \eqref{parametric model} holds. Then $\sigma(X,0)=\sigma(X,1)$ if and only if 
\[
\mathbb E \big[(Y- X'\tilde\beta_1-\tilde \beta_2 D)^2|X,Z=0\big]=\mathbb E \big[(Y- X'\tilde\beta_1-\tilde \beta_2 D)^2|X,Z=1\big]
\]where $\tilde \beta=(\tilde \beta_1',\tilde\beta_2)'$ satisfies $\mathbb E (Y- X'\tilde\beta_1-\tilde \beta_2 D|X,Z)=0$. 
\end{lemma}
\noindent
In \Cref{lemma1},  note that $\tilde\epsilon$ is the residual  from the nonparametric IV regression, and $\tilde \beta$ could be estimated by the  usual IV approach.

Another avenue for future research is to deal with a continuously supported endogenous treatment $D$. Nonparametric identification for this case has been established in \cite{chesher2003identification,chernozhukov2005iv,imbens2009identification,d2011identification,torgovitsky2011identification} in a general framework. For estimation, \cite{imbens2009identification}'s control function approach could be naturally extended to our semiparametric specification.

\bibliographystyle{econometrica}
\bibliography{bibles_ASF}

@article{maurin2009social,
	Author = {Maurin, Eric and Moschion, Julie},
	Journal = {American Economic Journal: Applied Economics},
	Number = {1},
	Pages = {251--72},
	Title = {The social multiplier and labor market participation of mothers},
	Volume = {1},
	Year = {2009}}

@article{rosenzweig1980life,
	Author = {Rosenzweig, Mark R and Wolpin, Kenneth I},
	Journal = {Journal of Political economy},
	Number = {2},
	Pages = {328--348},
	Publisher = {The University of Chicago Press},
	Title = {Life-cycle labor supply and fertility: Causal inferences from household models},
	Volume = {88},
	Year = {1980}}

@article{rosenzweig1980testing,
	Author = {Rosenzweig, Mark R and Wolpin, Kenneth I},
	Journal = {Econometrica: journal of the Econometric Society},
	Pages = {227--240},
	Publisher = {JSTOR},
	Title = {Testing the quantity-quality fertility model: The use of twins as a natural experiment},
	Year = {1980}}

@techreport{gangadharan1996effects,
	Author = {Gangadharan, Jaisri and Rosenbloom, Joshua and Jacobson, Joyce and Pearre III, James Wishart},
	Institution = {National Bureau of Economic Research},
	Title = {The effects of child-bearing on married women's labor supply and earnings: Using twin births as a natural experiment},
	Year = {1996}}

@article{bronars1994economic,
	Author = {Bronars, Stephen G and Grogger, Jeff},
	Journal = {The American Economic Review},
	Pages = {1141--1156},
	Publisher = {JSTOR},
	Title = {The economic consequences of unwed motherhood: Using twin births as a natural experiment},
	Year = {1994}}

@article{chen2014semi,
	Author = {Chen, Stacey H and Khan, Shakeeb},
	Journal = {Journal of Applied Econometrics},
	Number = {6},
	Pages = {901--919},
	Publisher = {Wiley Online Library},
	Title = {Semi-parametric estimation of program impacts on dispersion of potential wages},
	Volume = {29},
	Year = {2014}}

@article{angrist1998children,
	Author = {Angrist, Joshua D and Evans, William N},
	Journal = {The American Economic Review},
	Number = {3},
	Pages = {450},
	Publisher = {American Economic Association},
	Title = {Children and their parents' labor supply: Evidence from exogenous variation in family size},
	Volume = {88},
	Year = {1998}}

@article{vuong2015nonparametric,
	Author = {Feng, Qian and Vuong, Quang and Xu, Haiqing},
	Institution = {University of Texas at Austin and New York University},
	Journal = {arXiv preprint arXiv:1610.08899},
	Title = {Nonparametric estimation of heterogeneous individual treatment effects with endogenous treatments},
	Year = {2016}}

@article{blundell2007non,
	Author = {Blundell, Richard and Horowitz, Joel L},
	Journal = {The Review of Economic Studies},
	Number = {4},
	Pages = {1035--1058},
	Publisher = {Oxford University Press},
	Title = {A non-parametric test of exogeneity},
	Volume = {74},
	Year = {2007}}

@article{wan2015inference,
	Author = {Wan, Yuanyuan and Xu, Haiqing},
	Journal = {Journal of Econometrics},
	Number = {2},
	Pages = {347--360},
	Publisher = {Elsevier},
	Title = {Inference in semiparametric binary response models with interval data},
	Volume = {184},
	Year = {2015}}

@article{racine2004nonparametric,
	Author = {Racine, Jeff and Li, Qi},
	Journal = {Journal of Econometrics},
	Number = {1},
	Pages = {99--130},
	Publisher = {Elsevier},
	Title = {Nonparametric estimation of regression functions with both categorical and continuous data},
	Volume = {119},
	Year = {2004}}

@article{andrews1995nonparametric,
	Author = {Andrews, Donald WK},
	Journal = {Econometric Theory},
	Number = {03},
	Pages = {560--586},
	Publisher = {Cambridge Univ Press},
	Title = {Nonparametric kernel estimation for semiparametric models},
	Volume = {11},
	Year = {1995}}

@article{bierens1983uniform,
	Author = {Bierens, Herman J},
	Journal = {Journal of the American Statistical Association},
	Number = {383},
	Pages = {699--707},
	Publisher = {Taylor \& Francis Group},
	Title = {Uniform consistency of kernel estimators of a regression function under generalized conditions},
	Volume = {78},
	Year = {1983}}

@article{andrews1994asymptotics,
	Author = {Andrews, Donald WK},
	Journal = {Econometrica: Journal of the Econometric Society},
	Pages = {43--72},
	Publisher = {JSTOR},
	Title = {Asymptotics for semiparametric econometric models via stochastic equicontinuity},
	Year = {1994}}

@article{newey1994large,
	Author = {Newey, Whitney K and McFadden, Daniel},
	Journal = {Handbook of econometrics},
	Pages = {2111--2245},
	Publisher = {Elsevier},
	Title = {Large sample estimation and hypothesis testing},
	Volume = {4},
	Year = {1994}}

@article{ai2003efficient,
	Author = {Ai, Chunrong and Chen, Xiaohong},
	Journal = {Econometrica},
	Number = {6},
	Pages = {1795--1843},
	Publisher = {Wiley Online Library},
	Title = {Efficient estimation of models with conditional moment restrictions containing unknown functions},
	Volume = {71},
	Year = {2003}}

@article{barrett2003consistent,
	Author = {Barrett, Garry F and Donald, Stephen G},
	Journal = {Econometrica},
	Number = {1},
	Pages = {71--104},
	Publisher = {Wiley Online Library},
	Title = {Consistent tests for stochastic dominance},
	Volume = {71},
	Year = {2003}}

@article{abadie2002bootstrap,
	Author = {Abadie, Alberto},
	Journal = {Journal of the American statistical Association},
	Number = {457},
	Pages = {284--292},
	Publisher = {Taylor \& Francis},
	Title = {Bootstrap tests for distributional treatment effects in instrumental variable models},
	Volume = {97},
	Year = {2002}}

@article{fan1996consistent,
	Author = {Fan, Yanqin and Li, Qi},
	Journal = {Econometrica: Journal of the econometric society},
	Pages = {865--890},
	Publisher = {JSTOR},
	Title = {Consistent model specification tests: omitted variables and semiparametric functional forms},
	Year = {1996}}

@article{powell1989semiparametric,
	Author = {Powell, James L and Stock, James H and Stoker, Thomas M},
	Journal = {Econometrica: Journal of the Econometric Society},
	Pages = {1403--1430},
	Publisher = {JSTOR},
	Title = {Semiparametric estimation of index coefficients},
	Year = {1989}}

@book{PaganUllah1999,
	Author = {Pagan, Adrian and Ullah,Aman},
	Publisher = {Cambridge University Press},
	Title = {Nonparametric Econometrics},
	Year = {1999}}

@article{klein1993efficient,
	Author = {Klein, Roger W and Spady, Richard H},
	Journal = {Econometrica: Journal of the Econometric Society},
	Pages = {387--421},
	Publisher = {JSTOR},
	Title = {An efficient semiparametric estimator for binary response models},
	Year = {1993}}

@article{angrist1991does,
	Author = {Angrist, Joshua D and Krueger, Alan B},
	Journal = {The Quarterly Journal of Economics},
	Number = {4},
	Pages = {979--1014},
	Publisher = {The MIT Press},
	Title = {Does Compulsory School Attendance Affect Schooling and Earnings?},
	Volume = {106},
	Year = {1991}}

@article{heckman1997making,
	Author = {Heckman, James J and Smith, Jeffrey and Clements, Nancy},
	Journal = {The Review of Economic Studies},
	Number = {4},
	Pages = {487--535},
	Publisher = {Oxford University Press},
	Title = {Making the most out of programme evaluations and social experiments: Accounting for heterogeneity in programme impacts},
	Volume = {64},
	Year = {1997}}

@article{heckman2007econometric,
	Author = {Heckman, James J and Vytlacil, Edward J},
	Journal = {Handbook of econometrics},
	Pages = {4779--4874},
	Publisher = {Elsevier},
	Title = {Econometric evaluation of social programs, part I: Causal models, structural models and econometric policy evaluation},
	Volume = {6},
	Year = {2007}}

@article{lavergne1996nonparametric,
	Author = {Lavergne, Pascal and Vuong, Quang H},
	Journal = {Econometrica},
	Number = {1},
	Pages = {207--219},
	Publisher = {JSTOR},
	Title = {Nonparametric selection of regressors: The nonnested case},
	Volume = {64},
	Year = {1996}}

@article{vuong2014counterfactual,
	Author = {Vuong, Quang and Xu, Haiqing},
	Journal = {Quantitative Economics},
	Title = {Counterfactual mapping and individual treatment effects in nonseparable models with discrete endogeneity},
	Year = {2017}}

@article{vytlacil2002independence,
	Author = {Vytlacil, Edward},
	Journal = {Econometrica},
	Number = {1},
	Pages = {331--341},
	Publisher = {Wiley Online Library},
	Title = {Independence, monotonicity, and latent index models: An equivalence result},
	Volume = {70},
	Year = {2002}}

@article{guerre2000optimal,
	Author = {Guerre, Emmanuel and Perrigne, Isabelle and Vuong, Quang},
	Journal = {Econometrica},
	Number = {3},
	Pages = {525--574},
	Title = {Optimal nonparametric estimation of first--price auctions},
	Volume = {68},
	Year = {2000}}

@article{heckman2005structural,
	Author = {Heckman, James J and Vytlacil, Edward},
	Journal = {Econometrica},
	Number = {3},
	Pages = {669--738},
	Publisher = {Wiley Online Library},
	Title = {Structural equations, treatment effects, and econometric policy evaluation},
	Volume = {73},
	Year = {2005}}

@article{imbens1994identification,
	Author = {Imbens, Guido W and Angrist, Joshua D},
	Journal = {Econometrica},
	Number = {2},
	Pages = {467--475},
	Publisher = {JSTOR},
	Title = {Identification and estimation of local average treatment effects},
	Volume = {62},
	Year = {1994}}

@article{d2011identification,
	Author = {D'Haultf{\oe}uille, Xavier and F{\'e}vrier, Philippe},
	Journal = {Econometrica},
	Number = {3},
	Pages = {1199--1210},
	Publisher = {Wiley Online Library},
	Title = {Identification of nonseparable triangular models with discrete instruments},
	Volume = {83},
	Year = {2015}}

@article{torgovitsky2011identification,
	Author = {Torgovitsky, Alexander},
	Journal = {Econometrica},
	Number = {3},
	Pages = {1185--1197},
	Publisher = {Wiley Online Library},
	Title = {Identification of nonseparable models using instruments with small support},
	Volume = {83},
	Year = {2015}}

@article{chernozhukov2005iv,
	Author = {Chernozhukov, Victor and Hansen, Christian},
	Journal = {Econometrica},
	Number = {1},
	Pages = {245--261},
	Publisher = {Wiley Online Library},
	Title = {An IV model of quantile treatment effects},
	Volume = {73},
	Year = {2005}}

@article{cherno2004401kplan,
	Author = {Chernozhukov, Victor and Hansen, Christian},
	Journal = {The Review of Economics and Statistics},
	Number = {3},
	Pages = {735--751},
	Publisher = {MIT Press Journals},
	Title = {The effects of 401(K) participation on the wealth distribution: an instrumental quantile regression analysis},
	Volume = {86},
	Year = {2004}}

@article{jun2011tighter,
	Author = {Jun, Sung Jae and Pinkse, Joris and Xu, Haiqing},
	Journal = {Journal of Econometrics},
	Number = {2},
	Pages = {122--128},
	Publisher = {Elsevier},
	Title = {Tighter bounds in triangular systems},
	Volume = {161},
	Year = {2011}}

@article{imbens2009identification,
	Author = {Imbens, Guido W and Newey, Whitney K},
	Journal = {Econometrica},
	Number = {5},
	Pages = {1481--1512},
	Publisher = {Wiley Online Library},
	Title = {Identification and estimation of triangular simultaneous equations models without additivity},
	Volume = {77},
	Year = {2009}}

@article{chesher2005nonparametric,
	Author = {Chesher, Andrew},
	Journal = {Econometrica},
	Number = {5},
	Pages = {1525--1550},
	Publisher = {Wiley Online Library},
	Title = {Nonparametric identification under discrete variation},
	Volume = {73},
	Year = {2005}}

@article{chesher2003identification,
	Author = {Chesher, Andrew},
	Journal = {Econometrica},
	Number = {5},
	Pages = {1405--1441},
	Publisher = {Wiley Online Library},
	Title = {Identification in nonseparable models},
	Volume = {71},
	Year = {2003}}

\clearpage
\appendix
\small

\clearpage

\section{Proofs}

\subsection{Proof of \Cref{theorem1}}
\proof
By the definition of $\hat \beta$ and \eqref{parametric model}, 
\[
\hat \beta -\beta=\Big[\frac{1}{n}\sum_{i=1}^n\frac{T_{ni} W_i(X'_i, D_i)}{\hat S_i}\Big)\Big]^{-1}\times \frac{1}{n}\sum_{i=1}^n \frac{T_{ni}W_i\sigma(D_i,X_i)\epsilon_i}{\hat S_i}
\]
By \Cref{lemma3,lemma4}, $\frac{1}{n}\sum_{i=1}^n\frac{T_{ni}W_i'(X'_i, D_i)}{\hat S_i}$ converges in probability to $\mathbb E \big[\frac{W(X',D)}{S}\big]$ and $\frac{1}{n}\sum_{i=1}^n \frac{T_{ni}W_i\sigma(D_i,X_i)\epsilon_i}{\hat S_i}$ converges in probability to zero. 
By \Cref{ass9} and Slutsky's Theorem,  $\hat \beta-\beta\overset{p}{\rightarrow} 0$. 
\qed

\section{Proof of \Cref{theorem2}}\label{prooftheorem2}
\proof
By definition of $\hat \beta$ and \eqref{parametric model}, we have
\[
\sqrt n(\hat \beta -\beta)= \left[\frac{1}{n}\sum_{i=1}^n\frac{T_{ni}W_i(X'_i, D_i)}{\hat S_i}\right]^{-1} \frac{1}{\sqrt n}\sum_{i=1}^n \frac{T_{ni}W_i\sigma(D_i,X_i) \epsilon_i}{\hat S_i}.
\]
First, note that
\[
\frac{1}{n}\sum_{i=1}^n\frac{T_{ni}W_i'(X'_i, D_i)}{\hat S_i}
=\frac{1}{n}\sum_{i=1}^n\frac{T_{ni}W_i(X'_i, D_i)}{S_i}+\frac{1}{n}\sum_{i=1}^n\Big(\frac{S_i}{\hat S_i}-1\Big)\frac{T_{ni}W_i(X'_i, D_i)}{S_i}.
\]By \Cref{lemma3,lemma4}, 
\[
\frac{1}{n}\sum_{i=1}^n\frac{T_{ni}W_i'(X'_i, D_i)}{\hat S_i}\overset{p}{\rightarrow}\mathbb E [W(X',D)/S].
\]Hence, it suffices to derive the limiting distribution of  $\frac{1}{\sqrt n}\sum_{i=1}^n \frac{T_{ni}W_i\sigma(D_i,X_i) \epsilon_i}{\hat S_i}$.

Next, note that
\begin{multline*}
\frac{1}{\sqrt n}\sum_{i=1}^n \frac{T_{ni}W_i\sigma(D_i,X_i)\epsilon_i}{\hat S_i}=\frac{1}{\sqrt n}\sum_{i=1}^n\frac{T_{ni}W_i \epsilon_i}{\sqrt{|C(X_i)|}}+\frac{1}{\sqrt n}\sum_{i=1}^nT_{ni}\Big(\frac{S_i}{\hat S_i}-1\Big)\frac{W_i \epsilon_i}{\sqrt{|C(X_i)|}}\\
=\frac{1}{\sqrt n}\sum_{i=1}^n\frac{W_i \epsilon_i}{\sqrt{|C(X_i)|}}+\frac{1}{\sqrt n}\sum_{i=1}^nT_{ni}\Bigg[\frac{\sqrt{ |V_1(X)|}}{\sqrt{|\hat{V}_1(X)|}}-\frac{\sqrt{ |V_0(X)|}}{\sqrt{|\hat{V}_0(X)|}}\Bigg] \frac{W_i D_i\epsilon_i}{\sqrt{|C(X_i)|}}+o_p(1),
\end{multline*}
where the last step comes from \Cref{lemma5} and the fact that $\frac{S}{\hat S}-1= \frac{\sqrt{ |V_0(X)|}}{\sqrt{|\hat{V}_0(X)|}}-1 +\Bigg[\frac{\sqrt{ |V_1(X)|}}{\sqrt{|\hat{V}_1(X)|}}-\frac{\sqrt{ |V_0(X)|}}{\sqrt{|\hat{V}_0(X)|}}\Bigg] \times D$.   Applying a Taylor expansion, we have
\[
T_{ni}\frac{\sqrt{ |V_d(X_i)|}}{\sqrt{|\hat{V}_d(X_i)|}}=T_{ni}\left\{1- \frac{1}{2V_d(X_i)}   \big[\hat {V}_d(X_i)-V_d(X_i)\big]\right\}+o_p(n^{-1/2})
\]where the  $o_p$ term holds uniformly over $i$ by \Cref{theorem1}.   Hence, we have
\begin{multline}
\label{eq8}
\frac{1}{\sqrt n}\sum_{i=1}^n \frac{T_{ni}W_i\sigma(D_i,X_i)\epsilon_i}{\hat S_i}=\frac{1}{\sqrt n}\sum_{i=1}^n\frac{W_i \epsilon_i}{\sqrt{|C(X_i)|}}\\
  +\frac{1}{2\sqrt n}\sum_{i=1}^n \left\{\frac{W_iD_i\epsilon_i}{\sqrt{|C(X_i)|}}\times T_{ni}\Big[ \frac{\hat {V}_0(X_i)- V_0(X_i)}{V_0(X_i)}
-\frac{\hat {V}_1(X_i)- V_1(X_i)}{V_1(X_i)}\Big]\right\}+o_p(1).
\end{multline} 

Let $\tilde T_{ni}=\mathbbm 1\left(|\varphi_{ni}|\geq \tau_n; \ | {V}_0(X_i)|\geq \kappa_{0n}; \ | {V}_1(X_i)|\geq \kappa_{1n};\ X_i\in\mathscr X_n\right)$.  By a similar argument to \citet[Lemma B.7]{wan2015inference} and Bernstein's tail inequality, we have
\begin{multline}
\label{eq9}
\frac{1}{\sqrt n}\sum_{i=1}^n \left\{\frac{W_iD_i\epsilon_i}{\sqrt{|C(X_i)|}}\times T_{ni}\Big[ \frac{\hat {V}_0(X_i)- V_0(X_i)}{V_0(X_i)}
-\frac{\hat {V}_1(X_i)- V_1(X_i)}{V_1(X_i)}\Big]\right\}\\
=\frac{1}{\sqrt n}\sum_{i=1}^n \left\{\frac{W_iD_i\epsilon_i}{\sqrt{|C(X_i)|}}\times \tilde T_{ni}\Big[ \frac{\hat {V}_0(X_i)- V_0(X_i)}{V_0(X_i)}
-\frac{\hat {V}_1(X_i)- V_1(X_i)}{V_1(X_i)}\Big]\right\}+o_p(1).
\end{multline}
Let $A(X_i)=f_X(X_i)\text{Cov}(D_i,Z_i|X_i)$. By \Cref{lemma6}, we have
\begin{eqnarray*}
&&\frac{1}{\sqrt n}\sum_{i=1}^n \left\{\frac{W_iD_i\epsilon_i}{\sqrt{|C(X_i)|}}\times \tilde T_{ni}\Big[ \frac{\hat {V}_0(X_i)- V_0(X_i)}{V_0(X_i)}
-\frac{\hat {V}_1(X_i)- V_1(X_i)}{V_1(X_i)}\Big]\right\}\\
&=&-\frac{1}{\sqrt {n} (n-1)} \sum_{i=1}^n\sum_{j\neq i} \frac{ \tilde T_{ni}W_iD_i\epsilon_i  }{A(X_i)}\big[\Psi_{ji}-\mathbb E(\Psi_{i}|X_i)\big]\big[Z_j-\mathbb E(Z_i|X_i)\big]K_h\big(X_j-X_i\big)+o_p(1).
\end{eqnarray*}
Let further $ T^*_{ni}=\mathbbm 1\left(|\varphi_{i}|\geq \tau_n; \ | {V}_0(X_i)|\geq \kappa_{0n}; \ | {V}_1(X_i)|\geq \kappa_{1n};\ X_i\in\mathscr X_n\right)$, where $ \varphi_i=\phi_1(X_i) \phi_{DZ}(X_i)- \phi_{D}(X_i) \phi_{Z}(X_i)$. By \Cref{ass3}-(ii) and \Cref{ass4}, $T^*_{ni}=\mathbbm 1(X_i\in \mathscr X_n)$ for sufficiently large $n$. Thus, 
\begin{eqnarray*}
&&\frac{1}{\sqrt n}\sum_{i=1}^n \left\{\frac{W_iD_i\epsilon_i}{\sqrt{|C(X_i)|}}\times \tilde T_{ni}\Big[ \frac{\hat {V}_0(X_i)- V_0(X_i)}{V_0(X_i)}
-\frac{\hat {V}_1(X_i)- V_1(X_i)}{V_1(X_i)}\Big]\right\}\\
&=&-\frac{1}{\sqrt {n} (n-1)} \sum_{i=1}^n\sum_{j\neq i} \frac{ T^*_{ni}W_iD_i\epsilon_i  }{A(X_i)}\big[\Psi_{ji}-\mathbb E(\Psi_{i}|X_i)\big]\big[Z_j-\mathbb E(Z_i|X_i)\big]K_h\big(X_j-X_i\big)+o_p(1).
\end{eqnarray*}

Following the Hoeffding's Decomposition in \cite{powell1989semiparametric}, we have 
\begin{eqnarray*}
&& \frac{1}{\sqrt {n} (n-1)}\sum_{i=1}^n\sum_{j\neq i} \frac{ T^*_{ni}W_iD_i\epsilon_i  }{A(X_i)}\big[\Psi_{ji}-\mathbb E(\Psi_{i}|X_i)\big]\big[Z_j-\mathbb E(Z_i|X_i)\big]K_h\big(X_j-X_i\big)\\
&=&\frac{1}{\sqrt n }\sum_{j=1}^n \mathbb E\left\{ \frac{ T^*_{ni}W_iD_i\epsilon_i  }{A(X_i)}\big[\Psi_{ji}-\mathbb E(\Psi_{i}|X_i)\big]\big[Z_j-\mathbb E(Z_i|X_i)\big]K_h\big(X_j-X_i\big)\Bigg|\mathscr F_j\right\}+o_p(1)\\
&=&\frac{1}{\sqrt n }\sum_{j=1}^n \frac{ \mathbb E(W_jD_j\epsilon_j|X_j)  }{\text{Cov}(D_j,Z_j|X_j)}\big[\Psi_{j}-\mathbb E(\Psi_{j}|X_j)\big]\big[Z_j-\mathbb E(Z_j|X_j)\big]+o_p(1).
\end{eqnarray*} where the last step uses a  similar argument to \Cref{lemma5}.

Thus, we have
\[
\frac{1}{\sqrt n}\sum_{i=1}^n \frac{T_{ni}W_i\sigma(D_i,X_i)\epsilon_i}{\hat S_i}=\frac{1}{\sqrt n}\sum_{i=1}^n\frac{W_i \epsilon_i}{\sqrt{|C(X_i)|}} - \frac{1}{\sqrt n}\sum_{i=1}^n \zeta_i+o_p(1).
\]
The results then simply follow from the CLT and Slutsky's Theorem. \qed

\section{Technical Lemmas}
\begin{lemma}\label{lemma3}
Suppose the assumptions in \Cref{theorem1} hold. Then, 
\[
\frac{1}{n}\sum_{i=1}^n\frac{T_{ni}W_i(X'_i, D_i)}{S_i}=\mathbb E \Big[\frac{W(X',D)}{S}\Big]+o_p(1)
\]
\proof
Because
\begin{multline*}
\frac{1}{n}\sum_{i=1}^n\frac{T_{ni}W_i(X'_i, D_i)}{S_i}=\frac{1}{n}\sum_{i=1}^n\frac{W_i(X'_i, D_i)}{S_i}+ \frac{1}{n}\sum_{i=1}^n(T_{ni}-1)\frac{W_i(X'_i, D_i)}{S_i}\\
=\mathbb E \left[\frac{W(X',D)}{S}\right]+ \frac{1}{n}\sum_{i=1}^n(T_{ni}-1)\frac{W_i(X'_i, D_i)}{S_i}+o_p(1)
\end{multline*} where the last step comes from  the WLLN. By the Cauchy-Schwarz inequality, 
\[
\mathbb E\left\|\frac{1}{n}\sum_{i=1}^n(T_{ni}-1)\frac{W_i(X'_i, D_i)}{S_i}\right\|=\mathbb E\left\|(T_{ni}-1)\frac{W_i(X'_i, D_i)}{S_i}\right\|\leq\left\{\mathbb E\left[\frac{\|W_i(X'_i,D_i)\|^2}{S_i^2} \right]\times \mathbb E (T_{ni}-1)^2\right\}^{1/2}.
\]
Because of \Cref{ass4,ass10,ass11} and $\mathscr X_n\rightarrow \mathscr S_X$, we have
\[
\mathbb E (T_{ni}-1)^2\leq \Pr(|\phi_{D}(X_i)\phi_Z(X_i)|< \tau_n)+\Pr( |\hat {V}_0(X_i)|\geq\kappa_{0n})+\Pr( |\hat {V}_1(X_i)|\geq\kappa_{1n})+\mathbbm 1(X_i\in\mathscr X^c_n)\rightarrow 0.
\] By \Cref{ass8}, 
\[
\mathbb E\left\|\frac{1}{n}\sum_{i=1}^n(T_{ni}-1)\frac{W_i(X'_i, D_i)}{S_i}\right\|\rightarrow 0.\qed
\]
\end{lemma}

\begin{lemma}\label{lemma4}
 Suppose  the assumptions in \Cref{theorem1} hold. Then, 
\[
\frac{1}{n}\sum_{i=1}^nT_{ni}\Big(\frac{S_i}{\hat S_i}-1\Big)\frac{W_i(X'_i, D_i)}{S_i}=o_p(1)
\]
\proof By Cauchy Schwarz inequality, 
\[
\mathbb E \left\|\frac{1}{n}\sum_{i=1}^nT_{ni}\Big(\frac{S_i}{\hat S_i}-1\Big)\frac{W_i(X'_i, D_i)}{S_i}\right\|
\leq \left\{\mathbb E \left[T_{ni}\Big(\frac{S_i}{\hat S_i}-1\Big)^2\right]\times \mathbb E \left\|\frac{W_i(X'_i, D_i)}{S_i}\right\|^2\right\}^{-1/2}.
\]
By \Cref{lemma2} and \cref{ass11}-(ii),  $\mathbb E\left[T_{ni} \big(\frac{S_i}{\hat S_i}-1\big)^2\right] \rightarrow 0$.
 \qed

\end{lemma}

\begin{lemma}
\label{lemma6}
Suppose all the assumptions in \Cref{lemma2} and \Cref{ass12} hold. Then,
\begin{multline*}
\frac{\hat {V}_0(X_i)-V_0(X_i)}{V_0(X_i)}-\frac{\hat {V}_1(X_i)-V_1(X_i)}{V_1(X_i)}\\
=\frac{1}{A(X_i)}\times \frac{1}{n-1}  \sum_{j\neq i}\big[\big(\Psi_{ji}-\mathbb E (\Psi_{i}|X_i)\big)(Z_j-\mathbb E (Z_i|X_i))K_h\big(X_j-X_i\big)-\text{Cov} (\Psi_{i},Z_i|X_i)f_X(X_i)\big]  +o_p(n^{-1/2})\end{multline*}where the $o_p(\cdot)$ term holds uniformly over $i$, and $A(X_i)\equiv f_X(X_i)\text{Cov}(D_i,Z_i|X_i)$. 
\proof  Let $A(X_i)=f_X(X_i)\text{Cov}(D_i,Z_i|X_i)$. By Taylor expansion, we have 
\begin{eqnarray*}
&& \frac{\hat \phi_1(X_i)\hat \phi_{Y^2DZ}(X_i)-\hat \phi_{Y^2D}(X_i)\hat \phi_{Z}(X_i)}{\hat \phi_1(X_i)\hat \phi_{DZ}(X_i)-\hat \phi_{D}(X_i)\hat \phi_{Z}(X_i)}-\frac{\phi_1(X_i)\phi_{Y^2DZ}(X_i)- \phi_{Y^2D}(X_i) \phi_{Z}(X_i)}{ \phi_1(X_i) \phi_{DZ}(X_i)- \phi_{D}(X_i)\phi_{Z}(X_i)}\\
&=&\frac{1}{A(X_i)}\times \frac{1}{n-1}  \sum_{j\neq i}\big[Y_j^2 D_jZ_jK_h\big(X_j-X_i\big)-\mathbb E (Y^2_iD_iZ_i|X_i)f_X(X_i)\big] \\
&+&\frac{1}{A(X_i)}\times \frac{\mathbb E (Y_i^2D_iZ_i|X_i)}{n-1}  \sum_{j\neq i}\big[K_h\big(X_j-X_i\big)-f_X(X_i)\big] \\
&-&\frac{1}{A(X_i)}\times \frac{ \mathbb E (Z_i|X_i)}{n-1} \sum_{j\neq i}\big[Y_j^2 D_jK_h\big(X_j-X_i\big)-\mathbb E (Y^2_iD_i|X_i)f_X(X_i)\big] \\
&-&\frac{1}{A(X_i)}\times \frac{ \mathbb E (Y^2_iD_i|X_i)}{n-1} \sum_{j\neq i}\big[Z_jK_h\big(X_j-X_i\big)-\mathbb E (Z_i|X_i)f_X(X_i)\big] \\
&-&\frac{V_1(X_i)+\delta_1^2(X_i)}{A(X_i)}\times \frac{1}{n-1}  \sum_{j\neq i}\big[ D_jZ_jK_h\big(X_j-X_i\big)-\mathbb E (D_iZ_i|X_i)f_X(X_i)\big] \\
&-&\frac{V_1(X_i)+\delta_1^2(X_i)}{A(X_i)}\times\frac{ \mathbb E (D_iZ_i|X_i)}{n-1} \sum_{j\neq i}\big[K_h\big(X_j-X_i\big)-f_X(X_i)\big] \\
&+&\frac{V_1(X_i)+\delta_1^2(X_i)}{A(X_i)}\times\frac{ \mathbb E (Z_i|X_i)}{n-1} \sum_{j\neq i}\big[ D_jK_h\big(X_j-X_i\big)-\mathbb E (D_i|X_i)f_X(X_i)\big] \\
&+&\frac{V_1(X_i)+\delta_1^2(X_i)}{A(X_i)}\times\frac{ \mathbb E (D_i|X_i)}{n-1} \sum_{j\neq i}\big[Z_jK_h\big(X_j-X_i\big)-\mathbb E (Z_i|X_i)f_X(X_i)\big]
+o_p(n^{-1/2}),
\end{eqnarray*}where all higher order terms are of $o_p(n^{-1/2})$ uniformly over $i$ due to  a similar argument to \Cref{lemma2} and \Cref{ass12}. Similarly, we obtain Taylor expansions for 
\[
\frac{\hat \phi_1(X_i)\hat \phi_{Y^2(1-D)Z}(X_i)-\hat \phi_{Y^2(1-D)}(X_i)\hat \phi_{Z}(X_i)}{\hat \phi_1(X_i)\hat \phi_{DZ}(X_i)-\hat \phi_{D}(X_i)\hat \phi_{Z}(X_i)}-\frac{\phi_1(X_i)\phi_{Y^2(1-D)Z}(X_i)- \phi_{Y^2(1-D)}(X_i) \phi_{Z}(X_i)}{ \phi_1(X_i) \phi_{DZ}(X_i)- \phi_{D}(X_i)\phi_{Z}(X_i)}
\] and $\hat \delta_d(X_i)-\delta_d(X_i)$.

It follows that 
\begin{eqnarray*}
&&\frac{\hat {V}_1(X_i)-V_1(X_i)}{V_1(X_i)}\\
&=&\frac{1}{A(X_i)}\times \frac{1}{n-1}  \sum_{j\neq i}\big[\big(\Psi_{1ji}-\mathbb E (\Psi_{1i}|X_i)\big)(Z_j-\mathbb E (Z_i|X_i))K_h\big(X_j-X_i\big)-\text{Cov} (\Psi_{1i},Z_i|X_i)f_X(X_i)\big] \\
&-&\frac{1}{A(X_i)}\times \frac{1}{n-1}  \sum_{j\neq i}\big[\big(D_{j}-\mathbb E (D_i|X_i)\big)(Z_j-\mathbb E (Z_i|X_i))K_h\big(X_j-X_i\big)-\text{Cov} (D_i,Z_i|X_i)f_X(X_i)\big]\\
&+&\frac{\text{Cov} (\Psi_{1i},Z_i|X_i)-\text{Cov}(D_i,Z_i|X_i)}{A(X_i)}\times \frac{1}{n-1}  \sum_{j\neq i}\big[K_h\big(X_j-X_i\big)-f_X(X_i)\big]  +o_p(n^{-1/2}).
\end{eqnarray*} Similarly, we obtain $\frac{\hat {V}_0(X_i)-V_0(X_i)}{V_0(X_i)}$. Because $\text{Cov} (\Psi_{1i},Z_i|X_i)+\text{Cov} (\Psi_{0i},Z_i|X_i)=\text{Cov} (\Psi_{i},Z_i|X_i)=0$, $\text{Cov} (D_i,Z_i|X_i)+\text{Cov} (1-D_i,Z_i|X_i)=0$, and the result obtains. \qed
\end{lemma}

\begin{lemma}\label{lemma5}  
Suppose  the assumptions in \Cref{theorem2}   hold. Then, 
\[
\frac{1}{\sqrt n}\sum_{i=1}^nT_{ni}\left[\frac{\sqrt{ |V_0(X_i)|}}{\sqrt{|\hat{V}_0(X_i)|}}-1\right] \frac{W_i  \epsilon_i}{\sqrt{|C(X_i)|}}=o_p(1)
\]and
\[
\frac{1}{\sqrt n}\sum_{i=1}^n(T_{ni}-1)\frac{W_i \epsilon_i}{\sqrt{|C(X_i)|}}=o_p(1).
\]
\proof
Note that $\mathbb E [\frac{W  \epsilon}{\sqrt{|C(X)|}}\big|X]=0$. Then the result directly follows e.g. \cite{andrews1994asymptotics} or \citet[Theorem 8.1]{newey1994large}.\qed
\end{lemma}

\subsection{Proof of \Cref{lemma1}}\label{lemma1_proof}
\proof We first show the first half. It suffices to show the if part. By definition, 
\begin{align*}
&\tilde r_1(X)=\mu(1,X)-\mu(0,X)+[\sigma(1,X)-\sigma(0,X)]\times  \frac{\mathbb E (D\epsilon|Z=1)-\mathbb E \left(D\epsilon|Z=0\right)}{p(X,1)-p(X,0)};\\
&\tilde r_0(X)=\mu(0,X)- [\sigma(1,X)-\sigma(0,X)]\times\frac{\mathbb E (\epsilon D|X,Z=1)p(X,0)-\mathbb E (\epsilon D|X,Z=0)p(X,1)}{p(X,1)-p(X,0)}.
\end{align*}
Under the condition $\mathbb E (\tilde \epsilon^2|X,Z=1)=\mathbb E (\tilde \epsilon^2|X,Z=0)$, we have 
\[
\frac{\mathbb E\left\{ \big[Y-\tilde r_0(X)-\tilde r_1(X)D\big]^2\big|X,Z=1\right\}-\mathbb E\left\{ \big[Y-\tilde r_0(X)-\tilde r_1(X)D\big]^2\big|X,Z=0\right\}}{p(X,1)-p(X,0)}=0.
\]Plug \eqref{eq1} into the above equation, so that  
\begin{eqnarray*}
0&=&[\mu (1,X)-\mu(0,X)-\tilde r_1(X)]^2+[\sigma(1,X)-\sigma(0,X)]^2\times \xi_2(X)\\
&+&2[\mu(0,X)-\tilde r_0(X)] \times [\mu (1,X)-\mu(0,X)-\tilde r_1(X)]\\
&+& 2[\mu(0,X)-\tilde r_0(X)]\times [\sigma (1,X)-\sigma(0,X)]\times \xi_1(X)\\
&+& 2\sigma(0,X) \times  [\mu (1,X)-\mu(0,X)-\tilde r_1(X)]\times \xi_1(X)\\
&+&2[\mu (1,X)-\mu(0,X)-\tilde r_1(X)]\times [\sigma(1,X)-\sigma(0,X)]\times \xi_1(X)\\
&+&2\sigma(0,X)\times[\sigma(1,X)-\sigma(0,X)]\times \xi_2(X)\\
&=&[\sigma(1,X)-\sigma(0,X)]^2\times \xi^2_1(X)\\
&-&2 [\sigma(1,X)-\sigma(0,X)]^2\times\frac{\mathbb E (\epsilon D|X,Z=1)p(X,0)-\mathbb E (\epsilon D|X,Z=0)p(X,1)}{p(X,1)-p(X,0)}\times \xi_1(X)\\
&+&2[\sigma(1,X)-\sigma(0,X)]^2\times \frac{\mathbb E (\epsilon D|X,Z=1)p(X,0)-\mathbb E (\epsilon D|X,Z=0)p(X,1)}{p(X,1)-p(X,0)}\times \xi_1(X)\\
&-&2[\sigma(1,X)-\sigma(0,X)]^2\times\xi_1^2(X)+[\sigma^2(1,X)-\sigma^2(0,X)]\times\xi_2(X)\\
&=&[\sigma^2(1,X)-\sigma^2(0,X)]\times C(X).
\end{eqnarray*} Under \Cref{rank}, it follows that $\sigma(0,X)=\sigma(1,X)$.

We now show the second half. Again, the only if part is straightforward and it suffices to show the if part. Suppose $\mathbb E \big[(Y- X'\tilde\beta_1-\tilde \beta_2 D)^2|X,Z=0\big]=\mathbb E \big[(Y- X'\tilde\beta_1-\tilde \beta_2 D)^2|X,Z=1\big]$ holds for $\tilde \beta$ satisfying $\mathbb E (Y- X'\tilde\beta_1-\tilde \beta_2 D|X,Z)=0$. Then, it follows that 
\begin{eqnarray*}
0&=&\mathbb E\left\{ \big[X'(\beta_1-\tilde \beta_1)+(\beta_2-\tilde\beta_2) D+\sigma(0,X) \epsilon+(\sigma(1,X)-\sigma(0,X))D\epsilon\big]^2\big|X,Z=1\right\}\\
&- &\mathbb E\left\{ \big[X'(\beta_1-\tilde \beta_1)+(\beta_2-\tilde\beta_2) D+\sigma(0,X) \epsilon+ (\sigma(1,X)-\sigma(0,X))D\epsilon\big]^2\big|X,Z=0\right\}.
\end{eqnarray*}
Dividing both sides by $p(X,1)-p(X,0)$, we have 
\begin{eqnarray*}
0&=&(\beta_2-\tilde\beta_2)^2+[\sigma(1,X)-\sigma(0,X)]^2 \xi_2(X)\\
&+& 2X'(\beta_1-\tilde \beta_1) (\beta_2-\tilde\beta_2)+2  \big[X'(\beta_1-\tilde \beta_1)+(\beta_2-\tilde\beta_2) \big] [\sigma(1,X)-\sigma(0,X)]\xi_1(X)\\
&+& 2 \sigma(0,X)(\beta_2-\tilde\beta_2) \xi_1(X)+2\sigma(X,0)[\sigma(1,X)-\sigma(0,X)]\xi_2(X).
\end{eqnarray*}
Since  $\mathbb E (Y- X'\tilde\beta_1-\tilde \beta_2 D|X,Z)=0$,  
\begin{multline*}
\mathbb E \big[X'(\beta_1- \tilde\beta_1)+(\beta_2-\tilde \beta_2) D+(\sigma(1,X)-\sigma(0,X))D\epsilon|X,Z=1\big]\\
-\mathbb E \big[X'(\beta_1- \tilde\beta_1)+(\beta_2-\tilde \beta_2) D+(\sigma(1,X)-\sigma(0,X))D\epsilon|X,Z=0\big]=0.
\end{multline*}Therefore, $\beta_2-\tilde\beta_2=-[\sigma(1,X)-\sigma(0,X)]\times \xi_1(X)$. It follows that 
\begin{eqnarray*}
0&=&[\sigma(1,X)-\sigma(0,X)]^2\xi^2_1(X)+[\sigma(1,X)-\sigma(0,X)]^2 \xi_2(X)-2  [\sigma(1,X)-\sigma(0,X)]^2\xi^2_1(X)\\
&-& 2\sigma(X,0)[\sigma(1,X)-\sigma(0,X)]\xi_1(X)+2\sigma(X,0)[\sigma(1,X)-\sigma(0,X)]\xi_2(X) \\
&=&[\sigma^2(1,X)-\sigma^2(0,X)] C(X).
\end{eqnarray*}Under \Cref{rank}, we have $\sigma(0,X)=\sigma(1,X)$.\qed

\clearpage

\section{Tables and figures}

 \begin{table}[ht] 
\caption{Simulation Summary of IV  Estimation (seed=7480)}
\begin{center}
{\footnotesize
\begin{tabular}{lcllcccccc}
\hline \hline
Est. & Kernel  &  Sample size  &Parameter & MB &  MEDB & SD & RMSE   \\ \hline   \\     
IV    &  NA      &1000               & $\beta_0$    & \ 0.1129  &\ 0.1095   &0.0493    &0.1232\\
       &              &                      &$\beta_1$    & -0.0001    &-0.0018   &0.0234   &0.0233\\
       &              &                      &$\beta_2$    &  -0.0720  &-0.0706    &0.0850   &0.1113\\ \\ 
                                                   
       &              &2000              &$\beta_0$    & \ 0.1085   & \ 0.1089   & 0.0344  &0.1138\\
       &              &                      &$\beta_1$    & \  0.0003   &\ 0.0028     & 0.0167  &0.0167\\
       &              &                      &$\beta_2$    & -0.0670    &-0.0678   &0.0570    &0.0879\\
                   \\
                          
       &              &4000            &$\beta_0$    & \ 0.1101   &\ 0.1090    &0.0225   &0.1124\\
       &              &                    &$\beta_1$    & \ 0.0003   &\ 0.0008    &0.0122   &0.0122\\
       &              &                    &$\beta_2$    &  -0.0673   &-0.0673    &0.0389   &0.0777\\ \hdashline
                         \\    
EHIV&$K_G$  &1000  & $\beta_0$     &\  0.0242    &\  0.0140   & 0.0468    &  0.0526  \\
        &             &          & $\beta_1$     &\ 0.0017    &\  0.0024    & 0.0606    & 0.0605\\
        &             &          & $\beta_2$     & -0.0271    &-0.0199     &  0.0868   & 0.0909\\   \\

       &              &2000  & $\beta_0$    &\  0.0140    &\ 0.0096     & 0.0287    & 0.0319\\
       &              &          & $\beta_1$    &-0.0023      & -0.0048     & 0.0397       & 0.0397\\
       &              &          & $\beta_2$    &-0.0157       & -0.0133    & 0.0550       & 0.0572     \\   \\
                      
       &              &4000  & $\beta_0$    &\ 0.0077     &\ 0.0060     & 0.0158       & 0.0176\\
       &              &          & $\beta_1$    &-0.0004      &-0.0004      & 0.0245       & 0.0245\\
       &              &          & $\beta_2$    &-0.0099       &-0.0091     &0.0341        &  0.0354    \\   \\

       &$K_E$   &1000  & $\beta_0$    &\ 0.0190     &\ 0.0149    & 0.0420    & 0.0461\\
       &              &         & $\beta_1$    &- 0.0005     &-0.0020  &0.0590    & 0.0589\\
       &              &         & $\beta_2$    &-0.0208     &-0.0231  &0.0851    &0.0875\\   \\

       &              &2000  & $\beta_0$    &\ 0.0165     &\ 0.0132       &   0.0292 & 0.0335\\
       &              &          & $\beta_1$    &-0.0021      &    -0.0017    &   0.0396     & 0.0396\\
       &              &          & $\beta_2$    &-0.0230      &-0.0235        &   0.0592     &  0.0635    \\   \\
                      
       &              &4000  & $\beta_0$    &\ 0.0120     &\  0.0091       &    0.0201   & 0.0233\\
       &              &          & $\beta_1$    &\  0.0007    &\ 0.0033        &    0.0277    & 0.0277\\
       &              &          & $\beta_2$    & -0.0177      & -0.0158       &   0.0405     &   0.0442   \\                           
                         \hline \hline
       \end{tabular}}
\end{center}

\label{est: tab1}

\end{table}

\begin{figure}[ht] 
   \centering
   \includegraphics[width=2.5in,height=2in]{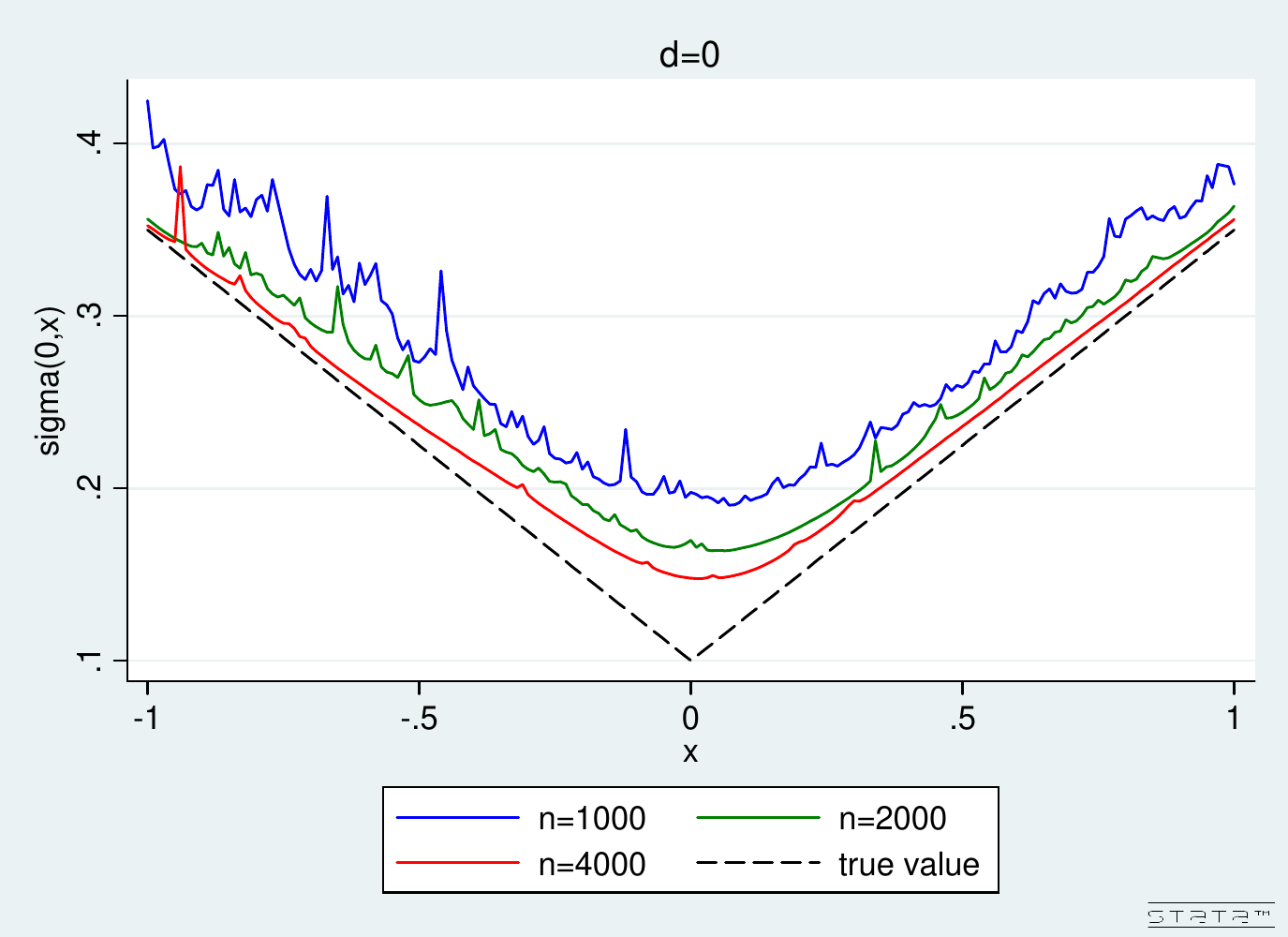}     \includegraphics[width=2.5in,height=2in]{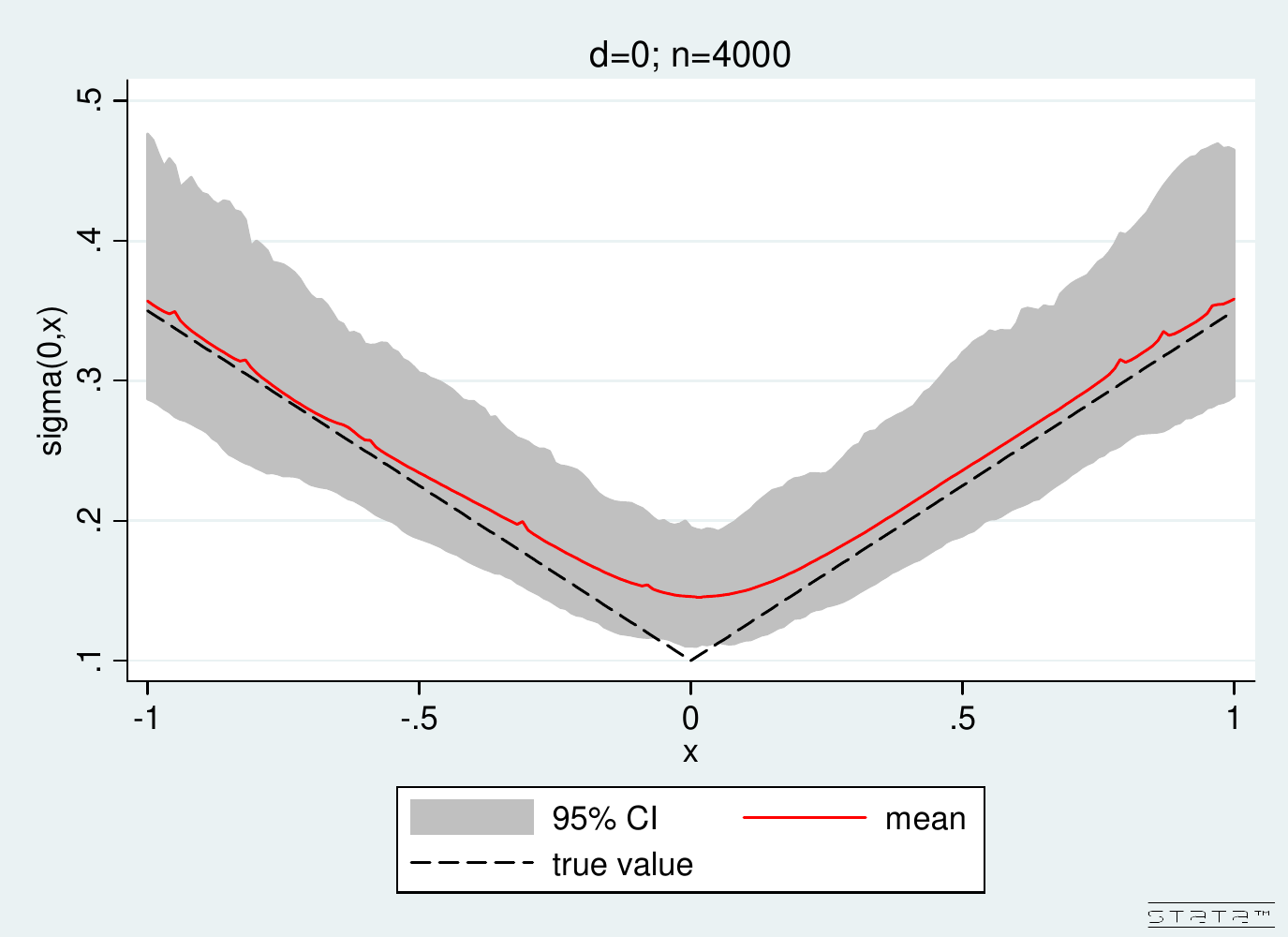}    \\
    \includegraphics[width=2.5in,height=2in]{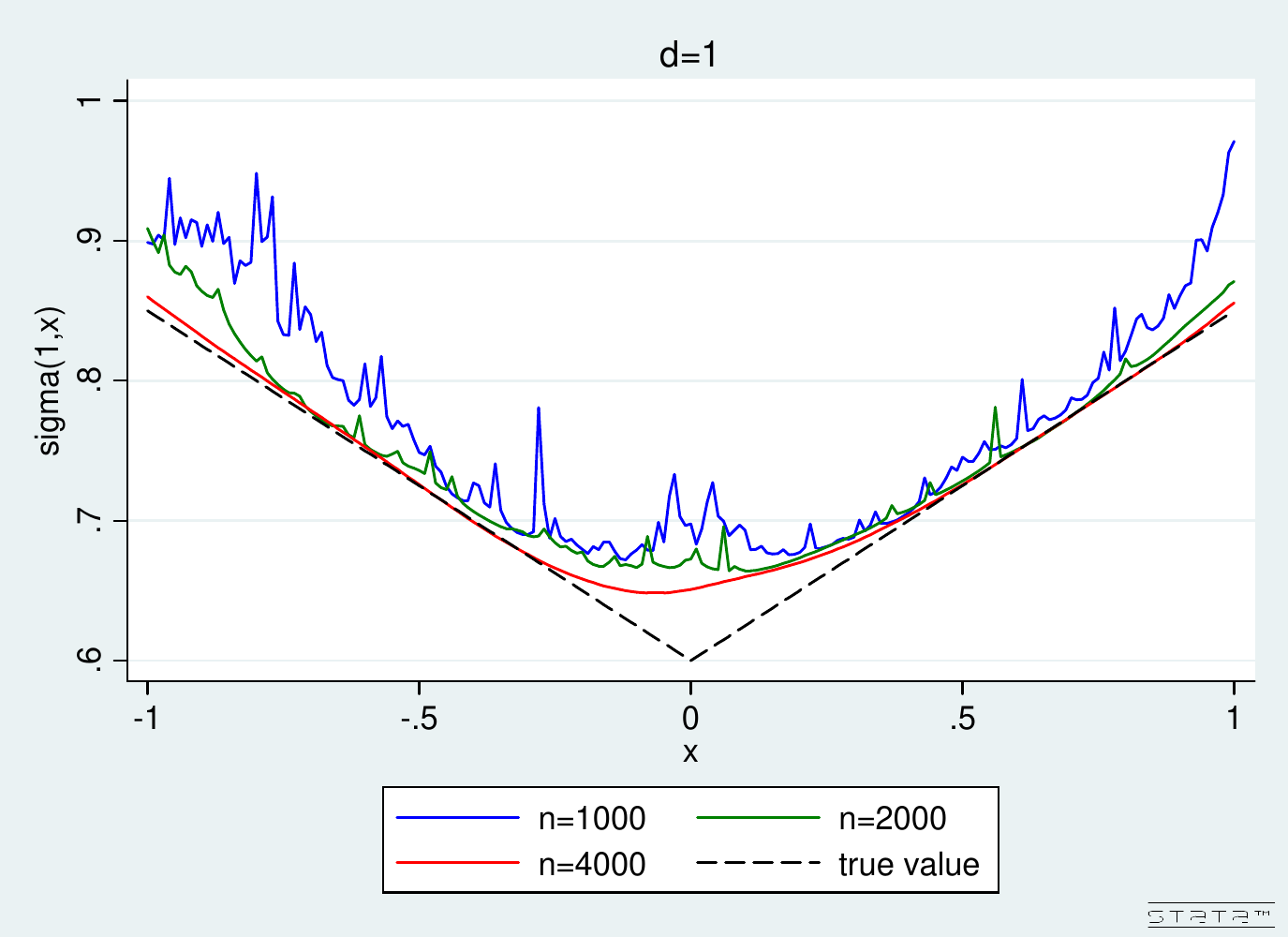}     \includegraphics[width=2.5in,height=2in]{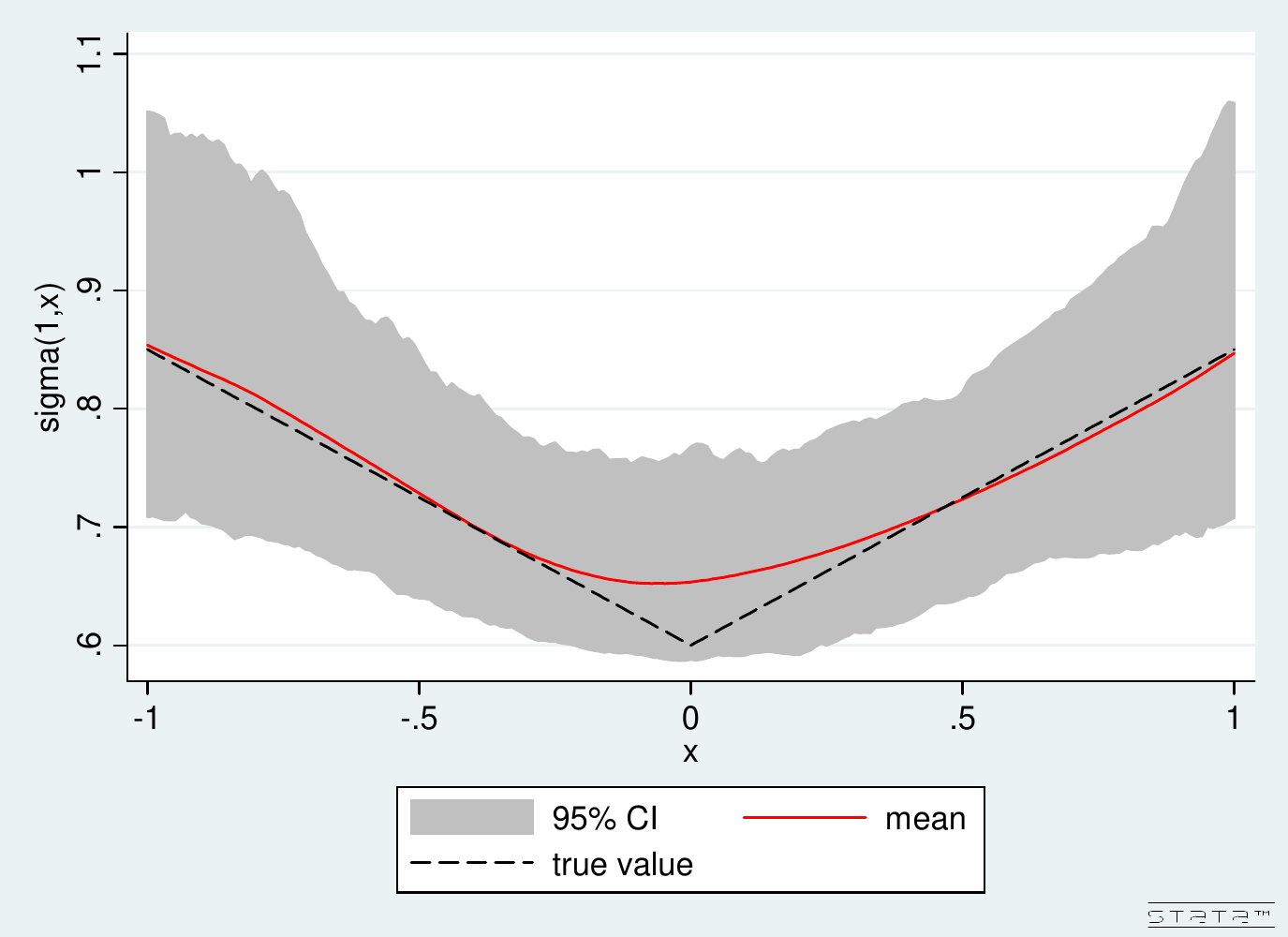}   \\
   \caption{Estimation of $\sigma(d,x)$}
   \label{fig1}
\end{figure}


\begin{figure}[ht] 
   \centering
   \includegraphics[width=2.5in,height=2in]{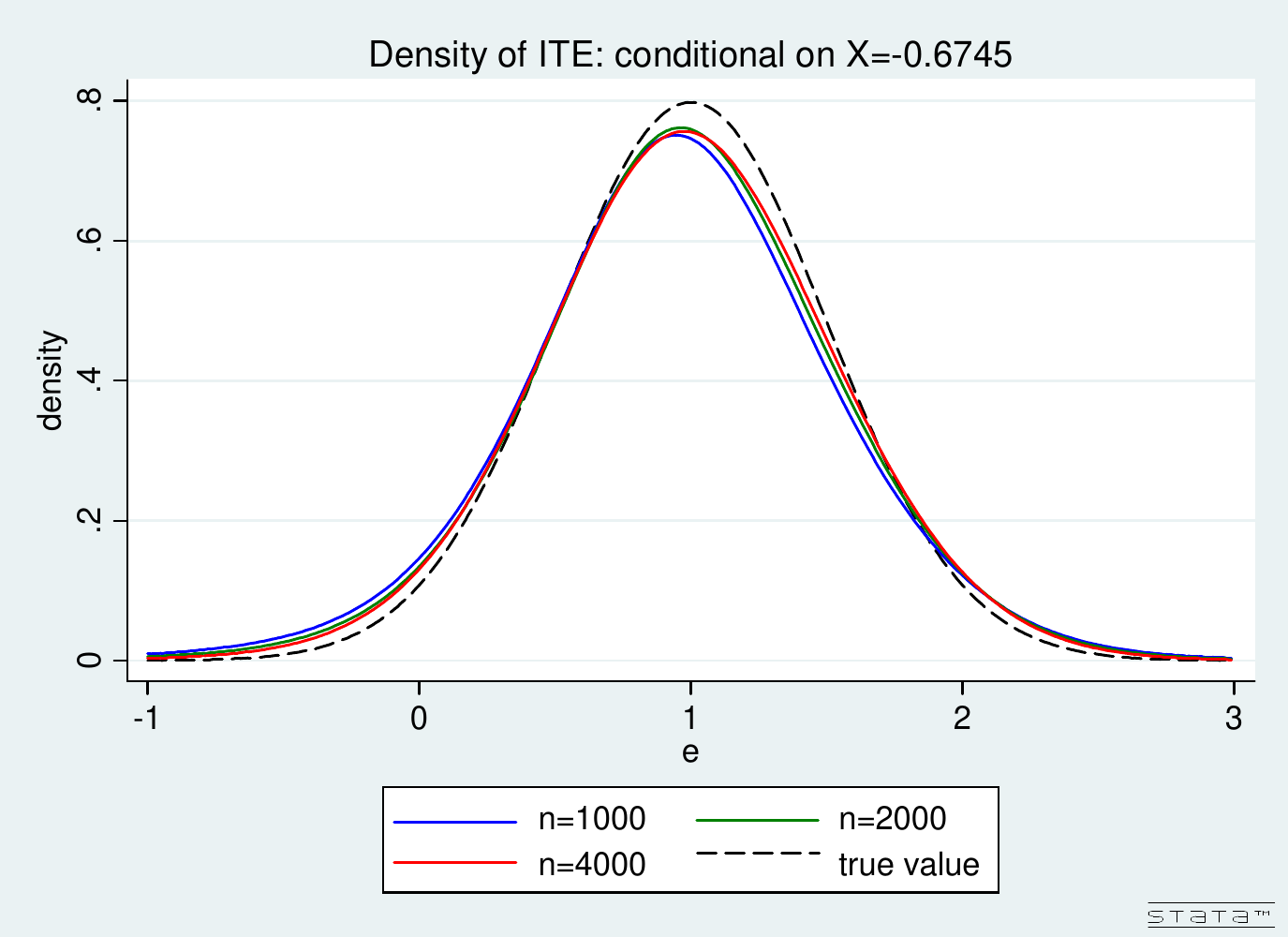}    \includegraphics[width=2.5in,height=2in]{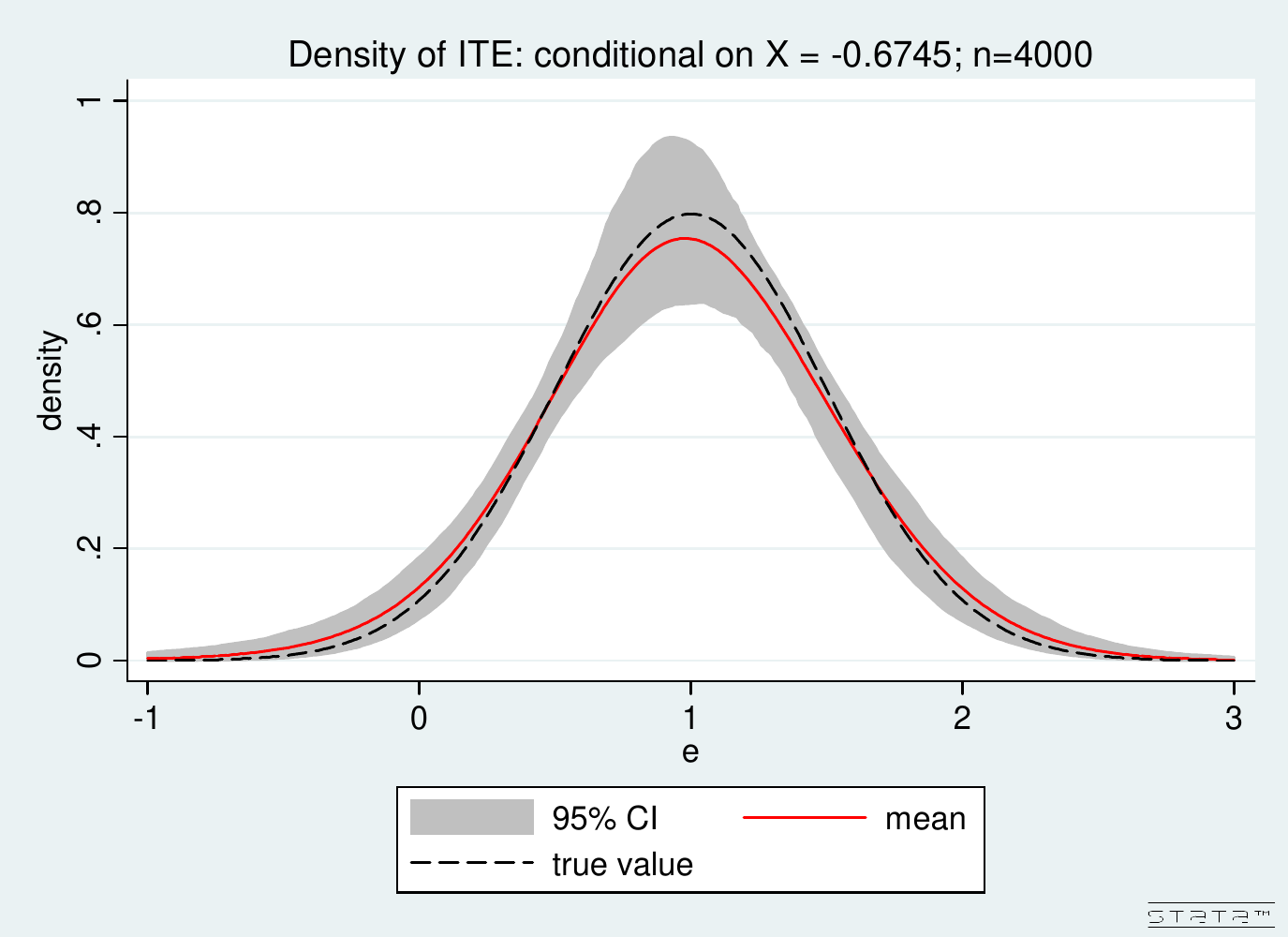} \\
      \includegraphics[width=2.5in,height=2in]{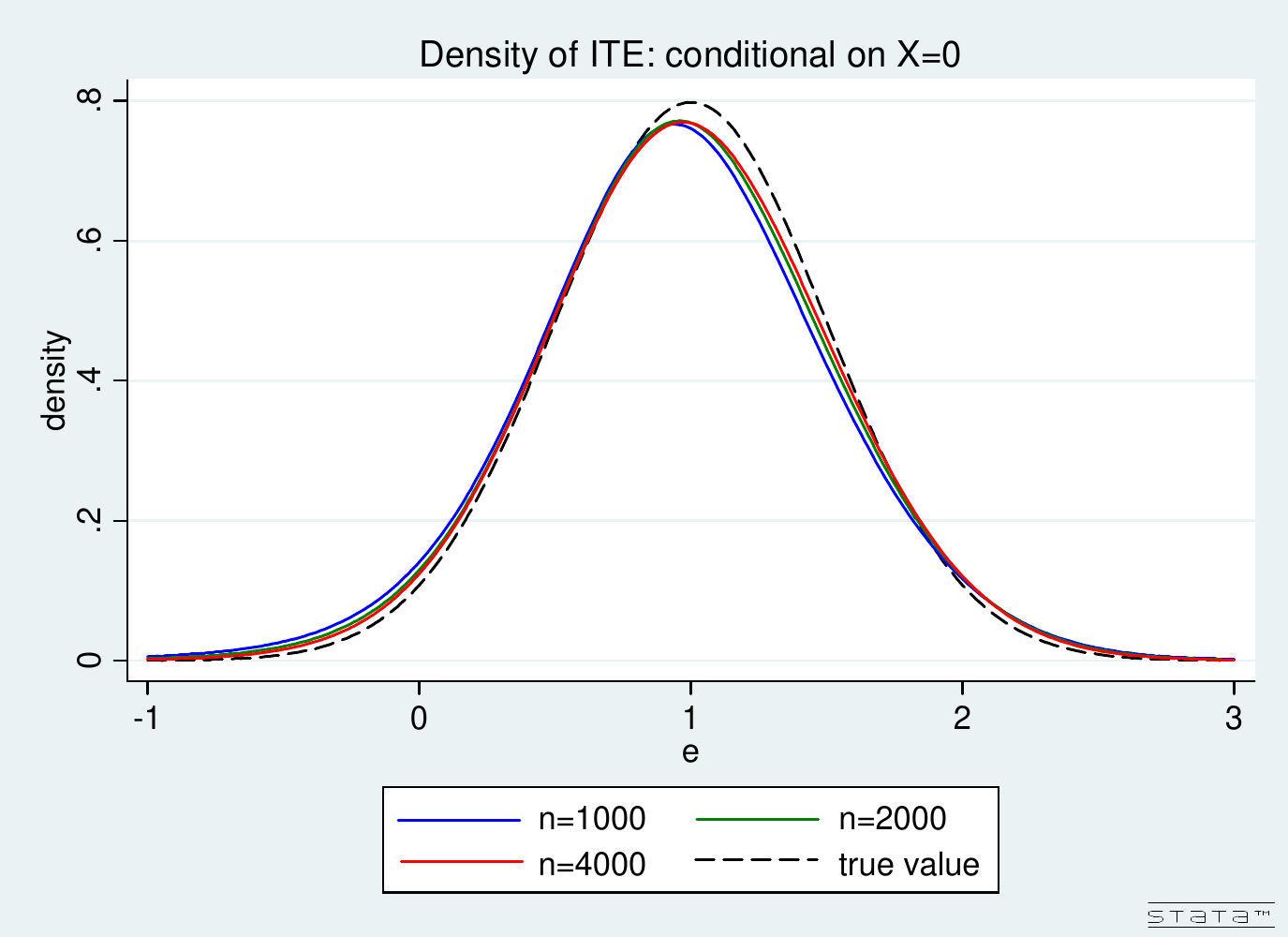}    \includegraphics[width=2.5in,height=2in]{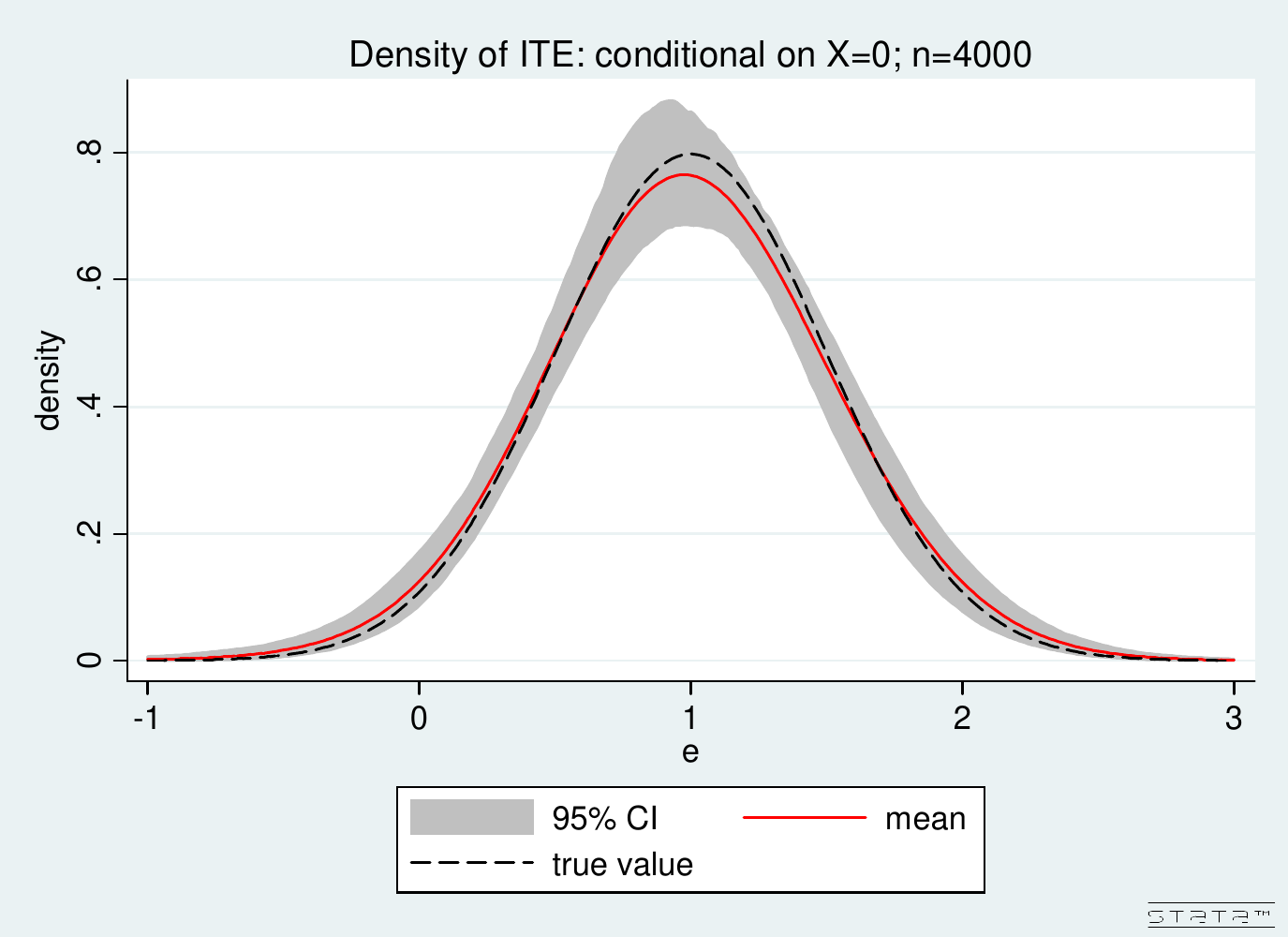} \\
         \includegraphics[width=2.5in,height=2in]{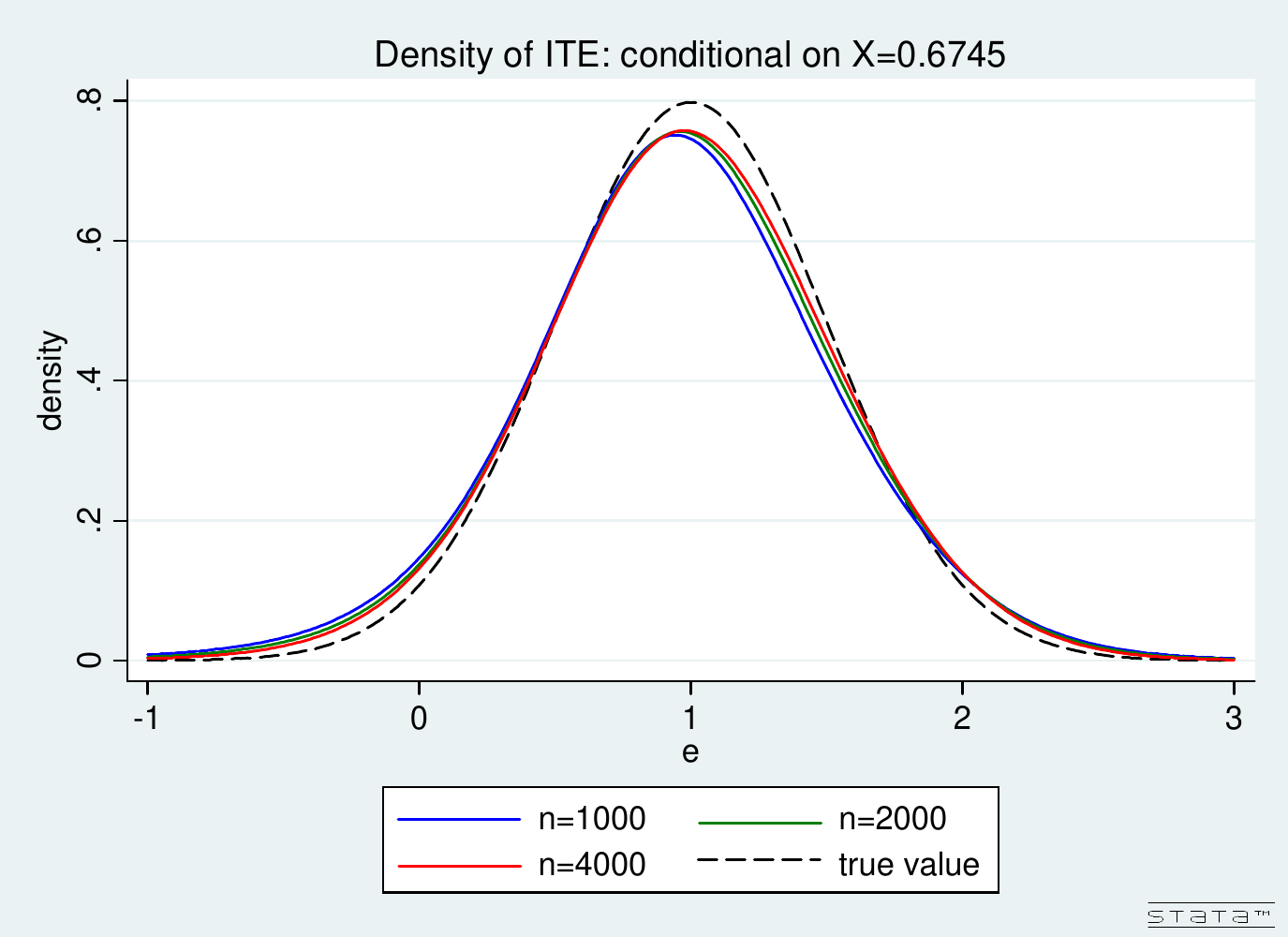}    \includegraphics[width=2.5in,height=2in]{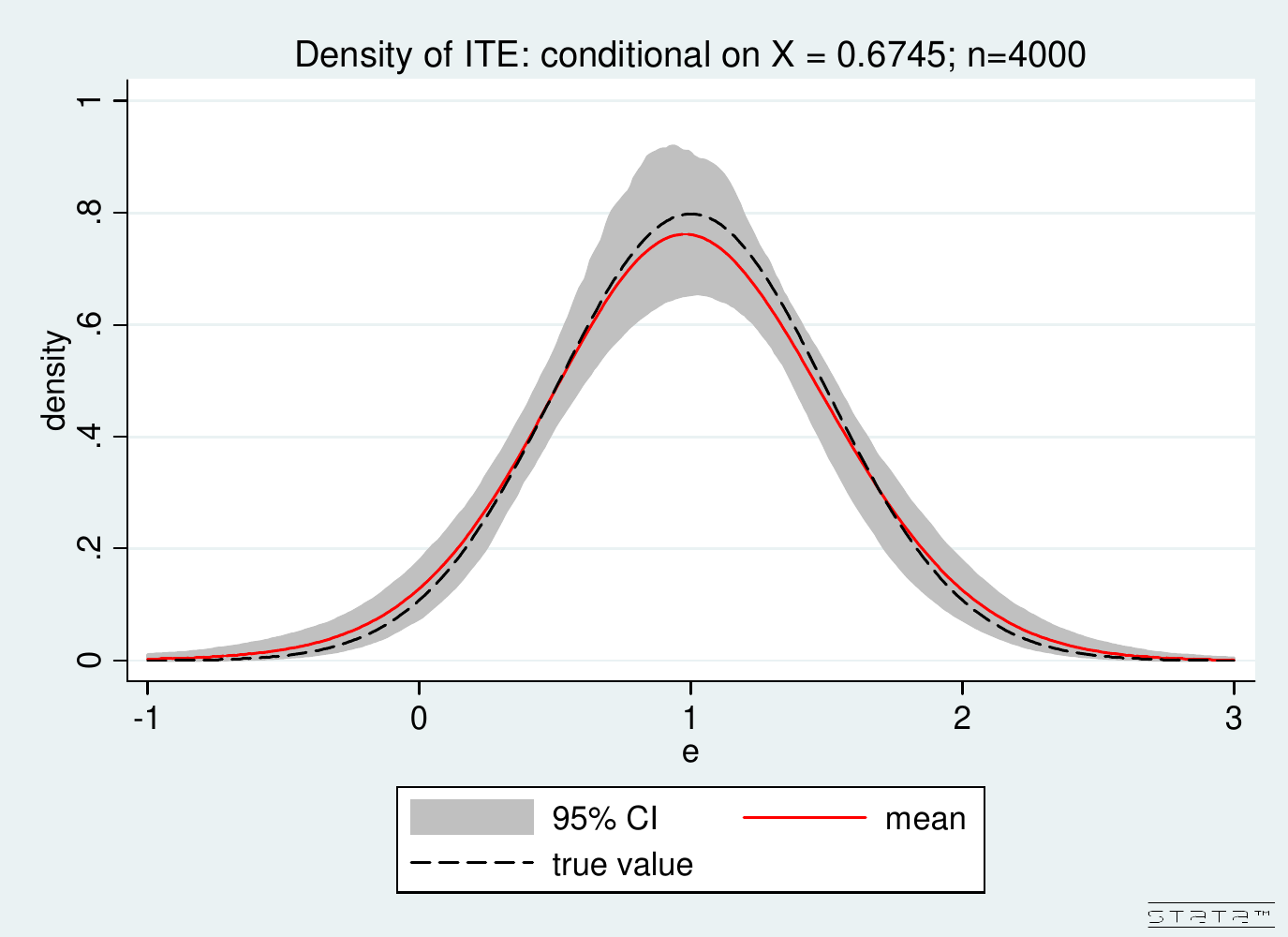} 
   \caption{Estimation of ITE's density}
   \label{fig2}
 \end{figure}

%

  \begin{table}[ht] 
\caption{ Robust check: $\hat \beta_2$,  $n=4000$ (seed=7480)}
\begin{center}
{\small
\begin{tabular}{llcccccc}
\hline \hline
$r_0$ &$\rho_0$& $\lambda_0$&& MB &  MEDB & SD & RMSE  \\ \hline\\

0.1     &     0.5           &  0.5 &&- 0.0001    &- 0.0029  &0.1433   & 0.1431\\
0.2     &                     &        &&-0.0508 &  -0.0477    &0.0918   & 0.1048  \\
0.3     &                     &        &&-0.0321  & -0.0250  & 0.0646  & 0.0721\\
0.4     &                     &        &&-0.0136  & -0.0085  &  0.0445  &0.0465   \\
0.5     &                  &       & & -0.0047  & -0.0035  & 0.0343   &0.0346\\\hdashline\\
   
   0.5     &       0.0         &   0.5     &&\ 0.0032  &\ 0.0047   & 0.0317   & 0.0318 \\
          &       0.1         &        &&\ 0.0022   &\ 0.0042   & 0.0317   &   0.0318\\
          &       0.2         &        &&\ 0.0010    & \ 0.0032    & 0.0315    &  0.0315\\  
          &       0.3         &        &&-0.0006    &   \ 0.0012  &0.0321   &0.0321\\  
          &       0.4         &        &&  -0.0022   & -0.0021    &0.0329   & 0.0329 \\
          &       0.6       &        && -0.0089     &  -0.0068   & 0.0383   & 0.0393  \\
          &       0.7       &        &&   -0.0145   & -0.0110   & 0.0438    &  0.0461\\ 
          &       0.8       &        && -0.0222    & -0.0177     &  0.0520    &  0.0564  \\
          &       0.9       &        && -0.0309  & -0.0259 & 0.0603   & 0.0677  \\                    
                  \hdashline\\
 0.5    &       0.5         &  0.00  &&  \ 0.0008    &  \ 0.0010     &  0.0180   & 0.0180\\ 
          &                    &  0.25  &&    -0.0046     &  -0.0040   & 0.0264  & 0.0268 \\ 
          &                     &  0.75  &&   -0.0045     & -0.0027   & 0.0422   & 0.0424\\ 
          &                     &  1.00  &&   -0.0042  & -0.0023 & 0.0503   & 0.0504\\ 
          &                     &  1.25  &&   -0.0040  & -0.0020  & 0.0586  & 0.0587\\
          &                     &  1.50  &&   -0.0038  & -0.0017  & 0.0673  & 0.0673 \\

          \hline \hline
       \end{tabular}}

\end{center}

\label{est: tab2}

\end{table}

\begin{table}[h] 
   \includegraphics[width=6in]{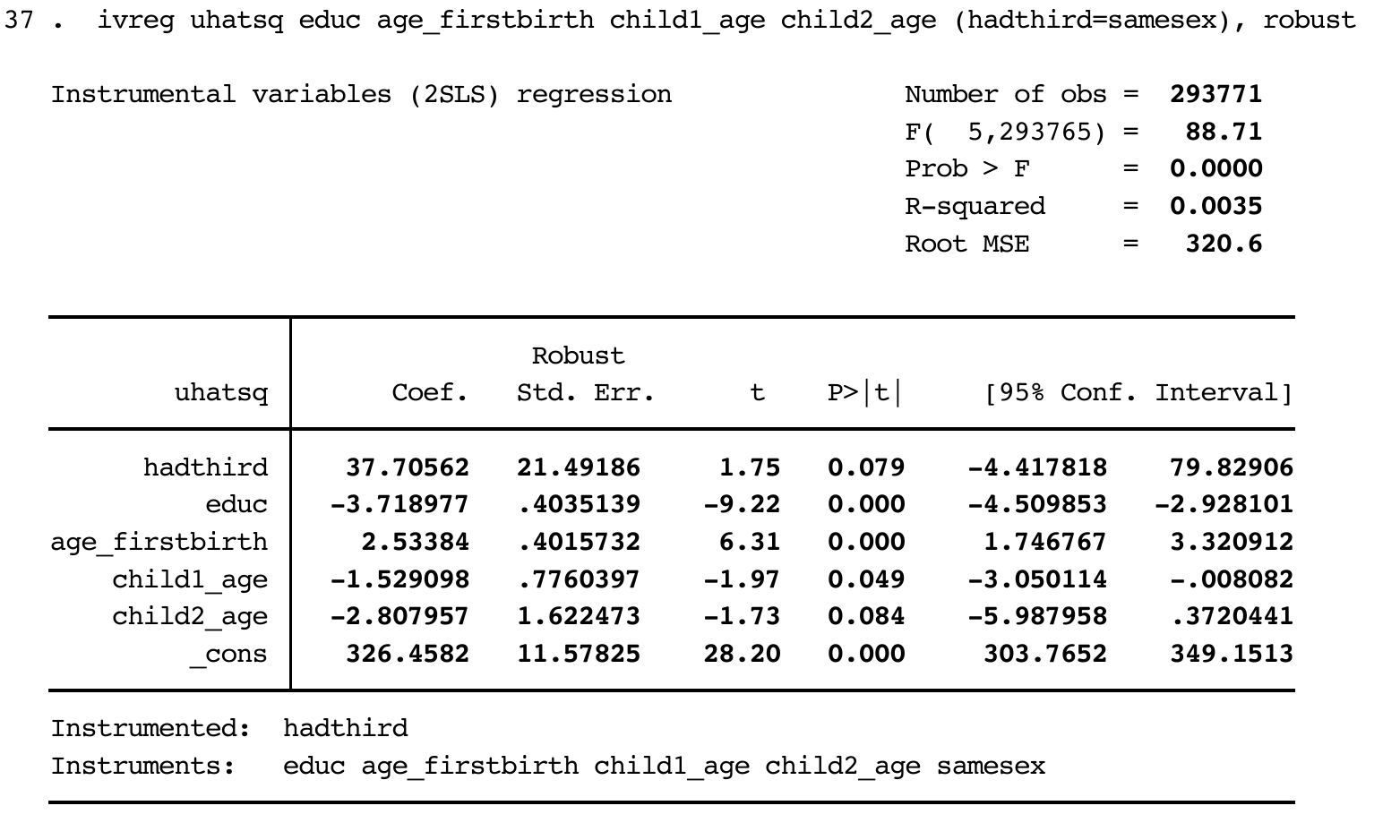} 
   \caption{\small Testing for exogenous heteroskedasticity }
   \label{fig3}
\end{table}

\end{document}